\documentstyle[12pt,bbox,epsf]{report} 
\begin{document}
\title{\LARGE Particle Interferometry\\ 
              for Relativistic Heavy-Ion Collisions\thanks{
	      Columbia University preprint CU-TP-931, 
              CERN report CERN-TH/99-15;
	      \hbox{\qquad submitted to Physics Reports} }}
\author{Urs Achim Wiedemann$^a$ and Ulrich Heinz$^{b,c}$\\
        \\
        $^a$Physics Department, Columbia University,\\
         New York, NY 10027, USA\\ 
        $^b$Theory Division, CERN, CH-1211 Geneva 23, Switzerland\\ 
        $^c$Institut f\"ur Theoretische Physik, Universit\"at Regensburg,\\
        D-93040 Regensburg, Germany}
\maketitle

\begin{abstract}
In this report we give a detailed account on Hanbury Brown/Twiss (HBT) 
particle interferometric methods for relativistic heavy-ion collisions. 
These exploit identical two-particle correlations to gain access to the 
space-time geometry and dynamics of the final freeze-out stage. The 
connection between the measured correlations in momentum space and the 
phase-space structure of the particle emitter is established, both with 
and without final state interactions. Suitable Gaussian parametrizations
for the two-particle correlation function are derived and the physical 
interpretation of their parameters is explained. After reviewing various 
model studies, we show how a combined analysis of single- and 
two-particle spectra allows to reconstruct the final state of 
relativistic heavy-ion collisions.
\end{abstract}

%
\newpage
\setcounter{page}{2}
\tableofcontents
\newpage

\chapter{Introduction}

By now, a large collection of experimental data exists from the
first relativistic collisions between truly heavy ions~\cite{QM96}, 
using the 11 GeV/nucleon gold beams from the Brookhaven AGS and the
160 GeV/nucleon lead beams from the CERN SPS. The first relativistic 
heavy-ion collider RHIC at BNL will soon start taking data at 
$\sqrt{s} = 200$ $A$ GeV, and in the next decade, the already approved 
LHC program at CERN will explore relativistic heavy-ion collisions at 
even higher energies ($\sqrt{s} = 5.5\, A$ TeV). The aim of this large 
scale experimental effort is to investigate the equilibration 
processes of hadronic matter and to test in this way the hadronic 
partition function at extreme energy densities and temperatures. 
Especially, one expects under sufficiently extreme conditions the
transition to a new state of hadronic matter, the quark gluon plasma
(QGP) in which the physical degrees of freedom of equilibration
processes are partonic rather than hadronic~\cite{S80,GPY81,S85,McL86}.
QCD lattice simulations predict this transition to occur at a
temperature of approximately 150 MeV~\cite{L96}. The experimental
confirmation of a possibly created QGP is, however, difficult, since
only very few particle species, mainly leptons, can provide direct
information about the initial partonic stage of the collision. The
much more abundant hadrons are substantially affected by secondary
interactions and decouple from the collision region only during the
final `freeze-out' stage. A successful dynamical model of relativistic
heavy-ion collisions should finally explain all these different
observables, their dependence on the incident energy, impact
parameter, and atomic number of the projectile and target nuclei.  

At the present stage, theoretical efforts concentrate on
discriminating between different models by comparing them with
characteristic observables~\cite{QM96,QM97,HM96,BGSG98}. The observed
enhancement of strange hadron and low-mass dilepton yields and the
measured $J/\Psi$-suppression provide strong indications that a dense
system was created in the collision whose extreme condition has
significantly affected particle production mechanisms. Furthermore,
various observations signal collective (hydro)dynamical behaviour in
the collision region which in turn indicates the importance of
equilibration processes for the understanding of the collision
dynamics. Especially, the hadronic momentum spectra show signs of both
radial and azimuthally directed flow, and two-particle correlations
indicate a strong transverse expansion of the source before
freeze-out. Despite the rich body of these and other observations, it
remains however controversial to what extent these data are indicative
for the creation of a QGP or can also be explained in purely hadronic
scenarios.  

To make further progress on this central issue, a more detailed
understanding of the space-time geometry and dynamics of the
evolving reaction zone is required. The systems created in
relativistic heavy-ion collisions are mesoscopic and shortlived, and
the geometrical and dynamical conditions of the cauldron play an
essential role for the particle production processes. For example, the
maximal energy density attained in the collision, the time-dependence
of its decrease, and the momenta of the produced particles relative to
the collectively expanding hadronic system will affect the observed
particle ratios. Two-particle correlations provide the only known way
to obtain directly information about the space-time structure
of the source from the measured particle momenta. The size and shape
of the reaction zone and the emission duration become thus
accessible. In combination with the analysis of single particle
spectra and yields, it is furthermore possible to separate the random
and collective contributions to the observed particle momenta. This
permits to also reconstruct the collective dynamical state of the
collision at freeze-out. These new pieces of information give powerful
constraints for dynamical model calculations; they can also be taken
as an experimental starting point for a dynamical back extrapolation
into the hot and dense initial stages of the collision. The present
work reviews the foundations of HBT interferometry in particle physics
and discusses the technical tools for its quantitative application
to relativistic heavy-ion collisions.

\section{Historical overview}
\label{sec1a}

HBT intensity interferometry was proposed and developed by the radio
astronomer Robert Hanbury Brown in the fifties, who was joined by
Richard Twiss for the mathematical analysis of intensity
correlations. Their original aim was to bypass the major constraint of
Michelson amplitude interferometry at that time: in amplitude
interferometry, the resolution at a given wavelength is limited by
the separation over which amplitudes can be compared. Hanbury
Brown started from the observation that "if the radiation received
at two places is mutually coherent, then the fluctuation in the
intensity of the signals received at those two places is also
correlated"~\cite{HBT1}. More explicitly, amplitude interferometry
measures the square of the sum of the two amplitudes $A_1$ and $A_2$
falling on two detectors 1 and 2: 
  \begin{equation}
    \vert A_1+A_2 \vert^2 = \vert A_1\vert^2 + \vert A_2\vert^2
            + (A_1^*\, A_2 + A_1\, A_2^*)\, .
    \label{1.1}
  \end{equation}
The last term, the `fringe visibility' $V$, is the part of the signal which 
is sensitive to the separation between the emission points. Averaged over 
random variations, its square is given by the product of the intensities 
landing on the two detectors~\cite{Baym98},
  \begin{equation}
    \langle V^2\rangle = 2\, \langle \vert A_1\vert^2
        \vert A_2\vert^2\rangle
        + \langle {A_1^*}^2 A_2^2\rangle
        + \langle A_1^2 {A_2^*}^2\rangle
        \longrightarrow 2 \langle I_1\, I_2\rangle\, .
    \label{1.2}
  \end{equation}
The last two terms of this expression vary rapidly and average to 
zero. According to (\ref{1.2}), intensity correlations between
different detectors contain information about the fringe visibility
and hence about the spatial extension of the source. To demonstrate
the technique, Hanbury Brown and Twiss measured in 1950 the diameter
of the sun, using two radio telescopes operating at 2.4 m wavelength,
and determined in 1956 the angular diameters of the radio sources
Cas A and Cyg A. Furthermore, they measured in a highly influential
experiment intensity correlations between two beams separated 
from a mercury vapor lamp. They thus demonstrated~\cite{HBT2} 
that photons in an apparently uncorrelated thermal beam tend to be
detected in close-by pairs. This photon bunching or HBT-effect, first
explained theoretically by Purcell~\cite{P56}, is one of the key
experiments of quantum optics~\cite{G65}. However, with the advent of
modern techniques which allow to compare radio amplitudes of separated
radio telescopes, Michelson interferometry has again completely
replaced intensity interferometry in astronomy. 

In particle physics, the HBT-effect was independently discovered by
G. Goldhaber, S. Goldhaber, W.Y. Lee and A. Pais~\cite{GGLP}. In 1960,
they studied at the Bevatron the angular correlations between
identical pions in $p\bar{p}$-annihilations. Their observation (the
``GGLP-effect''), an enhancement of pion pairs at small relative
momenta, was explained in terms of the finite spatial extension of the
decaying $p\bar{p}$-system and the finite quantum mechanical
localization of the decay pions~\cite{GGLP}. In the sequel of this
work, it was gradually realized that the correlations of identical
particles emitted by highly excited nuclei are sensitive not only to
the geometry of the system, but also to its lifetime~\cite{KP72,S73}. 
This point has become increasingly more important, and it was
supplemented by the later insight that the pair momentum dependence of
the correlations measured for relativistic heavy-ion collisions
contains information about the collision dynamics~\cite{P84}. The
origins of the wide field of applications to relativistic heavy-ion
collisions can be dated back to the works of Shuryak~\cite{S73},
Cocconi~\cite{C74}, Grishin, Kopylov and Podgoretski\u\i\
\cite{GKP71,KP72,KP74}, and to the seminal paper of Gyulassy,
Kauffmann and Wilson~\cite{GKW79}. Important contributions in the
eighties include a more detailed analysis of the role of final state 
interactions~\cite{K77,GK81,P86,B88,B91}, the development of a
parametrization~\cite{P83} taking into account the longitudinal
expansion of the system created in the collision~\cite{P84,P86} and
the first implementation of the HBT-effect in prescriptions for event
generator studies~\cite{Z87}. Also, the effect was seen in and
analyzed for high energy collisions (see the recent review by
L\"orstad~\cite{L89}). In addition, there is a wealth of experimental
data and theoretical work on correlations between protons and heavier
fragments (pp, pd, p${}^4$He) in lower energy ($< 1$ GeV) nuclear
collisions, which are summarized in the review article of Boal, Gelbke
and Jennings~\cite{bgj90}. 

With the advent of relativistic heavy-ion beams at CERN and 
Brookha\-ven, many of these concepts had to be  refined and extended
to the rapidly expanding particle emitting systems created in
heavy-ion collisions. The relativistic collision dynamics 
plays an important role in the derivation of the HBT 
two-particle correlator and of its modern parametrizations. It
is adequately reflected in recent model discussions of the
particle phase-space density from which the two-particle
correlator is calculated. Several smaller 
reviews~\cite{L89,BGP92,H94,prattrev,He96,Baym98} as well as a
selected reprint volume~\cite{Wei97} exist by now. The present 
work aims at a unified presentation of the underlying concepts and 
calculational techniques, and of the phenomenological applications
of HBT interferometry to the rapidly expanding sources
created in these relativistic heavy-ion collisions. It does not
provide a comprehensive review of the experimental data, for which
we refer to the overview given in~\cite{HJ99}.

\section{Outline}
\label{sec1b}

We start by discussing the relation between the single-particle 
Wigner phase-space density $S(x,K)$ of the particle emitting source,
the triple-differential one-particle spectrum $E_p\, dN/ d^3p$
and the two-particle correlation function $C(\bbox{q},\bbox{K})$
for pairs of identical bosons:
  \begin{eqnarray}
    E_p {dN\over d^3p} &=& \int d^4x\, S(x,p)\, ,
  \label{1.3}  \\
     C(\bbox{q},\bbox{K}) &\approx& 1 + 
     {\left\vert \int d^4x\, S(x,K)\, e^{iq{\cdot}x}\right\vert^2 
      \over
      \left\vert \int d^4x\, S(x,K)\right\vert^2 }
  \label{1.4}  \\
    &\approx& 1 + \lambda(\bbox{K})\, \exp\left[ 
      -\sum_{ij} R_{ij}^2(\bbox{K})\, \bbox{q}_i\, \bbox{q}_j\right]\, .
  \label{1.5}  
  \end{eqnarray}
The approximations are discussed in the main text; the notation used 
here and throughout this review is compiled at the end of this introduction. 
The main aim of particle interferometric methods is to extract as much
information as possible about the emission function $S(x,K)$, which
characterizes the particle emitting source created in the heavy-ion
collision. We discuss how the above expressions are modified to
include final state interactions and multiparticle symmetrization
effects and how they apply to numerical event simulations of 
relativistic heavy-ion collisions. Contact between theory and 
experiment is made with the help of Gaussian parametrizations 
(\ref{1.5}) of the correlator which we review in chapter~\ref{sec3}. 
We discuss the Cartesian Pratt-Bertsch parametrization as well as the 
Yano-Koonin-Podgoretski\u\i\ (YKP) parametrization where the latter
is particularly adapted to the description of systems with strong 
longitudinal expansion. We then turn to estimates of the pion phase
space density based on such Gaussian fits. Our main focus is on the
space-time interpretation of the HBT radius parameters
$R_{ij}^2(\bbox{K})$ which we establish in terms of space-time
variances of the Wigner phase-space density $S(x,K)$. 
Particle emission duration, average particle emission time, transverse 
and longitudinal extension of the source as well as position-momentum 
correlations in the source due to dynamical flow patterns are seen to 
be typical source characteristics to which identical
particle correlations are sensitive. While most of our discussion
is carried out for central collisions, we also review how this
framework can be extended to collisions at finite impact parameter 
where the HBT radius parameters depend on the azimuthal angle of the
emitted particles with respect to the reaction plane. Furthermore, we
discuss more advanced techniques which do not rely on a Gaussian
parametrization of the correlation function but require better
statistics of the experimental data. This concludes our review of
existing analysis tools.

 Chapter~\ref{sec5} is devoted to applications of the presented
framework within concrete model studies. We introduce a simple but
flexible class of models for particle emission in relativistic
heavy-ion collisions. These are motivated by hydrodynamical and
thermodynamical considerations and allow to illustrate the main
techniques discussed before. Different analytical and numerical
calculation schemes for the HBT radius parameters are contrasted, and
we explain which geometrical and dynamical model features
are reflected by which observables. Then we discuss how resonance decay
contributions to pion spectra modify these calculations, and we compare
the results of this model with various other model studies, focussing
on the qualitative and quantitative differences. All these results
are finally combined into an analysis strategy for the reconstruction
of the particle emitting source from the measured one- and two-particle
spectra. The method is illustrated on Pb+Pb data taken by
the NA49 Collaboration at the CERN SPS. 

\section{Notation and conventions}
\label{sec1c}

We use natural units $\hbar = c = k_B = 1$. Unless
explicitly stated otherwise, {\it pairs} or {\it sets of N
particles} are meant to be pairs or sets of identical spinless
bosons. In particular, we think of like-sign pions or kaons, the 
most abundant mesons in heavy-ion collisions. Most of our
discussion carries over to fermionic particles by replacing
the $+$ signs in (\ref{1.4}) and (\ref{1.5}) by $-$ signs
and changing from symmetrized to anti-symmetrized $N$-particle
states whereever they appear in derivations. In what follows,
we do not mention the fermionic case explicitly.

Most of our notation is introduced during the discussion. Variables in
{\bf bold face} denote 3-vectors. For simpler reference, we list here
some of the variables used most frequently.

\vspace{1cm}

\begin{tabular}{rcl}
$p_i = (E_i,\bbox{p}_i)$&\hspace{1.6cm}& detected final state particle 
 momenta, on-shell \\
$m_\perp{=}{\textstyle{\sqrt{m^2 + \bbox{p}_\perp^2}}}$
 & & single particle transverse mass \\
$\phi$ & & azimuthal angle of $\bbox{p}_\perp$ \\
${\rm y}{=}{\textstyle{1\over 2}} \ln{{E_p+p_l}\over {E_p-p_l}}$
 & & (roman y) single particle rapidity\\ 
$y$ & & ({\it italic} $y$) coordinate in configuration space \\
$\check{\bbox{p}}_i$ & & simulated particle momenta, e.g. from Monte\\
 & & Carlo simulations\\
$\hat{\bbox{p}}_i$ & & particle momentum operator\\
$\check{\bbox{r}}_i$ & & simulated particle positions\\
$\check{t}_i$ & & simulated particle emission times\\
$K{=}{\textstyle{1\over 2}}(p_1 + p_2)$ & & average pair momentum, off-shell\\
$M_\perp{=}{\textstyle{\sqrt{m^2 + \bbox{K}_\perp^2}}}$
 & & transverse mass associated with $K$ \\
$\Phi$ & & azimuthal angle of $\bbox{K}_\perp$ \\
$Y{=}{\textstyle{1\over 2}} \ln{{E_K + K_l}\over {E_K - K_l}}$
 & & rapidity associated with $K$\\ 
$q = (p_1 - p_2)$ & & relative pair momentum, off-shell\\
$\bbox{\beta} = \bbox{K}/K^0$ & & velocity of particle pair (approximately)\\
$C(\bbox{q},\bbox{K})$ & & two-particle correlation function, also denoted\\
 & & by $C(\bbox{p}_1,\bbox{p}_2)$ \\
$S(x,K)$ & & single-particle Wigner density, emission function\\
$\rho(x,\bbox{p})$ & & classical phase-space density\\
${\cal P}_1(\bbox{p})$ & & covariant one-particle spectrum\\
${\cal P}_2(\bbox{p}_1,\bbox{p}_2)$ & & covariant two-particle spectrum\\
${\cal N}$ & & normalization of the correlator\\ 
$\hat{N}$ & & number operator\\
$\sigma$ & & quantum mechanical wave packet width\\
\end{tabular}

\chapter{Particle correlations from phase-space distributions}
\label{sec2}

There are numerous derivations of identical two-particle correlations 
from a given boson emitting source. An (over)simplified argument
starts from the observation that, after weighting the emission points
of a two-particle Bose-Einstein symmetrized plane wave
$\Psi_{12}(\bbox{x}_1,\bbox{x}_2,\bbox{p}_1,\bbox{p}_2)  
= \textstyle{1\over 2} \left( e^{i\bbox{p}_1{\cdot}\bbox{x}_1 +
i\bbox{p}_2{\cdot}\bbox{x}_2}\right. $ 
$+ \left. e^{i\bbox{p}_1{\cdot}\bbox{x}_2 + 
i\bbox{p}_2{\cdot}\bbox{x}_1}\right)$ by a normalized spatial distribution 
of emission points $\rho(\bbox{x})$, the two-particle correlator
$C(\bbox{q})$ is given by the Fourier transform of the spatial distribution: 
  \begin{eqnarray}
    C(\bbox{q}) &=& \int d^3x_1\, d^3x_2\,
                   \rho(\bbox{x}_1)\, \rho(\bbox{x}_2)\,
                   \vert\Psi_{12}\vert^2
                   \nonumber \\
               &=& 1 + \vert\tilde{\rho}(\bbox{p}_1-\bbox{p}_2)\vert^2\, .
    \label{2.1}
  \end{eqnarray}
Extracting the spatial information $\rho(\bbox{x})$ from the 
measured momentum spectra is then a Fourier inversion problem. 
The solution is unique if we assume $\rho(\bbox{x})$ to be real
and positive.

Equation (\ref{2.1}) remains, however, unsatisfactory since it does 
not allow for a possible time-dependence of the emitter and cannot be
easily extended to sources with position-momentum correlations. 
Both properties are indispensable for an analysis of the boson 
emitting sources created in heavy-ion collisions. A sound starting
point is provided by the Lorentz invariant one- and two-particle 
distributions for each particle species 
  \begin{eqnarray}
    {\cal P}_1(\bbox{p}) 
      &=& E\, \frac{dN}{d^3p} 
        = E \, \langle\hat{a}^+_{\bbox{p}} \hat{a}_{\bbox{p}}\rangle \, ,
        \label{2.2} \\
   {\cal P}_2(\bbox{p}_1,\bbox{p}_2) 
      &=& E_1\, E_2\, \frac{dN}{d^3p_1\, d^3p_2}
      = E_1 \, E_2\, 
          \langle\hat{a}^+_{\bbox{p}_1} \hat{a}^+_{\bbox{p}_2}
                 \hat{a}_{\bbox{p}_2} \hat{a}_{\bbox{p}_1} \rangle \, .
      \label{2.3}
  \end{eqnarray}
These distributions involve expectation values $\langle\, ...\, \rangle$ 
which can be specified in terms of a density operator characterizing
the collision process. In most applications, 
$\langle\, ...\, \rangle$ involves an average over an ensemble
of events. The two-particle correlation function
of identical particles is defined, up to a proportionality
factor ${\cal N}$, as the ratio of the one- and two-particle spectra:
 \begin{equation}
 \label{2.4}
   C(\bbox{p}_1,\bbox{p}_2) = 
   {\cal N}
   {{\cal P}_2(\bbox{p}_1,\bbox{p}_2)
     \over {\cal P}_1(\bbox{p}_1)\, {\cal P}_1(\bbox{p}_2)}\, .
 \end{equation}
In section~\ref{sec2a}, we discuss its normalization as well as the 
experimentally used method of ``normalization by mixed pairs''. 
Sections~\ref{sec2b} and ~\ref{sec2c} then deal with two different 
derivations of the basic relation (\ref{1.4}) between the two-particle
correlation function and the Wigner phase-space density. Final
state interactions and multiparticle symmetrization effects
are discussed subsequently in sections~\ref{sec2d} and ~\ref{sec2e}.
We conclude this chapter by discussing the implementation of this
formalism into event generators.

\section{Normalization}
\label{sec2a}

The normalization ${\cal N}$ of the two-particle 
correlator (\ref{2.4}) can be specified by relating the particle 
spectra to inclusive differential cross sections, or by requiring
a particular behaviour for the correlator (\ref{2.4}) at 
large relative pair momentum $\bbox{q}$. Pair mixing algorithms 
used for the analysis of experimental data approximate these
normalizations.

\subsection{Differential and total one- and two-particle cross 
            sections}
\label{sec2a1}

The one- and two-particle spectra (\ref{2.2}/\ref{2.3}) are given in
terms of the one- and two-particle inclusive differential cross sections as
  \begin{eqnarray}
    {\cal P}_1(\bbox{p}) &=& E\, {1\over \sigma}\, 
    {d\sigma_{\pi}\over d^3\bbox{p}}\, ,
    \label{2.7} \\
    {\cal P}_2(\bbox{p}_1,\bbox{p}_2) &=& E_1\, E_2\, {1\over \sigma}\,
    {d\sigma_{\pi\pi}\over d^3\bbox{p}_1\, d^3\bbox{p}_2}\, .
    \label{2.8}
  \end{eqnarray}
They are normalized by
    \begin{eqnarray}
      \int {d^3p\over E}\, 
      {\cal P}_1(\bbox{p}) &=& \langle \hat{N}\rangle\, ,
      \label{2.9} \\
      \int {d^3p_1\over E_1}\, {d^3p_2\over E_2}\,
      {\cal P}_2(\bbox{p}_1,\bbox{p}_2) &=& 
      \langle \hat{N}(\hat{N} - 1)\rangle\, , 
      \label{2.10}
    \end{eqnarray}
where $\hat{N} = \int d^3k\, a_{\bf k}^\dagger a_{\bf k}$
is the number operator. Two natural choices for the normalization
${\cal N}$ in (\ref{2.4}) arise~\cite{GKW79,MV97,ZC97,W98,ZSH98} by
either taking directly the ratio of the measured spectra (\ref{2.9})
and (\ref{2.10}), which results in
  \begin{equation}
    {\cal N} = 1\, ,
    \label{2.6}
  \end{equation}
or by first normalizing the numerator and denominator of (\ref{2.4})
separately to unity, which gives
  \begin{equation}
    {\cal N} = \frac{\langle \hat{N}\rangle^2}
    {\langle \hat{N}(\hat{N}-1)\rangle}\, .
    \label{2.5}
  \end{equation}
Since in either case ${\cal N}$ is momentum-independent, it does not 
affect the space-time interpretation of the correlation function,
with which we will focus  primarily.

For ${\cal N} =1$, the correlator equals 1 whenever 
${\cal P}_2(\bbox{p}_1,\bbox{p}_2){=}{\cal P}_1(\bbox{p}_1) 
{\cal P}_1(\bbox{p}_2)$. Neglecting kinematical constraints
resulting from finite event multiplicities, one can often
assume this factorization property to be valid
for large relative momenta $\bbox{q}$. Since at small values of 
$\bbox{q}$ the correlation function is larger than unity, this
generally implies ${\langle \hat{N}(\hat{N}-1)\rangle} > 
{\langle \hat{N}^2\rangle}$, i.e., larger than Poissonian
multiplicity fluctuations. This is a natural consequence of
Bose-Einstein correlations.

The second choice (\ref{2.5}) permits to view the
correlator as a factor which relates the two-particle differential
cross section $d\sigma_{\pi\pi}^{\rm BE}/d^3p_1 d^3p_2$ 
of the   
real world (where Bose-Einstein symmetrization exists) to an idealized world 
in which Bose-Einstein final state correlations are absent,
  \begin{equation}
    d\sigma_{\pi\pi}^{\rm BE}/d^3p_1\, d^3p_2 
    = C(\bbox{q},\bbox{K})\, 
    d\sigma_{\pi\pi}^{\rm NO}/d^3p_1\, d^3p_2\, , 
    \label{2.11}
  \end{equation}
without changing the event multiplicities. Such an idealized world is
a natural concept in event generator studies which simulate
essentially the two-particle cross sections 
$d\sigma_{\pi\pi}^{\rm NO}/d^3p_1\, d^3p_2$. They are typically tuned
to reproduce the measured multiplicity distributions and thereby
account heuristically for the effects of Bose-Einstein statistics on 
particle production; the quantum statistical symmetrization of  
the final state, however, is not a part of the code. With the
normalization (\ref{2.5}), the factor $C(\bbox{q},\bbox{K})$ in (\ref{2.11})
preserves the total cross sections,
$\sigma_{\pi\pi}^{\rm BE, tot} = \sigma_{\pi\pi}^{\rm NO, tot}$.

\subsection{Experimental construction of the correlator}
\label{sec2a2}

From the data of relativistic heavy-ion experiments, the
two-particle correlator is usually constructed 
as a quotient of samples of so-called {\it actual pairs} and 
`mixed' pairs or {\it reference pairs}.

One starts by selecting events from the primary data set. 
Actual pairs are pairs of particles that belong
to the same event. Reference pair partners are picked
randomly from different events within the set of events that
yielded the actual pairs. The correlation function is then 
constructed by taking the ratio, bin by bin, of the distribution
$D_A$ of these actual pairs with the distribution $D_R$ 
of the reference pairs~\cite{Z82,Z84},
  \begin{eqnarray}
    D_A(\Delta \bbox{q},\Delta \bbox{K}) &=&
    {\hbox{ number of actual pairs in bin 
            $(\Delta \bbox{q},\Delta \bbox{K})$}
      \over 
     \hbox{ number of actual pairs in sample}}\, ,
    \label{2.13} \\
    D_R(\Delta \bbox{q},\Delta \bbox{K}) &=&
    {\hbox{ number of reference pairs in bin 
            $(\Delta \bbox{q},\Delta \bbox{K})$}
      \over 
     \hbox{ number of reference pairs in sample}}\, ,
    \label{2.14}\\
    C(\Delta \bbox{q},\Delta \bbox{K}) &=& 
    {D_A(\Delta \bbox{q},\Delta \bbox{K})\over
      D_R(\Delta \bbox{q},\Delta \bbox{K})}
    \label{2.15}\, .
  \end{eqnarray}
The number of reference pairs for each actual pair, the so-called
mixing factor, is typically between $10$ and $50$. It has to be
chosen sufficiently large to ensure a statistically independent
reference pair sample while for numerical implementations it is
of course favourable to keep the size of this sample as small
as possible. The two-particle correlator constructed in (\ref{2.15}) 
coincides with the
theoretical definition (\ref{2.4}) only if the reference pair
distribution $D_R$ coincides with an appropriately normalized
product of one-particle spectra. Since both actual and reference 
pair distributions are normalized to the corresponding total
particle multiplicity, this normalization of (\ref{2.15}) coincides
with the normalization ${\cal N} = \langle \hat{N}\rangle^2$
$/\langle \hat{N}(\hat{N} - 1)\rangle$. A different construction
of the correlator from experimental data has been proposed by
Mi\'skowiec and Voloshin~\cite{MV97} (see also~\cite{ZSH98}). 
Their proposal amounts to a 
modification of the number of pairs in the sample by which 
(\ref{2.13}) and (\ref{2.14}) is normalized and coincides with 
${\cal N} = 1$.

In general, the reference sample contains residual 
correlations, which are not of physical origin but stem 
typically from the restricted acceptance of experiments.
For these residual correlations, corrections can be employed~\cite{Z82,Z84}.

\section{Classical current parametrization}
\label{sec2b}

How does one calculate the momentum correlations for identical 
pions produced in a heavy-ion collision?
The pion production in a nuclear collision is described by the
field equations for the pion field $\phi(x)$~\cite{GKW79},
  \begin{equation}
    \Bigl( \Box + m^2 \Bigr) \hat \phi(x) = \hat{J}(x) \, .
    \label{2.16}
  \end{equation}
This equation is obviously intractable since the nuclear current
operator $\hat{J}(x)$ couples back to the pion field and is not 
explicitly known. The classical current parametrization~\cite{GKW79} 
approximates the nuclear current $\hat{J}(x)$ by a classical 
commuting space-time function $J(x)$. The underlying picture is
that at freeze-out, when the pions stop interacting,
the emitting source is assumed not to be affected by the 
emission of a single pion. This approximation can be justified
for high event multiplicities~\cite{GKW79}. For a classical source 
$J(x)$, the final pion state is then a coherent state $\vert J\rangle$ 
which is an eigenstate of the annihilation operator
  \begin{equation}
    \hat{a}_{\bbox{p}} \vert J \rangle = i\tilde{J}(\bbox{p})|J\rangle \, .
    \label{2.17} 
  \end{equation}
The Fourier transformed classical currents $\tilde{J}$ are on-shell. 
Using (\ref{2.17}), the one- and two-particle spectra (\ref{2.2}) and 
(\ref{2.3}) are then readily calculated. Usually, the classical 
current is taken to be a superposition of independent elementary 
source functions $J_0$: 
  \begin{equation}
    \tilde{J}(\bbox{p}) 
    = \sum_{i=1}^N e^{i\phi_i}\, e^{ip{\cdot}x_i}
      \tilde{J}_0(p-p_i)\, .
      \label{2.18}
  \end{equation}
If the phases $\phi_i$ are random, then
this ansatz characterizes uncorrelated ``chaotic''  particle
emission, and the intercept $\lambda$ in (\ref{1.5}) equals one. 
In more general settings, where the phases $\phi_i$
are not random, the intercept drops below unity. One distinguishes 
accordingly between a formalism for chaotic and partially chaotic sources.

\subsection{Chaotic sources}
\label{sec2b1}

Chaotic sources are given by a superposition (\ref{2.18}) of 
elementary sources $J_0$, centered around phase-space points
$x_i$, $p_i$ with random phases $\phi_i$. The corresponding
ensemble average specifying the particle spectra (\ref{2.2}/\ref{2.3}) 
is~\cite{CH94}
  \begin{equation}
    \langle\,\hat{O}\rangle = \sum_{N=0}^\infty P_N
    \prod_{i=1}^N\int d^4x_i\, d^4p_i\,\rho(x_i,p_i)
    \int_0^{2\pi}\frac{d\phi_i}{2\pi}\,
    \big\langle J \big\vert \hat O \big\vert J \big\rangle \, ,
   \label{2.19} 
  \end{equation}
where $P_N$ is the probability distribution for the number $N$
of sources, $\sum_{N=0}^{\infty}$ $P_N = 1$ and the normalized
probability $\rho(x_i,p_i)$ describes the distribution of the 
elementary sources in phase space. A direct consequence
of the ensemble average (\ref{2.19}) is the factori\-zation of the
two-particle distribution into two-point functions~\cite{CH94,He96} 
  \begin{eqnarray}
    {\cal P}_2(\bbox{p}_1,\bbox{p}_2) 
    &=& { {\langle N(N-1)\rangle}_P\over {\langle N\rangle}_P^2}
      \left( {\cal P}_1(\bbox{p}_1){\cal P}_1(\bbox{p}_2) 
      + |\bar{S}_J(\bbox{p}_1,\bbox{p}_2)|^2 \right) \, ,
   \label{2.20}\\
    \bar{S}_J(\bbox{p}_1,\bbox{p}_2) &\equiv& \sqrt{E_1 E_2}\,
    \langle \hat{a}^+_{\bbox{p}_1} \hat{a}_{\bbox{p}_2} \rangle
        = \sqrt{E_1 E_2}\, \langle \tilde{J}^*(\bbox{p}_1)
        \tilde{J}(\bbox{p}_2)\rangle\, ,
   \label{2.21}
  \end{eqnarray}
where ${\langle N(N-1)\rangle}_P$ $= \sum_{N=0}^\infty P_N\, N(N-1)$,
${\langle N\rangle}_P = $ $\sum_{N=0}^\infty P_N\, N$ and the 
expectation value in (\ref{2.21}) is evaluated according to the
prescription (\ref{2.19}). 
Here, the integer $N$ denotes the number of sources, not the 
number of emitted pions. For a Poissonian source multiplicity 
distribution the prefactor in (\ref{2.20}) equals unity.
In the derivation of (\ref{2.20}) a
term is dropped in which both final state particles come from
the same source. This term vanishes in the large $N$ 
limit~\cite{GKW79,He96}.

The factorization in (\ref{2.20}) follows from the commutation relations,
once independent particle emission and the absence of final state
interactions is assumed. Due to its generality, it is sometimes 
referred to as ``generalized Wick theorem''. The emission function
$S(x,K)$ which enters the basic relation (\ref{1.4}) can then be identified 
with the Fourier transform of the covariant quantity 
$\bar{S}_J(\bbox{p}_1,\bbox{p}_2)$~\cite{CH94,He96}. The latter
is given by the Wigner transform of the density matrix associated
with the classical currents, 
  \begin{equation}
    S_J(x,K) = \int\frac{d^4y}{2(2\pi)^3}\, e^{-iK{\cdot}y}
              \langle J^*(x+\textstyle{1\over 2} y)
              J(x-\textstyle{1\over 2} y)\rangle \, , 
    \label{2.22a}
  \end{equation}
for which the following folding relation can be derived~\cite{CH94}
  \begin{eqnarray}
    S_J(x,K) &=& \langle{N}\rangle_P \int d^4z\, d^4q \,\rho(z,q) \, 
    S_0(x-z,K-q) \, ,
    \label{2.22} \\
    S_0(x,K) &=& \int\frac{d^4y}{2(2\pi)^3}\, e^{-iK{\cdot}y}
              J_0^*(x+\textstyle{1\over 2} y)
              J_0(x-\textstyle{1\over 2} y) \, . 
    \label{2.23}
  \end{eqnarray}
Here $\langle {N} \rangle_P$ is fixed by normalizing the one-particle
spectrum to the mean pion multiplicity $\langle {N} \rangle_P =
\langle \hat{N} \rangle$; the distribution $\rho$ and elementary
source function $S_0$ are normalized to unity. The full emission
function is hence given by folding the distribution $\rho$ of the
elementary currents $J_0$ in phase-space with the Wigner density
$S_0(x,K)$ of the elementary sources. Wigner functions are quantum
mechanical analogues of classical phase-space distributions 
\cite{HOSW84}. In general they are real but not positive definite, but
when integrated over $x$ or $K$ they yield the observable particle
distributions in coordinate or momentum space, respectively. Averaging
the quantum mechanical Wigner function $S(x,K)$ over phase-space 
volumes which are large compared to the volume $(2\pi\hbar)^3$ of 
an elementary phase-space cell, one obtains a smooth and positive
function which can be interpreted as a classical phase-space
density.

From the particle distributions (\ref{2.20}/\ref{2.21}) one finds the
two-particle correlator~\cite{S73,GKW79,P84,BDH94,CH94} 
  \begin{equation}
    C(\bbox{q},\bbox{K}) = 1 + 
    {{\vert \int d^4x\, S(x,K)\, e^{iq\cdot x}\vert^2}\over
      \int d^4x\, S(x,p_1)\, \int d^4y\, S(y,p_2)}\, ,
    \label{2.24}
  \end{equation}
if the normalization ${\cal N}$ in (\ref{2.4}) is chosen to cancel the
prefactor in (\ref{2.20}). Adopting instead the normalization
prescription (\ref{2.5}) leads to a normalization of 
$C(\bbox{q},\bbox{K})$ at $\bbox{q} \to \infty$ which is smaller 
than unity, since (\ref{2.5}) is
not the inverse of the prefactor in (\ref{2.20}).

\subsection{The smoothness and on-shell approximation}
\label{sec2b2}

The smoothness approximation assumes that the emission function
has a sufficiently smooth momentum dependence such that one 
can replace
  \begin{equation}
    S(x,K - \textstyle{1\over 2}q)\, 
    S(y,K + \textstyle{1\over 2}q)\, 
    \approx S(x,K)\, S(y,K)\, .
    \label{2.25}
  \end{equation}
Deviations caused by this approximation are proportional to the 
curvature of the single-particle distribution in logarithmic 
representation~\cite{CSH95a} and were shown to be negligible for 
typical hadronic emission functions. Using this smoothness
approximation, the two-particle correlator (\ref{2.24})
reduces to the expression on the r.h.s. in (\ref{1.4}).

Eq. (\ref{1.4}) which uses the smoothness approximation, forms 
the basis for the interpretation of 
correlation measurements in terms of space-time variances of
the source as will be explained in section~\ref{sec3}.
For the calculation of the correlator from a given emission
function, the smoothness approximation can be released by
staring directly from (\ref{2.24}). In an analysis of 
measured correlation functions in terms of space-time 
variances of the source, one can correct for it
systematically~\cite{CSH95a} using information from the
single particle spectra.

 The emission function $S(x,K)$ depends in principle on the
{\it off-shell} momentum $K$, where $K^0 = \textstyle{1\over 2}
(E_1 + E_2)$. In many applications one uses the on-shell
approximation
  \begin{equation}
    S(x,K^0,\bbox{K}) \approx S(x,E_K,\bbox{K})\, ,
    \qquad \qquad \, E_K = \sqrt{m^2 + \bbox{K}^2}\, .
    \label{2.26}
  \end{equation}
Again, the corrections can be calculated systematically~\cite{CSH95a} 
but were shown to be small for typical model emission functions for
pions and heavier hadrons. The on-shell approximation (\ref{2.26}) is
instrumental in event generator studies, where one aims at associating
the emission function $S(x,K)$ with the simulated on-shell particle
phase-space distribution at freeze-out, see section~\ref{sec2f}. It is
also used heavily in analytical model studies, see section~\ref{sec5}.

\subsection{The mass-shell constraint}
\label{sec2b3}

Although the correlator (\ref{2.24}) is obtained as a Fourier 
transform of the emission function $S(x,K)$, this emission 
function cannot be reconstructed uniquely from the momentum 
correlator (\ref{2.24}). Note that since the Wigner density
$S(x,K)$ is always real, the reconstruction of its phase is
not the issue. The reason is rather the mass-shell constraint
  \begin{equation}
    K\cdot q = p_1^2 - p_2^2 = m_1^2 - m_2^2 = 0\, ,
     \label{2.27} 
  \end{equation}
which implies that only three of the four relative momentum components
are kinematically independent. Hence,
the $q$-dependence of $C(\bbox{q},\bbox{K})$ allows to test only
three of the four independent $x$-directions of the emission function.
This introduces an unavoidable model-dependence in the
reconstruction of $S(x,K)$, which can only be removed by additional
information not encoded in the two-particle correlations between
identical particles. Eq. (\ref{2.27}) suggests that this
ambiguity may be resolvable by combining correlation data from
unlike particles with different mass combinations, if they are
emitted from the same source. Unlike particles do
not exhibit Bose-Einstein correlations, but are correlated via
final state interactions and therefore also contain information 
about the source emission function. In this review,
we do not discuss unlike particle correlations, although this
is presently a very active field of 
research~\cite{LL82,AL96,LL96,LLEN96,VLPX97,Setal97,Misk98}. 
It is still an open question to what extent a combined analysis of 
like and unlike particles allows to bypass the mass-shell constraint
(\ref{2.27}). 

The consequences of the mass-shell constraint are discussed extensively in
sections~\ref{sec3} and ~\ref{sec5}.

\subsection{The relative distance distribution}
\label{sec2b3a}
   
For several applications of two-particle interferometry it is
useful to reformulate the correlator (\ref{1.4}) in terms of
the so-called normalized relative distance distribution 
  \begin{eqnarray}
    d(x,K) &=& \int d^4X\, s(X+{\textstyle{x\over 2}},K) \,
                           s(X-{\textstyle{x\over 2}},K) \, ,
    \label{4.01} \\
    s(x,K) &=& {S(x,K)\over \int d^4x\, S(x,K)}\, ,
    \label{4.02}
  \end{eqnarray}
constructed from the normalized emission function $s(x,K)$.
Note that $d(x,K)$ $=d(-x,K)$ is an even function of $x$. This
allows to rewrite the correlator (\ref{1.4}) as
  \begin{equation}
        C(\bbox{q},\bbox{K}) - 1 = \int d^4x\, \cos(q\cdot x)\,
        d(x,K)\, ,
        \label{4.03}
  \end{equation}
where the smoothness and on-shell approximations were used.
With the mass-shell constraint in the form $q^0 = \bbox{\beta}
\cdot \bbox{q}$ this can be further rewritten in terms of 
the ``relative source function'' $S_{\bbox{K}}(\bbox{r})$:
  \begin{eqnarray}
        C(\bbox{q},\bbox{K}) - 1 &=& \int d^3x\, 
        \cos(\bbox{q}\cdot \bbox{x})\,
        \int dt\, d(\bbox{x}+\bbox{\beta}t, t; K)
  \nonumber \\
        &=& \int d^3x\, \cos(\bbox{q}\cdot\bbox{x})\, 
            S_{\bbox{K}}(\bbox{x})\, .
  \label{4.04}
  \end{eqnarray}
In the rest frame of the particle pair where $\bbox{\beta} = 0$,
the relative source function $S_{\bbox{K}}(\bbox{x})$ is a simple  
integral over the time argument of the relative distance distribution 
$d(\bbox{x},t;K)$. In this particular frame the time structure of 
the source is completely integrated out. This illustrates in the
most direct way the basic limitations of any attempt to reconstruct
the space-time structure of the source from the correlation function.

\subsection{Partially coherent sources}
\label{sec2b4}

It is well-known from quantum optics~\cite{G65} that, in spite of
Bose-Einstein statistics, the HBT-effect does not exist for particles 
emitted with phase coherence, but only for chaotic sources. This is why 
in (\ref{2.18}) a chaotic superposition of independent elementary source 
functions $J_0(x)$ was adopted. The question of possible phase coherence
in pion emission from high energy collisions was raised by Fowler and 
Weiner in the seventies~\cite{FW77,FW78,FW79}; so far the dynamical 
origin of such phase coherence effects has however remained
speculative. Their consequences for
HBT interferometry can be studied by adding a coherent component to
the classical current discussed above,
  \begin{equation}
    J(\bbox{p}) =   J_{\rm coh}(\bbox{p}) +  J_{\rm cha}(\bbox{p})\, .
    \label{2.28}
  \end{equation}
An analysis similar to the one presented in section~\ref{sec2b1}
shows that as the number $n_{\rm coh}(\bbox{p})$ of coherently
emitted particles increases, the strength of the correlation is
reduced~\cite{GKW79}:
  \begin{eqnarray}
    \lambda(\bbox{K}) &=& 1 - D^2(\bbox{K})\, ,
    \label{2.29} \\
    D(\bbox{K}) &=& {n_{\rm coh}(\bbox{K})\over 
                   {n_{\rm coh}(\bbox{K}) + n_{\rm cha}(\bbox{K})}}\, .
    \label{2.30}
  \end{eqnarray}
For this reason the intercept parameter $\lambda(\bbox{K})$ is often
referred to as the coherence parameter. In practice various other
effects (e.g. particle misidentification, resonance decay 
contributions, final state Coulomb interactions) can decrease the
measured intercept parameter significantly. Although experimentally
it is always found smaller than unity, $0 < \lambda(\bbox{K}) < 1$,
this can thus not be directly attributed to a coherent field component.
For a detailed account of the search for coherent particle emission
in high energy physics we refer to the reprint collection~\cite{Wei97}. 
Recent work~\cite{HZ97} shows that the strength $D(\bbox{K})$ of the 
coherent component can be determined independent of resonance decay 
contributions and contaminations from misidentified particles if
two- and three-pion correlations are compared, see section~\ref{sec4c}.
A coherent component would also affect the size of the HBT radius parameters
and their momentum dependence~\cite{GKW79,Wei89,APW91,APW93,HZ97}.
The ansatz (\ref{2.28}) is only one possibility to describe 
partially coherent emission. Alternatively, one may e.g. choose a 
distribution of the phases $\phi_i$ in (\ref{2.18}) which is not 
completely random, thereby mimicking partial coherence~\cite{pscotto}.
The equivalence of these two approaches still remains to be studied.

\section{Gaussian wave packets}
\label{sec2c}

It has been suggested 
repeatedly~\cite{PGG90a,PGG90b,MP97,MKFW96,Weal97,ZWSH97,CZ97,ZC97,W98}
that due to the smallness of the source in high energy and relativistic
heavy-ion physics, particle interferometry should be based on finite
size wave packets rather than plane waves. This leads to an
alternative derivation of the basic relations (\ref{1.3}), (\ref{1.4})
which replaces the classical currents from the previous section by the
more intuitive notion of quantum mechanical wave packets, at the
expense of giving up manifest Lorentz covariance in intermediate steps
of the derivation. One starts from a definition of the boson emitting 
source by a discrete set of $N$ phase-space points
$(\check{\bbox{r}}_i, \check{t}_i,\check{\bbox{p}}_i)$ 
or by a continuous distribution 
$\rho(\check{\bbox{r}}_i, \check{t}_i,\check{\bbox{p}}_i)$.
These emission points are associated with the centers of $N$ Gaussian
one-particle wave packets $f_i$,~\cite{MP97,Weal97,ZC97} 
  \begin{equation}
    (\check{\bbox{r}}_i, \check{t}_i,\check{\bbox{p}}_i) \longrightarrow
    f_i(\bbox{x},\check{t}_i) 
    = {1\over (\pi\sigma^2)^{3/4} }
    e^{-{1\over 2\sigma^2}(\bbox{x}-\check{\bbox{r}}_i)^2}\,
    e^{i\bbox{x}\cdot\check{\bbox{p}}_i}\, .
    \label{2.31}
  \end{equation}
The wave packets $f_i$ are quantum mechanically best localized
states, i.e., they saturate the Heisenberg uncertainty relation
with $\Delta x_l = \sigma/\sqrt{2}$ and 
$\Delta p_l = 1/\sqrt{2}\sigma$ for all three spatial
components $l = 1,2,3$. Here, $(\Delta x_l)^2 \equiv$ 
$\langle f_i,\hat{x}_l^2 f_i\rangle$  $- \langle f_i,\hat{x}_l f_i\rangle^2$,
$\hat{x}_l$ being the position operator, and analogously for $ \Delta p_l$.
We consider the free time evolution of these wave packets determined
by the single particle hamiltonian $\hat{H}_0$,
  \begin{eqnarray}
    f_i(\bbox{x},t) &=& 
    \left( e^{-i \hat{H}_0 (t-\check{t}_i)} f_i\right) (\bbox{x},t)
  \nonumber \\
    &=& {1\over (2\pi)^3} \int d^3k\, \tilde{f}_i(\bbox{k})\, 
    e^{i(\bbox{k}\cdot\bbox{x}- E_k(t-\check{t}_i))}\, .
  \label{2.32}
  \end{eqnarray}
In momentum space, the free non-relativistic and relativistic
time evolutions differ only by the choices
$E_k = \bbox{k}^2/2m$ and $E_k = \sqrt{\bbox{k}^2 + m^2}$,
respectively. For the non-relativistic case, the integral (\ref{2.32})
can be done analytically.

\subsection{The pair approximation}
\label{sec2c1}

Here, we derive the correlator in the so-called 
pair approximation in which two-particle symmetrized wave
functions $\Phi_{ij}(\bbox{x},\bbox{y},t)$ are associated with
all boson pairs $(i,j)$ constructed from the set 
of emission points~\cite{Weal97,W98}:
  \begin{equation}
    \Phi_{ij}(\bbox{x},\bbox{y},t)
    = {1\over \sqrt{2}} \Big(
    f_i(\bbox{x},t) f_j(\bbox{y},t)
      + f_i(\bbox{y},t)\, f_j(\bbox{x},t)\Big) \, .
    \label{2.33}
  \end{equation}
The norm of this two-particle state differs from unity by terms
proportional to the wave packet overlaps $\langle f_i,f_j\rangle$,
but in the pair approximation this difference is neglected,
$\langle f_i,f_j\rangle \approx \delta_{ij}$.
In section~\ref{sec2d} we will release this approximation and
instead start from properly normalized $N$-particle wavefunctions.
It is then seen that the pair approximation is equivalent
to approximating the two-particle correlator from an $N$-particle 
symmetrized wavefunction by a sum of contributions involving only
two-particle terms $\Phi_{ij}$. 

The two-particle Wigner phase-space density 
${\cal W}_{ij}(\bbox{x},\bbox{y},\bbox{p}_1,\bbox{p}_2,t)$ 
associated with $\Phi_{ij}$ reads~\cite{HOSW84} 
  \begin{eqnarray}
   {\cal W}_{ij}(\bbox{x},\bbox{y},\bbox{p}_1,\bbox{p}_2,t)
   &=& \int d^3x'\, d^3y' \,
         \Phi_{ij}\left(\bbox{x}+ {\textstyle{ {\bbox{x}}'\over 2}},
         \bbox{y} + {\textstyle{{\bbox{y}}'\over 2}},t\right)\,
         e^{i\bbox{p}_1{\cdot}{\bbox{x}}'}\, 
         \nonumber \\
   && \quad \times\,
         e^{i\bbox{p}_2{\cdot}{\bbox{y}}'}\, 
         \Phi_{ij}^*\left(\bbox{x}- {\textstyle{ {\bbox{x}}'\over 2}},
         \bbox{y} - {\textstyle{{\bbox{y}}'\over 2}},t\right) .
    \label{2.34}
  \end{eqnarray}
Integrating this Wigner function over the positions $\bbox{x}$,
$\bbox{y}$, we obtain the positive definite probability ${\cal
  P}_{ij}(\bbox{p}_1, \bbox{p}_2, t)$ to measure the bosons of the
state $\Phi_{ij}$ at time $t$ with momenta $\bbox{p}_1$,
$\bbox{p}_2$. As long as final state interactions are ignored, this
probability is independent of the detection time. For Gaussian wave
packets it takes the explicit form (the energy factors ensure that 
${\cal P}_{ij}$ transforms covariantly)
  \begin{eqnarray}
    {{\cal P}_{ij}(\bbox{p}_1,\bbox{p}_2)\over 
        E_1\, E_2} &=& {1\over (2\pi)^6}\, \int d^3x\, d^3y\, 
        {\cal W}_{ij}(\bbox{x},\bbox{y},\bbox{p}_1,\bbox{p}_2,t) 
  \nonumber \\
       &=& {\textstyle{1\over 2}}
        w_i(\bbox{p}_1,\bbox{p}_1)\, w_j(\bbox{p}_2,\bbox{p}_2)
        + {\textstyle{1\over 2}}
        w_i(\bbox{p}_2,\bbox{p}_2)\, w_j(\bbox{p}_1,\bbox{p}_1)
  \nonumber \\
        &&+ w_i(\bbox{p}_1,\bbox{p}_2)\, w_j(\bbox{p}_1,\bbox{p}_2)\,
  \nonumber \\
        &&\ \ \times
          \cos\left((\check{\bbox{r}}_i{-}\check{\bbox{r}}_j)\cdot 
                    (\bbox{p}_1{-}\bbox{p}_2)
                  - (\check{t}_i{-}\check{t}_j)(E_1{-}E_2)\right) ,
  \label{2.35} \\
      w_i(\bbox{p}_1,\bbox{p}_2) &=& 
      s_i\left({\textstyle{1\over 2}}(\bbox{p}_1+ \bbox{p}_2)\right)\,
      \exp\left[ -{\sigma^2\over 4}(\bbox{p}_1- \bbox{p}_2)^2 \right]\, , 
  \label{2.36} \\
      s_i(\bbox{K}) &=& \pi^{-3/2}\, \sigma^3\, 
      \exp\left[-\sigma^2(\check{\bbox{p}}_i - \bbox{K})^2\right] \, .
  \label{2.37}
  \end{eqnarray}
Here, the integral over $s_i(\bbox{K})$ is normalized to unity, and
the two-particle probability is normalized such that its momentum
integral equals one for pairs which are well-separated in phase-space.
To relate this formalism to the emission function
(\ref{2.22}/\ref{2.23}) of the classical current parametrization,
we rewrite the two-particle probability (\ref{2.35})
in terms of the Wigner densities $s_i(x,\bbox{K})$ of the 
wave packets,
  \begin{eqnarray}
    s_i(x,\bbox{K}) &=& \delta(t-\check{t}_i)\,
        \int {d^3y\over (2\pi)^3}\,  
        e^{-i\bbox{K}\cdot \bbox{y}}\, 
        f_i(\bbox{x}+{\textstyle{\bbox{y}\over 2}},\check{t}_i)\,
        f_i^*(\bbox{x}-{\textstyle{\bbox{y}\over 2}},\check{t}_i)
  \nonumber \\
    &=& {1\over \pi^3}\, \delta(t-\check{t}_i)\,
    e^{-(\bbox{x}-\check{\bbox{r}}_i)^2/\sigma^2
    -\sigma^2 (\bbox{K}-\check{\bbox{p}}_i)^2}\, ,
  \label{2.38}\\
    {{\cal P}_{ij}(\bbox{p}_1,\bbox{p}_2)\over E_{p_1}\, E_{p_2}} 
    &=& \phantom{+}{1\over 2}
    \int d^4x\, s_i(x,\bbox{p}_1) \int d^4y\, s_j(y,\bbox{p}_2)
  \nonumber \\
    && + {1\over 2}
    \int d^4x\, s_i(x,\bbox{p}_2) \int d^4y\, s_j(y,\bbox{p}_1)
  \nonumber \\
    && + {1\over 2} \int d^4x\, s_i(x,\bbox{K})\, e^{iq\cdot x}
         \int d^4y\, s_j(y,\bbox{K})\, e^{-iq\cdot y}
  \nonumber \\
    && + {1\over 2} \int d^4x\, s_i(x,\bbox{K})\, e^{-iq\cdot x}
         \int d^4y\, s_j(y,\bbox{K})\, e^{iq\cdot y}\, .
  \label{2.39}
  \end{eqnarray}
In the pair approximation, the two pion spectrum 
${\cal P}_2(\bbox{p}_1,\bbox{p}_2)$ for an event with $N$ pions emitted 
from phase-space points $(\check{\bbox{r}}_i,\check{t}_i,\check{\bbox{p}}_i)$ 
is a sum over the probabilities ${\cal P}_{ij}$ of all 
$\textstyle{1\over 2}N(N-1)$ pairs $(i,j)$. The corresponding
expression for a continuous distribution  
$\rho(\check{\bbox{r}}_i,\check{t}_i,\check{\bbox{p}}_i)$ of wave packet
centers is obtained by an integral over (\ref{2.39}). Defining
  \begin{equation}
        {\cal D}\rho_i = 
        d^3\check{p}_i\, d^3\check{r}_i\, d\check{t}_i\,
        \rho(\check{\bbox{r}}_i,\check{t_i},\check{\bbox{p}}_i)\, ,
        \qquad \int {\cal D}\rho_i = 1\, ,
        \label{2.79}
  \end{equation}
we find 
  \begin{eqnarray}
    &&{\cal P}_2(\bbox{p}_1,\bbox{p}_2) = 
      \int {\cal D}\rho_i\, {\cal D}\rho_j\, 
      {\cal P}_{ij}(\bbox{p}_1,\bbox{p}_2)
    \nonumber \\
    && = \int d^4x\, S_{\rm wp}(x,\bbox{p}_1)\,  
               \int d^4y\, S_{\rm wp}(y,\bbox{p}_2)
        + \left\vert \int d^4x\, S_{\rm wp}(x,\bbox{K})\, 
         e^{iq\cdot x}\right\vert^2 \!\!.
    \label{2.40}
  \end{eqnarray}
The index on $S_{\rm wp}$ indicates that this emission function is
constructed from a superposition of wavepackets while the emission
function $S_J$ in (\ref{2.22}) was generated from the classical source
currents. Similar to (\ref{2.22}/\ref{2.23}), $S_{\rm wp}(x,\bbox{K})$
is given by a folding relation between the classical distribution
$\rho$ of wave packet centers and the elementary source Wigner
function $s_0(x,\bbox{K})$,
  \begin{eqnarray}
    S_{\rm wp}(x,\bbox{K}) &=& \int d^3\check{p}_i\, d^3\check{r}_i\, 
    d\check{t}_i\,
    \rho(\check{\bbox{r}}_i,\check{t}_i,\check{\bbox{p}}_i)\, 
    s_0(\bbox{x}{-}\check{\bbox{r}}_i,t{-}\check{t}_i,
        \bbox{K}{-}\check{\bbox{p}}_i)\, ,
    \label{2.41}\\
    s_0(x,\bbox{K}) &=& 
    {E_K\over \pi^3}\,
    \exp\left(- {\bbox{x}^2 \over
        \sigma^2}-{\sigma^2}\bbox{K}^2\right)
    \, \delta(t)\, .
    \label{2.42}
  \end{eqnarray}
The normalization of (\ref{2.41}) is consistent with the interpretation of 
the integral (\ref{1.3}) as the one-particle spectrum. 

To determine the normalization $N(\bbox{p}_1,\bbox{p}_2)$ of the
two-particle correlation function 
  \begin{equation}
    C(\bbox{p}_1,\bbox{p}_2) =
             {{\cal P}_2(\bbox{p}_1,\bbox{p}_2)\over 
              N(\bbox{p}_1,\bbox{p}_2)}\, ,
    \label{2.43}
  \end{equation}
we proceed in analogy to the experimental practice of 
``normalization by mixed pairs'': An uncorrelated 
(mixed) pair is described by an unsymmetrized product state
  \begin{equation}
    \Phi_{ij}^{\rm uncorr}(\bbox{x},\bbox{y},t) =
    f_i(\bbox{x},t) f_j(\bbox{y},t)\, ,
    \label{2.44}
  \end{equation}
for which the two particle Wigner phase-space density and the
corresponding detection probability 
${\cal P}_{ij}^{\rm uncorr}(\bbox{p}_1,\bbox{p}_2)$ can be
calculated~\cite{Weal97} according to (\ref{2.34})-(\ref{2.37}). 
Taking both distinguishable states $\Phi_{ij}^{\rm uncorr}$ and
$\Phi_{ji}^{\rm uncorr}$ into account and averaging over the
distribution $\rho$ of wave packet centers, the normalization 
$N(\bbox{p}_1,\bbox{p}_2)$ coincides with the first two terms in (\ref{2.40}),
  \begin{equation}
     N(\bbox{p}_1,\bbox{p}_2) =  
     \int d^4x\, S_{\rm wp}(x,\bbox{p}_1)\,  
     \int d^4y\, S_{\rm wp}(y,\bbox{p}_2)\, .
  \label{2.45}
  \end{equation}
The two-particle correlator then coincides with the basic relation 
(\ref{2.24}) after identifying $S_{\rm wp}\equiv S$. We note already
here that starting from a discrete finite set of $N$ emission points 
$\lbrace(\check{\bbox{p}}_i, \check{\bbox{r}}_i,\check{t}_i)
\rbrace_{i\in [1,N]}$, rather than averaging over a smooth
distribution $\rho$, the expression for the two-particle correlator
(\ref{2.24}) receives finite multiplicity
corrections~\cite{Weal97}. These will be discussed in
section~\ref{sec2f}. 

\subsection{An example: the Zajc model}
\label{sec2c3}

We illustrate the consequences of the above folding relation 
(\ref{2.41}/\ref{2.42}) with a simple model emission function 
first proposed by Zajc~\cite{Z93}:
  \begin{eqnarray}
  \label{2.46}
    S(\bbox{r},t,\bbox{p}) &=& {\cal N}_s\, 
    \exp\left[-\frac{1}{2(1-s^2)} 
        \left( \frac{\bbox{r}^2}{R_0^2} 
            -2s\frac{\bbox{r}\cdot\bbox{p}}{R_0 P_0}
            +  \frac{\bbox{p}^2}{P_0^2} \right) \right]\, \delta(t)\, ,
               \nonumber \\
    {\cal N}_s  &=& E_p\, 
    {N\over (2\pi R_s P_0)^3},\quad R_s \equiv R_0\sqrt{1-s^2}\, . 
  \end{eqnarray}
This emission function is normalized to a total event multiplicity $N$.
The parameter $s$ smoothly interpolates between completely uncorrelated 
($s \to 0$) and completely position-momentum correlated ($s \to 1$)
sources. In the limit $s\to 0$, this emission function can be
considered as a quantum mechanically allowed Wigner function as long
as $R_0\, P_0 \geq \hbar/2$. In the opposite limit,
  \begin{equation}
  \label{2.47}
    \lim_{s\to 1}  S(x,\bbox{p}) \sim \delta^{(3)}
    \left(\frac{\bbox{x}}{R_0} - \frac{\bbox{p}}{P_0}\right)\,\delta(t)\, ,
  \end{equation}
the position-momentum correlation is perfect, and the phase-space 
localization described by the model is no longer consistent with the 
Heisenberg uncertainty relation. Inserting the model emission 
function (\ref{2.46}) into the general expression (\ref{2.24}) for the 
two-particle correlator one finds~\cite{Z93}
 \begin{eqnarray}
    C(\bbox{q},\bbox{K}) &=& 1 + \exp 
    \Bigl( -R_{\rm HBT}^2\, \bbox{q}^2  \Bigr)\, ,
 \label{2.48}\\
    R_{\rm HBT}^2 &=& R_s^2 \left(1-{1\over(2 R_s P_0)^2}\right)\, .
 \label{2.49}
 \end{eqnarray}
For sufficiently large $s$ this leads to an unphysical rise of the 
correlation function with increasing $\bbox{q}^2$. One can argue
\cite{ZWSH97,GEHW98} that the sign change in (\ref{2.49}) is directly
related to the violation of the uncertainty relation by the emission
function (\ref{2.46}). 

If one does not interpret (\ref{2.46}) directly as the emission function
$S(x,\bbox{p})$, but as a classical phase-space distribution 
$\rho(\check{\bbox{r}},\check{\bbox{p}})$ of Gaussian wave packets centers, 
then the correlator is readily calculated via (\ref{2.41}) \cite{GEHW98}:
  \begin{eqnarray}
    C(\bbox{q},\bbox{K}) &=& 1 + \exp \Bigl( -R_{\rm HBT,\sigma}^2\,
                                      \bbox{q}^2 \Bigr)\, ,
  \label{2.50} \\
    R_{\rm HBT, \sigma}^2 &=& R^2 \left(1-{1\over (2RP)^2}\right)\, ,
  \label{2.51}\\
    R^2 &\equiv& R_s^2+{\sigma^2\over 2},\quad
    P^2 \equiv P_0^2 + {1\over 2\sigma^2}\, .
  \end{eqnarray}
Now $2RP\geq 1$ independent of the value of $\sigma$, and the radius
parameter is always positive. Even if the classical distribution
$\rho(\check{\bbox{r}},\check{\bbox{p}})$ is sharply localized in
phase-space, its folding with minimum-uncertainty wave packets leads
to a quantum mechanically allowed emission function $S(x,\bbox{p})$
and a correlator with a realistic fall-off in $\bbox{q}$. 

\subsection{Spatial localization of wave packets}
\label{sec2c4}

Both the two-particle correlator and the one-particle spectrum 
calculated from (\ref{2.41}) depend on the initial spatial localization 
$\sigma$ which is a free para\-me\-ter. One easily sees that both limits
$\sigma \to 0$ and $\sigma \to \infty$ lead to unrealistic physical
situations: 

In the limit $\sigma \to 0$, the wave-packets is sharply localized in
coordinate space, and the momenta $\check{\bbox{p}}_i$ drop out of all
physical observables. The one-particle spectrum $E_p\, dN/d^3p$ comes
out momentum-independent irrespective of the range of the wave packet
momenta $\check{p}_i$. The momentum correlations read~\cite{Weal97,ZWSH97}
  \begin{eqnarray}
    && \lim_{\sigma\to 0} C(\bbox{p}_1,\bbox{p}_2) = 1 +
    \nonumber \\
    && \int {\cal D}\rho_i\, {\cal D}\rho_j
     \cos{\left[{ (\check{\bbox{r}}_i{-}\check{\bbox{r}}_j)\cdot 
          (\bbox{p}_1{-}\bbox{p}_2) - (\check{t}_i{-}\check{t}_j)\cdot 
          (E_1{-}E_2) }\right]}\, .
    \label{2.52}
  \end{eqnarray}
Due to the cosine term, the dependence of the two-particle correlator 
on the {\it measured} relative energy $E_1-E_2$ and momentum 
$\bbox{p}_1 - \bbox{p}_2$ gives information on the {\it initial} 
spatial and temporal relative distances in the source. This is the HBT
effect. On the other hand (\ref{2.52} shows that in this limit 
the correlator does not depend on the pair momentum $\bbox{K}$,
since position eigenstates cannot carry momentum information.
 
In the other limiting case $\sigma \to \infty$, the wave packets
are momentum eigenstates which contain no information about the 
emission points $\check{\bbox{r}}_i$. In this limit, nothing can be 
said about the spatial extension of the source, since the wave packets
show an infinite spatial delocalization. A calculation shows that also 
the temporal information is lost in this case: 
  \begin{equation}
    \lim_{\sigma\to \infty} C(\bbox{p}_1,\bbox{p}_2) = 1 +
           \delta_{\bbox{p}_1,\bbox{p}_2}\, .
    \label{2.53}
  \end{equation}

Clearly, physical applications of the wave packet formalism require
finite values of $\sigma$. For example, one can use the Gaussian 
(\ref{2.46}) with $s = 0$ to generate the distribution of wave packet
centers. Writing $P_0^2 = 2mT$ to allow for an intuitive interpretation 
of its momentum dependence in terms of a non-relativistic thermal
distribution of temperature $T$, the one-particle spectrum shows again
thermal behaviour $\sim \exp\left( -\bbox{p}^2/ 2mT_{\rm eff}\right)$,
but with a shifted temperature~\cite{MP97,Weal97,ZWSH97}
  \begin{equation}
    T_{\rm eff} = T + {1\over 2m\sigma^2}\, .
    \label{2.54} 
  \end{equation}
The corresponding HBT radius parameter reads
  \begin{equation}
    R_{\rm HBT, \sigma}^2 = R_0^2 + {\sigma^2\over 2}
     {2mT\sigma^2\over {1 + 2mT\sigma^2}}\, .
    \label{2.55}
  \end{equation}
The second terms in these equations reflect the contributions from
the intrinsic momenta and spatial extension of the wave packets.

This shows that repairing possible violations of the uncertainty relation
in a given classical phase space distribution by smearing it with
Gaussian wave packets of finite size $\sigma$, one changes both
the single-particle momentum spectra and two-particle correlations.
While large values of $\sigma$ strongly affect the source size and thereby
the HBT-radii but have little effect on the slope of the single-particle 
spectrum, the opposite is true for small values of $\sigma$. With 
both quantities fixed by experiment, one has therefore limited 
freedom in the choice of $\sigma$.

Different attempts to give physical meaning to the parameters $\sigma$
can be found in the literature. For example, Goldhaber et 
al.~\cite{GGLP} argued that the HBT-radius measured in
$p\bar{p}$ annihilation at rest can be interpreted in terms of
the pion Compton wavelength. Baym recently tried
to associate $\sigma$ with the coherence length for phase
coherence in the source~\cite{Baym98}. On the other hand, 
(\ref{2.54}), (\ref{2.55}) show that (at least in Gaussian models)
the physical observables have a functional dependence on only
two independent combinations of the three paramters $T$, $R_0$ and 
$\sigma$. In practice, this allows to reabsorb the wave packet width
in a redefinition of the source parameters~\cite{H98,WFH98}.

\section{Multiparticle symmetrization effects}
\label{sec2d}

Multiparticle symmetrization effects are contributions to the
spectra of Bose-Einstein symmetrized $N$-particle states which 
cannot be written in terms of simpler pairwise ones. In 
many-particle systems with high phase-space density, the 
single- and two-particle spectra receive non-negligible
contributions from multiparticle symmetrization effects.
This complicates the interpretation of the emission function
$S(x,K)$ as reconstructed from the data.

Based on strategies proposed by Zajc~\cite{Z84,Z87} and Pratt
\cite{P93,PZ94}, there exists by now an extensive literature on these
effects consisting of numerical~\cite{P93,PZ94,Z84,Z87,Zh93} and
analytical~\cite{BK95,AL96,FW97,Wos97,Z97,W98,ZSH98} model
studies. Multiparticle symmetrization effects have been considered
essentially in two different settings. Either one starts from
events which at freeze-out have a fixed particle multiplicity
$N$~\cite{Z87,W98,Z98fix}, encoded e.g. in the model assumptions 
by choosing sets of $N$ phase-space points $(\check{\bbox{p}}_i,
\check{\bbox{r}}_i,\check{t}_i)$. Bose-Einstein correlations in the final state
then lead to an enhancement of the two-particle correlator at
small relative pair momentum, but they do not affect the particle
multiplicity. A second approach~\cite{P93,CGZ95,CZ97,ZC97,Z97} does not only 
calculate the HBT enhancement effect of identical particles, but
aims at accounting for the effects of Bose-Einstein statistics
during the particle production processes as well. As a result,
modifications of the multiplicity distribution of event samples
are calculated.

Here, we first review the formalism for fixed event multiplicities,
which is tailored to calculate the final state HBT effect only.
Then we discuss shortly how this formalism can be adapted to 
calculate changes of multiplicity distributions.

\subsection{The Pratt formalism}
\label{sec2d1}

In his original calculation~\cite{P93,prattrev}, Pratt starts from
the (unnormalized) probability $\tilde{\cal P}_N(\vec{\bbox{p}})$ for 
detecting $N$ particles with momenta $\vec{\bbox{p}} = (\bbox{p}_1,
\dots,\bbox{p}_N)$. It is expressed through single particle
production amplitudes $T_{a}(x)$ for particles with quantum numbers
$a$ and $N$-particle symmetrized plane waves
$U(x_1,\dots,x_N;p_1,\dots,p_N)$ as follows: 
  \begin{eqnarray}
     && \tilde{\cal P}_N(\vec{\bbox{p}}) =
  \label{2.56}\\
     && \quad\sum_{\lbrace a_i\rbrace}
        \left\vert \int d^4x_1 \cdots  d^4x_N\, 
        T_{a_1}(x_1)\cdots  T_{a_N}(x_N)\,
     U(x_1,\dots,x_N;p_1,\dots,p_N)\right\vert^2 ,  
  \nonumber\\
     && U(x_1,\dots,x_N;p_1,\dots,p_N) = {1\over \sqrt{N!}}\, 
        \sum_{s\in {\cal S}_N}
     \prod_{j=1}^N \exp\left[i\, p_j \cdot x_{s_j}\right]\, . 
     \label{2.57}
  \end{eqnarray}
The sum runs over all permutations $s$ of $N$ particles.
The main assumptions entering here are (i) the absence of final
state interactions which allows the plane wave ansatz (\ref{2.57}),
and (ii) the assumption of independent particle emission which
allows to factorize the $N$-particle production amplitude into
$N$ one-particle production amplitudes $T_a(x)$.
It is technically convenient to change from these 
to the corresponding Wigner transformed emission 
function~\cite{P93,prattrev}
  \begin{equation}
    S_T(x,p) = \sum_a\int d^4y\, e^{ip{\cdot}y}\, 
    T_a^*(x+\textstyle{y\over 2})\, T_a(x-\textstyle{y\over 2})\, .
    \label{2.58}
  \end{equation}
Calculating $\tilde{\cal P}_2(\bbox{p}_1, \bbox{p}_2)$, one 
recovers in this formalism up to a normalization factor the usual 
expression (\ref{2.20}) for the two-particle spectrum. 
This is, however, not the relevant calculation
because it gives only the two-particle spectrum from events with
exactly two particles. The aim of Pratt's formalism is to compute
the one- and two-particle spectra for events with multiplicity $N$, 
including all multiparticle symmetrization effects. They are 
obtained by integrating (\ref{2.56}/\ref{2.57}) over $N-1$ or $N-2$ 
momenta, respectively. 
We use the notation $\overline{\cal P}_N$ for $N$-particle momentum
distributions which, in contrast to (\ref{2.7}/\ref{2.8}), are 
normalized to unity. Using the following building
blocks~\cite{P93} 
  \begin{eqnarray}
    G_1(p_1,p_2) &=& \int d^4x\, S_T(x,K)\, e^{-iq\cdot x}\, ,
    \label{2.59} \\
    G_n(p_1,p_2) &=& \int {d^3k_2\over E_2} \cdots 
    {d^3k_n\over E_n}\, G_1(p_1,k_2)\, G_1(k_2,k_3) \cdots
    G_1(k_n,p_2)\, ,
    \label{2.60} \\
    C_n &=& \int {d^3p\over E_p}\, G_n(p,p)\, ,
    \label{2.61}
  \end{eqnarray}
one obtains the desired spectra by the following
algorithm~\cite{Z87,P93,CGZ94,CZ97,ZC97,W98}: 
  \begin{eqnarray}
    &&\overline{\cal P}_N^{1}(\bbox{p}) = 
    {1\over N} {1\over \omega(N)} \sum_{m=1}^N 
    \omega(N-m)\, G_m(p,p)\, ,
    \label{2.62} \\
    &&\overline{\cal P}_N^{2}(\bbox{p}_1,\bbox{p}_2) = 
        {1\over N(N-1)} {1\over \omega(N)}
    \sum_{J=2}^N \omega(N-J)
    \nonumber \\ 
    && \ \ \times \sum_{i=1}^{J-1} 
    \Big\lbrack
         G_i(p_1,p_1)G_{J-i}(p_2,p_2)
    + G_i(p_1,p_2)G_{J-i}(p_2,p_1)
    \Big\rbrack \, ,
    \label{2.63}\\
    &&\omega(N) = {1\over N!} 
        \int d^3p_1\cdots d^3p_N\, \tilde{\cal P}_N(\vec{\bbox{p}})
        = \sum_{(n,l_n)_N} 
    { C_1^{l_1}\, C_2^{l_2} \cdots C_n^{l_n} \over 
        {\prod_n n^{l_n} (l_n!)}}
        \nonumber \\
    && \quad \quad  = {1\over N} \sum_{m=1}^N \omega(N-m)\, C_m\, .
    \label{2.64}
  \end{eqnarray}
While the sum in (\ref{2.57}) runs over $N!$ terms,
this algorithm involves only sums over all partitions 
$(n,l_n)_N$ of $N$ elements; this reduces 
the complexity of the problem considerably. The algorithm 
(\ref{2.62})-(\ref{2.64}) is sometimes referred to 
as ``ring algebra''~\cite{CZ97,ZC97}, 
since the building blocks $C_m$ and $G_m$ have a very simple diagrammatic 
representation in terms of closed and open rings~\cite{P93,R98,W98}. 
It is also referred to as Zajc-Pratt algorithm, since Zajc had
analyzed essential parts of the above combinatorics in~\cite{Z87}.
The definition of the $C_m$ sometimes differs by a factor $m$
from the one given here, which results in appropriately modified
combinatorial factors in (\ref{2.62})-(\ref{2.64}). 

While the set of equations (\ref{2.62}) - (\ref{2.64})
constitutes a great simplification over a direct evaluation of
(\ref{2.56}), 
the high-dimensional integrations required to determine $G_m$ in
(\ref{2.60}) still limit its applications significantly. Numerical
Monte Carlo techniques have been proposed~\cite{P93,PZ94,FWW98} to 
calculate (\ref{2.60}). An alternative strategy can be applied
to a small class of simple (Gaussian) models, where one can control 
the $m$-dependence
of $G_m$ analytically or via simple recursion schemes. Especially
for Gaussian emission functions, (\ref{2.60}) allows for simple
one-step recursion relations~\cite{P93,PZ94,CGZ95,Z97} between 
$G_{n+1}$ and $G_n$ which can be solved analytically~\cite{ZC97}.

\subsection{Multiparticle correlations for wave packets}
\label{sec2d2}

There have been several recent attempts to combine the Pratt
formalism with an explicit parametrization of the source in
terms of $N$-particle Gaussian wave packets~\cite{ZC97,CZ97,W98}.
The strategy in these studies is to associate with each {\it event}
$\lbrace(\check{\bbox{r}}_i,\check{t}_i,\check{\bbox{p}}_i)
\rbrace_{i\in [1,N]}$
a properly symmetrized $N$-particle wave function~\cite{ZC97,CZ97,W98}
  \begin{equation}
    \lbrace(\check{\bbox{r}}_i,\check{t}_i,\check{\bbox{p}}_i)
    \rbrace_{i\in [1,N]}
    \longrightarrow
    \Psi_N(\vec{\bbox{x}},t) = {1\over \sqrt{N!}}\, 
                          \sum_{s\in {\cal S}_N} 
              {\left({ \prod_{i=1}^N f_{s_i}(\bbox{x}_i,t) }\right)}\, ,
    \label{2.65}
  \end{equation}
where the $f_i$ are the Gaussian wave packets of (\ref{2.31}). Note
that the normalization 
${\cal N}_{\Psi} = 1/ {\langle{\Psi_N\vert\Psi_N}\rangle}$ of 
$\Psi_N$ depends on the positions $\lbrace(\check{\bbox{r}}_i,
\check{t}_i,\check{\bbox{p}}_i)\rbrace_{i\in [1,N]}$ of the wave
packet centers in phase-space. As we shall see, this prevents a  
straightforward application of the Pratt formalism.

The normalized probability 
$\overline{\cal P}_N(\vec{\bbox{p}};{\lbrace \check{z}\rbrace})$ for
detecting $N$ particles with momenta 
$\vec{\bbox{p}} = (\bbox{p}_1,\dots,\bbox{p}_N)$ in the specific wave
packet configuration $\lbrace \check{z}\rbrace \equiv \lbrace
(\check{\bbox{r}}_i,\check{t}_i,\check{\bbox{p}}_i)\rbrace_{i\in [1,N]}$ 
takes the following form~\cite{W98}
  \begin{equation}
    \overline{\cal P}_N(\vec{\bbox{p}};{\lbrace \check{z}\rbrace}) 
    = {{\cal N}_{\Psi}\over N!}\, \sum_{s,s'\in {\cal P}_n}
    \prod_{l=1}^N {\cal F}_{s'_l\, s_l}(\bbox{p}_l)\, ,
    \label{2.69} 
  \end{equation}
where the building blocks ${\cal F}_{ij}(\bbox{p})$ are given in
terms of the Fourier transforms $\tilde{f}_i$ of the 
single-particle wave packets as follows:
  \begin{eqnarray}
    {\cal F}_{ij}(\bbox{p}) &=&
    D_i(\bbox{p},t)\, D_j^*(\bbox{p},t)\, ,
    \label{2.70}\\
    D_i(\bbox{p},t) &=& e^{-iE_p(t-\check{t}_i)}
                     \tilde{f}_i(\bbox{p},\check{t}_i)\, .
                     \label{2.71} 
  \end{eqnarray}
The time dependence of (\ref{2.71}) drops out in ${\cal F}_{ij}$.
From (\ref{2.69}) the normalized one- and two-particle momentum 
distributions are obtained by integrating over the unobserved 
momenta~\cite{Z87,W98}:
  \begin{eqnarray}
    \overline{\cal P}^1_N(\bbox{p}_1;{\lbrace \check{z}\rbrace}) 
    &=& {\cal N}_{\Psi}
    {1\over N!} \sum_{s,s'\in {\cal S}_N} 
    {\cal F}_{s'_1\, s_1}(\bbox{p}_1)
    \prod_{l=2}^N f_{s'_l\, s_l}\, , 
  \label{2.72} \\
    \overline{\cal P}^2_N(\bbox{p}_1,\bbox{p}_2;
                          {\lbrace \check{z}\rbrace}) 
    &=& {\cal N}_{\Psi}
    {1\over N!} \sum_{s,s'\in {\cal S}_N} {\cal F}_{s_1's_1}(\bbox{p}_1)
    {\cal F}_{s_2's_2}(\bbox{p}_2)
    \prod_{l=3}^N f_{s_l's_l}\, .
    \label{2.73} \\
    f_{ij} &=& \int d^3p\, {\cal F}_{ij}(\bbox{p})\, .
    \label{2.74}
  \end{eqnarray}
Similarly, higher order particle spectra $\overline{\cal P}_N^m$
contain $m$ factors ${\cal F}_{s'_i\, s_i}$ in each term. 

The factors $\prod f_{ij}$ occurring in (\ref{2.72}/\ref{2.73})
reflect the multiparticle symmetrization effects on the one- and
two-particle spectra. They involve the overlap between pairs of
wave packets $f_i$, $f_j$, which for the simple case of instantaneous
emission, $\check{t}_i = \check{t}_j$, take the simple form 
  \begin{eqnarray}
    \vert f_{ij}\vert &=&  
    \exp \left[{-\textstyle{1\over 4} \vert 
      \check{\bbox{z}}_i-\check{\bbox{z}}_j\vert^2}\right]\, , 
    \label{2.75} \\ 
    \check{\bbox{z}}_j &=& {1\over \sigma}\check{\bbox{r}}_j 
                  + i\sigma \check{\bbox{p}}_j\, .
    \label{2.76}
  \end{eqnarray}
This overlap equals 1 for $i = j$ and decreases like a Gaussian
with increasing phase-space distance 
$\vert \check{\bbox{z}}_i-\check{\bbox{z}}_j\vert$ between the wave packet
centers. According to (\ref{2.76}), this distance depends on the 
wave packet width $\sigma$, and in the limiting cases $\sigma \to 0$
and $\sigma \to \infty$, the overlap functions reduce to what
is known as the 
  \begin{equation}
    \hbox{pair approximation: }\, \, f_{ij} = \delta_{ij}\, .
    \label{2.77}
  \end{equation}
As we will see, these limits correspond to the case of infinite
phase-space volume, i.e., vanishing phase-space density of the
source. Then all sums over ${\cal S}_N$ 
in (\ref{2.72}/\ref{2.73}) are trivial and the two-particle 
spectrum (\ref{2.73}) reduces to a sum over all particle
pairs, involving only two-particle symmetrized contributions,
thus coinciding~\cite{W98} with the correlator derived in 
section~\ref{sec2c}. 

In order to apply the Zajc-Pratt algorithm (\ref{2.62})-(\ref{2.64}),
the distributions given above must be averaged over the phase-space
positions $(\check{\bbox{r}}_i,\check{t}_i,\check{\bbox{p}}_i)$ of the
wave packet centers:
  \begin{eqnarray}
    \overline{\cal P}_N^{n}(\bbox{p}_1,\dots,\bbox{p}_n)
    &=& \int {\left({ \prod_{i=1}^N {\cal D}\rho_i }\right)} \,
        {\cal P}_N^{n}(\bbox{p}_1,\dots,\bbox{p}_n;
                       {\lbrace\check{z}\rbrace})\, .
        \label{2.78}
  \end{eqnarray}
where ${\cal D}\rho_i$ is defined in (\ref{2.79}). This is the
analogue of the sum over quantum numbers $a_i$ in (\ref{2.56}). At
this point, one encounters the problem that the normalization 
${\cal N}_{\Psi}$ of the $N$-particle wave packet does not
factorize. This destroys the factorization property (\ref{2.60}) of
the Pratt algorithm. However, an analytical calculation becomes
possible if a different averaging procedure is used instead of
(\ref{2.78}): 
  \begin{eqnarray}
    &&\prod_{i=1}^N {\cal D}\rho_i
    \longrightarrow {\cal D}\tilde{\rho} =
    {\langle\Psi_N\vert\Psi_N\rangle \over \omega(N)}\, 
    \prod_{i=1}^N 
    \rho(\check{\bbox{r}}_i,\check{t}_i,\check{\bbox{p}}_i)\, ,
  \label{2.80}\\
    &&\omega(N) = \int \left( \prod_{i=1}^N
      d^3\check{p}_i\, d^3\check{r}_i\, d\check{t}_i\,
       \rho(\check{\bbox{r}}_i,\check{t}_i,\check{\bbox{p}}_i)\right)
     \langle \Psi_N\vert\Psi_N\rangle\, .
  \label{2.81}
  \end{eqnarray}
This modification (\ref{2.80}) is equivalent to working with
unnormalized $N$-particle wave packet states, as done e.g. in
Ref.~\cite{CGZ95}. Zim\'anyi and Cs\"org\H{o} \cite{ZC97} have tried
to give this modification a simple physical interpretation by noting
that the factor ${\langle\Psi_N\vert\Psi_N\rangle}$ can be interpreted
as an enhanced emission probability for bosons (described by
normalized wave packets) which are emitted close to each other in
phase-space. According to (\ref{2.80}), which no longer factorizes, 
this version of ``stimulated emission'' leads to specific correlations
among the emission points
$(\check{\bbox{r}}_i,\check{t}_i,\check{\bbox{p}}_i)$, i.e., the
particles are not emitted independently.

For the average (\ref{2.80}), the spectra $\overline{\cal P}_N^n$
can be obtained from the Zajc-Pratt algorithm 
(\ref{2.62})-(\ref{2.64}) using the building blocks
  \begin{eqnarray}
    G_m(\bbox{p}_1,\bbox{p}_2) &=& \int {\left( \prod_{l=1}^m
    {\cal D}{\rho}_{i_l} \right)}\, D_{i_1}^*(\bbox{p}_1) 
    f_{i_1i_2}\, f_{i_2i_3} \, ...\,
    f_{i_{m-1}i_m} D_{i_m}(\bbox{p}_2)\, ,
    \label{2.82} \\
    C_m &=& \int d^3p\,  G_m(\bbox{p},\bbox{p})\, .
    \label{2.83}
  \end{eqnarray}
One only needs to replace in (\ref{2.59}) Pratt's definition (\ref{2.58})
for the single-particle Wigner density, $S_T(x,K)$, by the wave packet
analogue $S_{\rm wp}(x,\bbox{K})$ given in (\ref{2.41}).

\subsection{Results of model studies}
\label{sec2d3}

Explicit numerical~\cite{P93,PZ94,Z97} and analytical~\cite{ZC97,CZ97,W98}
calculations of multiparticle symmetrization effects
have so far only been performed for Gaussian source models.
In this case the m-th order Pratt terms $G_m$, see (\ref{2.60}),
can be calculated analytically. Writing them in the form
  \begin{equation}
    G_m(\bbox{K}+{\textstyle{1\over 2}}\bbox{q},
    \bbox{K}-{\textstyle{1\over 2}}\bbox{q}) 
    = C_m\, (g_K^{(m)}/\pi)^{3/2}
    \exp\left[ -g_q^{(m)}\bbox{q}^2 - g_K^{(m)}\bbox{K}^2\right]\, ,
    \label{2.84}
  \end{equation}
the coefficients $g_q^{(m)}$ and $g_K^{(m)}$ can then be obtained
from simple recursion relations~\cite{ZC97}.
Two generic features are observed in all 
these studies~\cite{Z87,P93,CZ97,Z97,W98}:
  \begin{enumerate}
    \item
      For increasing $m$ the factors $g_K^{(m)}$ become larger.
      This leads to steeper
      local slopes of the one-particle spectrum for small momenta.
      Multiparticle symmetrization effects thus enhance the particle
      occupation at low momentum. 
    \item
      For increasing $m$ the factors $g_q^{(m)}$ decrease. This 
      broadens the width of the two-particle correlator, indicating
      that multiparticle symmetrization effects 
      lead to an enhanced probability of finding particles close together 
      in configuration space.
  \end{enumerate}
While these observations are generic, their quantitative aspects 
are model dependent and sensitive to the particle phase-space 
density. Writing the single-particle spectrum (\ref{2.62}) in the
form~\cite{W98}
  \begin{equation}
    \overline{\cal P}_N^{1}(\bbox{p}) =
    \sum_{m=1}^N v_m G_m(\bbox{p},\bbox{p})/C_m\, ,\qquad
     v_m = {\omega(N-m)\over \omega(N)}\, C_m\,  ,
     \label{2.86}
  \end{equation}
the weights $v_m$, which satisfy  $\sum_{m=1}^N\, v_m = 1$,
can be analyzed in the limit of large phase-space volumes.
For a Gaussian source with width parameters $R_0$ and $P_0$ in
coordinate and momentum space, the phase-space
density is given by $\rho_{\rm ps} = N/(R_0 P_0)^3$. In the limit
of large phase-space volume $(R_0 P_0)^3 \gg 1$ one finds 
for fixed (but not necessarily small) multiplicity $N$~\cite{W98}: 
  \begin{equation}
    v_m \approx {\rho_{\rm ps}^{m-1}
         \over {(1 + \rho_{\rm ps})^m}} \, .
    \label{2.87}
  \end{equation}
Similarly, the two-particle spectrum can be written as~\cite{W98} 
  \begin{eqnarray}
    \overline{\cal P}_N^{2}(\bbox{p}_1,\bbox{p}_2) &=&
    \sum_{m=2}^N u_m 
    \sum_{i=1}^{m-1}
    H_{i,m-i}(\bbox{p}_1,\bbox{p}_2)\, ,
    \label{2.89}\\
    H_{i,m-i}(\bbox{p}_1,\bbox{p}_2) &=& 
         {G_i(\bbox{p}_1,\bbox{p}_1)\over C_i}
         {G_{m-i}(\bbox{p}_2,\bbox{p}_2)\over C_{m-i}}
         \nonumber \\
    && + {G_i(\bbox{p}_1,\bbox{p}_2)\over C_i}
      {G_{m-i}(\bbox{p}_2,\bbox{p}_1)\over C_{m-i}} \, .
      \label{2.90}
  \end{eqnarray}
Again the weights $u_m$ are normalized to unity, $\sum_{m=2}^N
u_m = 1$, and in the same limit as above their leading behaviour
is given by
  \begin{equation}
    u_m \approx
        { \rho_{\rm vol}^{m-2} \over 
          (1 + \rho_{\rm vol})^m}
        = { v_{m-1} \over 
          (1 + \rho_{\rm vol})}\, .
        \label{2.91}
  \end{equation}
We finally remark, that the correlator obtained from (\ref{2.86})
and (\ref{2.89}) takes the generic form~\cite{W98,ZSH98}
  \begin{equation}
        C(\bbox{p}_1,\bbox{p}_2) = 
        { \overline{\cal P}_N^2(\bbox{p}_1,\bbox{p}_2)\over
        \overline{\cal P}_N^1(\bbox{p}_1) \overline{\cal P}_N^1(\bbox{p}_2)}
        = {\cal N} \Bigl( 1 + \lambda\, F(\bbox{p}_1,\bbox{p}_2) \Bigr)\, ,
        \label{2.91b}
  \end{equation}
where $\lambda = 1 $, ${\cal N} \approx 1 - (R_0 P_0)^{-3}$ for large
phase-space volumes, and $F(\bbox{p}_1,\bbox{p}_2)$ approaches 1 and
0 in the limits $\bbox{q} \to 0$ and $\vert\bbox{q}\vert \to \infty$, 
respectively. 

\subsection{Bose-Einstein effects and multiplicity distributions}
\label{sec2d4}

So far we only discussed multiparticle symmetrization effects
for events with fixed multiplicity $N$. The question to what extent 
multipion correlations are also reflected in the multiplicity 
distributions was asked already in the 
seventies~\cite{GK78}. Recently, it was
revived in the context of Pratt's formalism~\cite{P93,CZ97,Z97}
with the aim to calculate the effect of Bose-Einstein statistics
on the particle production process. These applications typically 
start~\cite{P93,ZC97,Z97} from a  multiplicity distribution 
$p_n^{(0)}$ in the absence of Bose-Einstein statistics, for example 
a Poisson distribution 
$p_n^{(0)} = \textstyle{n_0^n\over n!}\exp\lbrack -n_0\rbrack$
with average multiplicity $n_0$. For this case the probability 
$p_n$ of finding events with multiplicity $n$ after having accounted 
for Bose-Einstein correlations is then computed as~\cite{P93,CZ97}
  \begin{equation}
    p_n = \omega(n) \left( \sum_{k=0}^\infty \omega(k)\right)^{-1}\, ,
    \label{2.92}
  \end{equation}
where $\omega(k)$ is given in (\ref{2.64}). For this particular
multiplicity distribution, the multiplicity averaged one- and
two-particle spectra are given by the simple 
expressions~\cite{CGZ94,CGZ95,ZC97,ZSH98}
  \begin{eqnarray}
        {\cal P}_1(\bbox{p}) &=& H(\bbox{p},\bbox{p})\, ,
        \label{2.92b} \\
        {\cal P}_2(\bbox{p}_1,\bbox{p}_2) 
        &=&  H(\bbox{p}_1,\bbox{p}_1)\,  H(\bbox{p}_2,\bbox{p}_2) +
             H(\bbox{p}_1,\bbox{p}_2)\,  H(\bbox{p}_2,\bbox{p}_1)\, , 
        \label{2.92c} 
  \end{eqnarray}
where
  \begin{equation}
        H(\bbox{p}_1,\bbox{p}_2) = \sum_{m=1}^{\infty}
        G_m(\bbox{p}_1,\bbox{p}_2)
         \equiv \int d^4x\, S(x,\bbox{K})\,
        e^{iq\cdot x}\, .
        \label{2.92d}
  \end{equation}
With the effective source distribution $S(x,\bbox{K})$ introduced in
the second step, the correlator again takes the simple form (\ref{1.4}).
We expect that this source distribution coincides with the one
defined in (\ref{2.22a}) in the context of the covariant classical
current formalism. The reasons are that (i) both satisfy (\ref{2.24}
and (ii) that the coherent states $\vert J\rangle$ of (\ref{2.17})
generate a Poissonian multiplicity distribution. 

According to the first equation (\ref{2.92d}), the emission function
$S(x,\bbox{K})$ contains all multiparticle symmetrization effects.
Expressed in terms of the single-particle Wigner density $S_{\rm wp}$
in (\ref{2.41}), it takes a complicated form. Model 
studies~\cite{P93,PZ94,CGZ94,CGZ95,CZ97,Z97,ZSH98} indicate that
irrespective of the particular multiplicity distribution the
general features discussed below (\ref{2.84}) persist: compared to
the input distribution $S_{\rm wp}(x,\bbox{K})$, the multiparticle
symmetrized emission function $S(x,\bbox{K})$ is more strongly
localized in both coordinate and momentum space. For the 
intercept parameter $\lambda$ one finds results which depend
on the specific choice for the multiplicity distribution. Cases
are known where $\lambda$ decreases
strongly with increasing phase-space density~\cite{P93,prattrev,CZ97,Z97}. 

This discussion illustrates that for sources with high phase-space 
density, where multiparticle symmetrization effects cannot be
neglected, the interpretation of the emission function $S(x,K)$
reconstructed from the one- and two-particle momentum spectra
by analyzing (\ref{1.3})-(\ref{1.5}) is not straightforward.
The question how Bose-Einstein effects on the multiplicity distribution
and on the phase-space distribution can be disentangled is still open.
In the remainder of this review we therefore concentrate on the
reconstruction of $S(x,K)$ and do not further consider its possible
contamination by multiparticle effects. 

\section{Final state interactions}
\label{sec2e}

Momentum correlations between identical particles can originate
not only from quantum statistics but also from conservation laws
and final state interactions.

Energy-momentum conservation constrains the momentum distribution
of produced particles near the kinematical boundaries. In high
multiplicity heavy-ion collisions its effects on two-particle
correlations at low relative momenta are negligible. Similarly,
constraints from the conservation of quantum numbers (e.g. charge or 
isospin) become less important with increasing event
multiplicity. Strong correlations exist between the decay products
of resonances, but since resonance decays rarely lead to the production 
of identical particle pairs, they do not matter in practice.  

This leaves final state interactions as the most important source 
of dynamical correlations. For the small relative momenta $q < 100$
MeV which are sampled in the two-particle correlator, effects of
the strong interactions are negligible for pions. For protons,
however, they dominate the two-particle correlations. On the other
hand, for pions, the long-range Coulomb interactions distort 
significantly the observed momentum correlations, dominating over the
Bose-Einstein effect for small relative momenta.
Here we discuss how Coulomb correlations are calculated for a given
source function and how they can be corrected for in the data.
The aim of Coulomb corrections is to modify the measured two-particle
correlations in such a way that the resulting correlator contains
only Bose-Einstein correlations, while the effects of final state 
interactions have been subtracted. For this, several simplified
procedures have been used in the literature, which we review in
what follows.

\subsection{Classical considerations}
\label{sec2e1}

The Coulomb interaction between particle pairs accelerates them
relative to each other, thus depleting (enhancing) the two-particle
correlation function at small relative momenta for like-sign 
(unlike-sign) pairs. In a high multiplicity
environment this final state interaction can be reduced by screening
effects until the particle pair has separated sufficiently from
the rest of the system. Both these effects can be taken into
account in a classical toy model which neglects the Coulomb
interaction between the pairs for separations less than $r_0$ and
includes it for larger separations~\cite{Baym96}. The initial and
the finally observed relative momenta $\bbox{q}_0$ and $\bbox{q}$ are 
then related by
  \begin{equation}
    { {(\bbox{q}/2)^2}\over 2\, \mu}
    = { {(\bbox{q}_0/2)^2}\over 2\, \mu}
    \pm {e^2\over r_0}\, ,
    \label{2.93}
  \end{equation}
where $\mu$ is the reduced mass. For two pions,
$\mu = m_{\pi}/2$ and a radius $r_0 = 10$ fm, this results in
a shift $(\bbox{q}/2)^2 = (\bbox{q}_0/2)^2 \pm 20\, \hbox{MeV}^2$.
The modification of the two-particle correlator is then given
by the Jacobian $\vert d^3q_0/d^3q \vert = q_0/q$ and
reads~\cite{GK81,Baym96} 
  \begin{equation}
    C(\bbox{q}) = \left\vert{d^3q_0\over d^3q}\right\vert 
    C_0(\bbox{q}_0)
    = C_0(\bbox{q}_0) \sqrt{ 1 \mp {2\mu e^2\over r_0 (q/2)^2}}\, ,
  \label{2.94}
  \end{equation}
where $C_0(\bbox{q}_0)$ denotes the two-particle correlator in the
absence of Coulomb interactions. When comparing (\ref{2.94})
with the data, the radius $r_0$ can be used to accomodate 
for the dependence of 
the Coulomb final state interaction on the source size. This toy
model reproduces the qualitative features of experimental data
surprisingly well but fails to account quantitatively for the
correct $q$-dependence of the correlator at very small relative 
momenta~\cite{Baym96,Baym98}.

\subsection{Coulomb correction for finite sources}
\label{sec2e2}
 
For a quantum mechanical discussion of final state
Coulomb interactions, we associate to the emitted
particle pairs a relative Coulomb wavefunction,
given analytically by the confluent hypergeometric function $F$,
  \begin{eqnarray}
    \Phi_{\bbox{q}/2}^{\rm coul}(\bbox{r}) &=& \Gamma(1+i\eta)\, 
    e^{-{1\over 2}\pi\eta}\,
    e^{{i\over 2}\bbox{q}\cdot \bbox{r}}\, 
    F \left(-i\eta; 1; z_-\right)\, ,
    \label{2.95}\\
    z_{\pm} &=& {\textstyle{1\over 2}}(q r \pm \bbox{q}\cdot \bbox{r})
    = {\textstyle{1\over 2}}q r (1 \pm \cos\theta)\, .
    \label{2.96}
  \end{eqnarray}
Here, $r = |\bbox{r}|$, $q = |\bbox{q}|$, and $\theta$ denotes the
angle between these vectors. The Sommerfeld parameter 
$\eta = \alpha/ (v_{\rm rel}/c) $ contains the dependence on the 
particle mass $m$ and the electromagnetic coupling strength $e$; 
we write 
  \begin{equation}
    \eta_\pm = 
    \pm {e^2\over 4\pi} {\mu\over q/2}
           = \pm {m\, e^2\over 4\pi q}\, ,
    \label{2.97}
  \end{equation}
where the plus (minus) sign is for pairs of unlike-sign (like-sign)
particles.

\begin{figure}[ht] 
\epsfxsize=11.50cm 
\centerline{\epsfbox{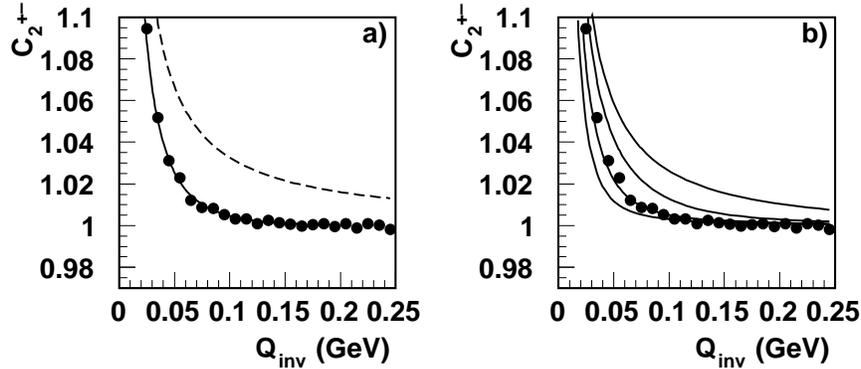} }
\caption{Unlike-sign correlations for $0 < K_\perp < 100 \,{\rm MeV/c}$ 
   and $4.0 < Y < 5.0$.  
a) Dashed line: Gamow function, corresponding to vanishing source 
   radius. Solid line: fit of the NA49 Pb+Pb data to the function
   \protect(\ref{2.100}).
b) The same data compared to calculations based on \protect(\ref{2.98})
   with a spherically symmetric Gaussian source, with $R =$ 0.5 fm, 2
   fm, 4.6 fm, and 8 fm (from top to bottom). The best fit is obtained
   for $R=4.6$ fm. Figure taken from \protect\cite{schoen}.}
\label{coulombcorr}
\end{figure}

To illustrate the influence of a finite source size in a simple case,
we take recourse to the relative source function
$S_{\bbox{K}}(\bbox{r})$ defined in (\ref{4.04}). This function
describes the probability that a particle pair with pair momentum
$\bbox{K}$ is emitted from the source at initial relative distance
$\bbox{r}$ in the pair rest frame. 
For sources without $x$-$K$-correlations and neglecting the time
structure of the particle emission process in the pair rest frame,
the corresponding two-particle correlation for non-identical 
charged particle pairs reduces to~\cite{BD97,Baym98}
  \begin{equation}
    C^{(+-)}(\bbox{q},\bbox{K}) = \int d^3r\, S_{\bbox{K}}(\bbox{r})\, 
                 \vert \Phi_{\bbox{q}/2}^{\rm coul}(\bbox{r})\vert^2\, .
    \label{2.98}
  \end{equation}
Corrections to this expression are discussed in section~\ref{sec2e4}.
For a pointlike source $S_{\bbox{K}}(\bbox{r}) =
\delta^{(3)}(\bbox{r})$, the correlator (\ref{2.98}) is given by the
Gamow factor $G(\eta)$ which denotes the square of the Coulomb
wavefunction $\Phi_{\bbox{q}/2}^{\rm coul}(\bbox{r})$ at vanishing
pair separation $\bbox{r} = 0$, 
  \begin{equation}
    G(\eta) = \Bigg\vert \Gamma(1+i\eta)\, 
              e^{-{1\over 2}\pi\eta}\Bigg\vert^2
    = {2\pi\eta\over e^{2\pi\eta} - 1}\, .
    \label{2.99}
  \end{equation}
For a Gaussian ansatz $S_{\bbox{K}}(\bbox{r}) \propto \exp\left[ 
-\bbox{r}^2/ 4R^2\right]$, the dependence of the Cou\-lomb
correlations on the size $R$ of the source is then determined via
(\ref{2.98}), see Fig.~\ref{coulombcorr}. If the particles are emitted
with finite separation $\bbox{r}$, their Coulomb interaction is weaker
and the Gamow factor overestimates the final state interaction
significantly, see Fig.~\ref{coulombcorr}. The source size thus enters
estimates of the Coulomb correction in a crucial way, and its
selfconsistent inclusion in the correction procedure can lead to
significantly modified source size estimates \cite{NA35HBT,NA35newcoulomb}.

\subsection{Coulomb correction by unlike sign pairs}
\label{sec2e3}

Large acceptance experiments can measure like-sign and unlike-sign
particle correlations simultaneously. The latter do not show Bose-Einstein
enhancement but depend on final state interactions as well.  
This opens the possibility to correct for the Coulomb correlations
in like-sign pairs by using the information contained in unlike-sign
pairs. The $\bbox{q}$-dependence of the unlike-sign correlations
$C_{\rm meas}^{(+-)}(\bbox{q},\bbox{K})$ is often parametrized
by a $q_{\rm inv}$-dependent simple function~\cite{NA35newcoulomb,lasiuk} 
  \begin{equation}
    \label{2.100}
    F(q_{\rm inv}) = 1 + \bigl(G(\eta_+) -1\bigr)\, 
    e^{-q_{\rm inv}/Q_0}\, .
  \end{equation}
where $q_{\rm inv}^2 = \bbox{q}^2 - (q^0)^2$ and $Q_0$ is a
fit parameter. $q_{\rm inv}^2$ is the square of the spatial
relative momentum in the pair rest frame, where $q^0 = 0$.
For small $q_{\rm inv}$, this function approaches
the Gamow factor (\ref{2.99}) for a pointlike source, while it
includes a phenomenological finite-size correction for large
relative momentum. More recently, one has started to avoid this
intermediate step by constructing the corrected correlator 
$C_{\rm corr}^{(--)}(\bbox{q},\bbox{K})$ 
for like-sign pairs directly using bin by bin the experimental data
from the measured like- and unlike-sign correlators~\cite{schoen,appels}:
  \begin{equation}
    C_{\rm corr}^{(--)}(\bbox{q},\bbox{K}) =
    C_{\rm meas}^{(+-)}(\bbox{q},\bbox{K})\, 
    C_{\rm meas}^{(--)}(\bbox{q},\bbox{K})\, .
    \label{2.101}
  \end{equation}
In the absence of Bose-Einstein correlations and for pointlike
sources, the lefthand side of this equation reduces to unity
while the righthand side becomes a product of Gamow factors
  \begin{equation}
    G(\eta_+)\, G(\eta_-) =
    {1\over {1 + (\pi^2/ 3)\eta^2 + O(\eta^4)}}\, .
    \label{2.102}
  \end{equation}
This expression provides an estimate for the accuracy of the
correction procedure (\ref{2.101}). It deviates from unity by
less than five percent for relative momenta 
$q > 8 \textstyle{m\over 137}$~\cite{schoen}. For pions, only the 
region $q < 10$ MeV is affected significantly, while for the more 
massive kaons the whole region $q < 25$ MeV shows an error larger 
than five percent. Calculating the
correction factors in (\ref{2.101}) for extended sources,
this picture does not change since the main difference
between like-sign and unlike-sign correlations is due to
the different Gamow factors, and not to the $\bbox{r}$-dependent
confluent hypergeometric function in the relative 
wavefunctions~\cite{SLAPE98}.
As a consequence, one can obtain an improved 
Coulomb correction for heavier particles by dividing
out these Gamow factors~\cite{SLAPE98},
  \begin{equation}
    C_{\rm corr, improved}^{(--)}(\bbox{q},\bbox{K}) =
    {C_{\rm meas}^{(+-)}(\bbox{q},\bbox{K}) \, 
      C_{\rm meas}^{(--)}(\bbox{q},\bbox{K})
    \over G(\eta_+)\, G(\eta_-)} \, .
    \label{2.103}
  \end{equation}
This was shown to work with excellent accuracy for a wide range
of source parameters~\cite{SLAPE98}.

\subsection{General formalism for final state interactions}
\label{sec2e4}

We now discuss a general formalism for the discussion of the effects
of final state interactions, starting from an arbitrary two-particle 
symmetrized wave function $\Psi(\bbox{x}_1,\bbox{x}_2,t)$ which we expand in
terms of plane waves $\phi_{\bbox{p}_1,\bbox{p}_2}$,
  \begin{eqnarray}
    \Psi(\bbox{x}_1,\bbox{x}_2,t) &=& e^{-i\hat{H}(t-t_0)}
    \Psi(\bbox{x}_1,\bbox{x}_2,t_0)
    \nonumber \\
    &=& 
    \int {d^3p_1\over (2\pi)^3}\, {d^3p_2\over (2\pi)^3}\,
    {\cal A}_{\Psi}(\bbox{p}_1,\bbox{p}_2,t)\, 
    \phi_{\bbox{p}_1,\bbox{p}_2}(\bbox{x}_1,\bbox{x}_2,t)\, ,
    \label{2.104} \\
    \phi_{\bbox{p}_1,\bbox{p}_2}(\bbox{x}_1,\bbox{x}_2,t) &=&
    e^{-i\hat{H}_0(t-t_0)}
    \phi_{\bbox{p}_1,\bbox{p}_2}(\bbox{x}_1,\bbox{x}_2,t_0)
    = e^{-i(E_1+E_2)t}\,
      e^{i\bbox{p}_1\bbox{x}_1 + i\bbox{p}_2\bbox{x}_2 }
      \nonumber \\
    &=& \phi_{\bbox{K}}(\bbox{x},t)\, \phi_{\bbox{q}/2}(\bbox{r},t)
    = e^{-i(E_1+E_2)t}\, e^{2i\bbox{K}\cdot \bbox{x}}
      \, e^{{i\over 2}\bbox{q}\cdot \bbox{r}}\, .
    \label{2.105} 
  \end{eqnarray}
In the last step, we have changed to center of mass coordinates 
$\bbox{x}= \textstyle{1\over 2}(\bbox{x}_1+\bbox{x}_2)$
and relative coordinates $\bbox{r} = (\bbox{x}_1-\bbox{x}_2)$.
The two-particle state $\Psi$ is evolved with the interacting
hamiltonian $\hat{H}$ while the plane waves in which we 
expand follow a free time evolution, determined by 
$\hat{H}_0$, $\hat{H}_0^x$, 
  \begin{eqnarray}
    \hat{H} &=& \hat{H}_0^x + \hat{H}_0 
                + \hat{H}_{\rm int}^r\, ,
    \nonumber \\
    \hat{H}_0^x &=& - {{\Delta}_x\over 2 M}\, ,
    \qquad 
    \hat{H}_0 = - {{\Delta}_r\over 2 \mu}\, ,
    \qquad
    \hat{H}^r_{\rm int} = V(r)\, .
    \label{2.106}
  \end{eqnarray}
Here $M = 2m$ and $\mu = m/2$ are the pair and reduced mass,
respectively. The two-particle state $\Psi$ determines the
two-particle Wigner phase-space density and hence the two-particle
correlator. The probability ${\cal P}_{\Psi}(\bbox{p}_1,\bbox{p}_2,t)$
of detecting the bosons at time $t$ with momenta $\bbox{p}_1$,
$\bbox{p}_2$ is 
  \begin{equation}
    {\cal P}_{\Psi}(\bbox{p}_1,\bbox{p}_2,t)
    = {\cal A}^*_{\Psi}(\bbox{p}_1,\bbox{p}_2,t)\,
      {\cal A}_{\Psi}(\bbox{p}_1,\bbox{p}_2,t)\, .
      \label{2.107}
  \end{equation}
Let us assume that from a time $t_0$ onwards final state interactions
have to be taken into account in the description of the time evolution
of $\Psi$. The time evolution of ${\cal A}_{\Psi}(\bbox{p}_1,\bbox{p}_2,t)$
then reads 
  \begin{eqnarray}
    {\cal A}_{\Psi}(\bbox{p}_1,\bbox{p}_2,t) &=&
    \int d^3x\, d^3r\, \phi_{\bbox{K}}^*(\bbox{x},t_0)
    \left[ e^{i(\hat{H}_0 + \hat{H}_{\rm int}^r)
                      (t-t_0)}\, e^{-i\hat{H}_0(t-t_0)}
                      \phi_{\bbox{q}/2}(\bbox{r},t_0)\right]^*\, 
    \nonumber \\
    && \qquad \qquad \times
    \Psi(\bbox{x}+ \textstyle{\bbox{r}\over 2},
    \bbox{x}- \textstyle{\bbox{r}\over 2},t_0)\, .
    \label{2.108}
  \end{eqnarray}
We are interested in the limit $t \to \infty$ of this expression.
To this end, we use the M\o ller operator
  \begin{equation}
    \Omega_+ = \lim_{t\to\infty}
    e^{i(\hat{H}_0 + \hat{H}_{\rm int}^r)
                      (t-t_0)}\, e^{-i\hat{H}_0(t-t_0)}\, ,
    \label{2.109}
  \end{equation}
which determines the solution of the Lippmann-Schwinger equation for
the corresponding stationary scattering problem,
  \begin{eqnarray}
    &&\left( {\Delta_r\over 2{\mu}} + V(r)\right)
    \Phi_{\bbox{q}/2}^{\rm scatt}(\bbox{r}) 
    = E\, \Phi_{\bbox{q}/2}^{\rm scatt}(\bbox{r})\, ,
    \nonumber \\
    && \qquad \Phi_{\bbox{q}/2}^{\rm scatt}(\bbox{r}) 
    = \Omega_+\, \phi_{\bbox{q}/2}(\bbox{r},t_0)\, .
    \label{2.110}
  \end{eqnarray}
Irrespective of the form of $V(r)$, once $\Phi_{\bbox{q}/2}^{\rm scatt}$
is determined, the two-particle detection probability 
${\cal P}_{\psi}(\bbox{p}_1,\bbox{p}_2,t=\infty)$ is known from
(\ref{2.107}). The corresponding two-particle spectrum is then
given by summing over all pair wave functions of the event:
  \begin{equation}
    {\cal P}_2(\bbox{p}_1,\bbox{p}_2) = \sum_{{\rm pairs}\, \Psi}
    {\cal A}_{\Psi}^*(\bbox{p}_1,\bbox{p}_2, t=\infty)\, 
    {\cal A}_{\Psi}(\bbox{p}_1,\bbox{p}_2, t=\infty)\, .
    \label{2.111}
  \end{equation}
%

\subsubsection{Coulomb correlations for instantaneous sources}

We now illustrate the use of the two-particle spectrum (\ref{2.111}),
starting from the Gaussian wavefunction introduced in section 
\ref{sec2c}. The sum $\sum_{{\rm pairs}\, \Psi}$ in (\ref{2.111})
is then a sum over all pairs $(i,j)$ of the set
$(\check{\bbox{r}}_i,\check{t}_i,\check{\bbox{p}}_i)$, or an average
over some distribution 
$\rho(\check{\bbox{r}}_i,\check{t}_i,\check{\bbox{p}}_i)$. We restrict
the calculation to instantaneous emission at time $t_i = t_j =
t_0$. For non-identical particles (e.g. unlike-sign pions) the
two-particle wave function at emission reads then
  \begin{eqnarray}
    \Psi_{ij}(\bbox{r},\bbox{x},t_0) &=& \Psi_{\rm rel}(\bbox{r},t_0)\,
             \Psi_{\rm pair}(\bbox{x},t_0)
  \nonumber \\
    &=& (\pi\sigma^2)^{-3/2} e^{-{1\over \sigma^2}
        (\check{\bbox{x}}- \bbox{x})^2
         + 2i\check{\bbox{K}}\cdot\bbox{x}}\, 
        e^{-{1\over 4\sigma^2}(\check{\bbox{r}} - \bbox{r})^2
                 +{i\over 2} \check{\bbox{q}}\cdot \bbox{r}}\, ,
  \label{2.112}
  \end{eqnarray}
and the corresponding amplitude entering the two-particle spectrum
(\ref{2.111}) is
  \begin{eqnarray}
    {\cal A}_{\Psi}(\bbox{p}_1,\bbox{p}_2,t=\infty)
    &=& \langle \phi_{\bbox{K}}\vert \Psi_{\rm pair}\rangle
        \langle \Phi_{\bbox{q}/2}^{\rm scatt}\vert \Psi_{\rm rel}\rangle
    \label{2.113} \\
     \langle \Phi_{\bbox{q}/2}^{\rm scatt}\vert \Psi_{\rm rel}\rangle
     &=& \int d^3r\, 
     {\Phi_{\bbox{q}/2}^{\rm scatt}}^*(\bbox{r})\,
          \Psi_{\rm rel}(\bbox{r})\, ,
     \label{2.114} \\
     \langle \Phi_{\bbox{K}}\vert \Psi_{\rm pair}\rangle &=&
     (\pi\sigma^2)^{-3/4}
     e^{-\sigma^2(\bbox{K}-\check{\bbox{K}})^2}
     \, e^{2i\check{\bbox{x}}\cdot(\bbox{K}-\check{\bbox{K}})}\, .
     \label{2.115}
  \end{eqnarray}
For pairs of identical charged particles, the state (\ref{2.112})
and the corresponding amplitude (\ref{2.113}) must be symmetrized
properly, adding the missing $\bbox{q} \leftrightarrow -\bbox{q}$
terms and replacing e.g. $\Phi_{\bbox{q}/2}$ by 
${\textstyle{1\over \sqrt{2}}}(\Phi_{\bbox{q}/2} \pm \Phi_{-\bbox{q}/2})$.
Further analytical simplifications of the amplitude (\ref{2.113})
depend on the functional form assumed for the two-particle state
$\Psi$. A study with Gaussian wave packets was presented in
Ref.~\cite{WFH98}. In the plane wave limit
$\sigma \to 0$ one recovers Pratt's result~\cite{P86} 
  \begin{eqnarray}
    {\cal P}_{\Psi}(\bbox{q},\bbox{K}) &\propto& G(\eta)
    \Bigl[ F(-i\eta;1;iz_-)\, F^*(-i\eta;1;iz_-) 
  \nonumber \\
    &&\qquad + F(-i\eta;1;iz_+)\, F^*(-i\eta;1;iz_+) 
  \nonumber \\
    &&\qquad \pm e^{iqr\cos\theta}
           F(-i\eta;1;iz_-)\, F^*(-i\eta;1;iz_+)
  \nonumber \\
    &&\qquad \pm e^{-iqr\cos\theta}
           F(-i\eta;1;iz_+)\, F^*(-i\eta;1;iz_-) \Bigr] .
  \label{2.116}
  \end{eqnarray}
The first two lines are the Born probabilities 
$\vert \Phi_{\bbox{q}/2}(\bbox{r})\vert^2$ and 
$\vert \Phi_{-\bbox{q}/2}(\bbox{r})\vert^2$; the exchange or
interference terms in the last two lines exist only for pairs of
identical particles. Weighting (\ref{2.116}) with the distribution
$S_{\bbox{K}}(\bbox{r})$, one obtains the properly symmetrized
generalization of (\ref{2.98}) for pairs of identical bosons.

\subsubsection{Coulomb correlations for time-dependent sources}
    
In general, identical bosons interfering in the final state of a
relativistic heavy-ion collision are produced at different emission
times $t_i \not= t_j$. This temporal structure is neglected in
the ansatz for the two-particle wavefunction (\ref{2.112}) and 
does not appear in the corresponding result (\ref{2.116}).
A formalism appropriate for the calculation of two-particle
spectra from arbitrary space-time dependent emission functions
$S(x,K)$ was developed in Ref.~\cite{AHR97} (for a relativistic
approach using the Bethe-Salpeter ansatz see Ref.~\cite{LL82}):  
  \begin{eqnarray}
    {\cal P}_2(\bbox{p}_1,\bbox{p}_2) &=& \int {d^4p\over (2\pi)^4}\,
    d^4x\, d^4y\, S(x,K+p)\, 
    \nonumber \\
    && \qquad \times W_q(x-y,p)\, S(y,K-p)\, ,
    \label{2.117} \\
    W_{q/2}(x-y,p) &=& \int {d^4Q\over (2\pi)^4}\, 
    e^{-iQ\cdot (x-y)}\, \chi_{q/2}(p+\textstyle{Q\over 2})\,
    \chi_{q/2}^*(p-\textstyle{Q\over 2})\, .
    \label{2.118}
  \end{eqnarray}
Here $\chi_{q/2}$ denotes essentially a Fourier transformed relative
wavefunction times a propagator~\cite{AHR97}, and the function 
$W_{q/2}$ can be interpreted as the Wigner density associated
with the (symmetrized) distorted wave describing final state interactions. 
For a free time-evolution in the final state, the two-particle
spectrum (\ref{2.117}) coincides with the appropriately normalized
spectrum (\ref{2.20}). From the result (\ref{2.117}) for the general 
interacting case, a simplified expression can be obtained by
expanding the temporal component of the phase factor in $W_{q/2}$ 
in leading order of the small energy transfer caused by the final
state interaction~\cite{AHR97}
 \begin{eqnarray}
  {\cal P}_2(\bbox{p}_1,\bbox{p}_2) &=&
  \int d^4x\,d^4y\,S\left(x+{\textstyle{y\over 2}},p_1\right)\,
       S\left(x-{\textstyle{y\over 2}},p_2\right) 
 \nonumber\\
  &&\ \times \left[ \theta(y^0) 
  \left\vert \Phi_{\bbox{q}/2}(\bbox{y}{-}\bbox{v}_2 y^0) \right\vert^2
  + \theta(-y^0) 
  \left\vert \Phi_{\bbox{q}/2}(\bbox{y}{-}\bbox{v}_1 y^0) \right\vert^2 
  \right]
 \nonumber\\
  &\pm& \int d^4x\,d^4y\,S\left(x+{\textstyle{y\over 2}},K\right)\,
       S\left(x-{\textstyle{y\over 2}},K\right)
 \nonumber\\
  && \qquad \quad\  \times\, 
     \Phi^*_{-\bbox{q}/2}(\bbox{y}{-}\bbox{v} y^0) \, 
     \Phi_{\bbox{q}/2}(\bbox{y}{-}\bbox{v} y^0) \, ,
 \label{2.119}\\
   \bbox{v} &=& {\bbox{K}\over E_K}\, ,\quad
   \bbox{v}_1 = {\bbox{p}_1\over E_K}\, ,\quad
   \bbox{v}_2 = {\bbox{p}_2\over E_K}\, .
 \label{2.120}
 \end{eqnarray}
The velocities $\bbox{v}_1$, $\bbox{v}_2$, and $\bbox{v}$ are associated 
with the observed particle momenta $\bbox{p}_1$, $\bbox{p}_2$, and their 
average $\bbox{K}$. In all three cases the argument of the FSI-distorted 
wave $\Phi$ can be understood as the distance between the two
particles in the pair rest frame at the emission time of the second
particle. Equation (\ref{2.119}) is obtained without invoking the
smoothness approximation. Employing also the latter, both terms in 
(\ref{2.119}) are associated with the same combination of emission
functions which can then be written in terms of the (unnormalized)
relative distance distribution, see also (\ref{4.01}),
 \begin{equation}
   D(y,K) \equiv \int d^4x \, 
   S\left( x+{\textstyle{y\over 2}},K\right) \,
   S\left( x-{\textstyle{y\over 2}},K\right) \, .
 \label{2.121}
 \end{equation}
This function denotes the distribution of relative space-time distances $y$ 
between the particles in pairs emitted with momentum $K$. A particularly 
simple expression due to Koonin is then obtained~\cite{K77,BD97a,AHR97} 
in the pair rest frame, $\bbox{v}=0=\bbox{K}$:
 \begin{equation}
    {\cal P}_2(\bbox{p}_a,\bbox{p}_b) \approx \int d^3y\,
    \left\vert \Phi^{\rm sym}_{\bbox{q}/2}(\bbox{y})\right\vert^2
    \int dy^0\, D(y,K) \, ,
 \label{2.122}
 \end{equation}
where $\Phi^{\rm sym}_{\bbox{q}/2}={\textstyle{1\over \sqrt{2}}}
(\Phi_{\bbox{q}/2} \pm \Phi_{-\bbox{q}/2})$ for identical particle
pairs. As explained in section~\ref{sec2b3a}, the last factor 
in (\ref{2.122}) coincides up to normalization with the relative
source function $S_{\bbox{K}}(\bbox{r})$. With the help of the
smoothness approximation the two-particle spectrum can thus be
expressed by the relative source function weighted with the Born
probability of the Coulomb relative wavefunction, as given before in
(\ref{2.98}). 

We finally mention that first attempts have been made to include
in the analysis of Coulomb final state effects the role of a central
Coulomb charge~\cite{B96,SKS98} or effects due to high particle 
multiplicity~\cite{AZZ96}. 
It is an important open question to what extent these effects modify
the analysis presented here.

\section{Bose-Einstein weights for event generators}
\label{sec2f}

Numerical event simulations of heavy-ion collisions provide one
important method to simulate realistic phase-space distributions. Many
such event generators exist nowadays. In principle, their output
should be a set of observable momenta $\bbox{p}_i$ with all momentum
correlations (and hence the complete space-time information) built
in. However, none of the existing event generators propagates properly
symmetrized $N$-particle amplitudes from some initial condition. As a
consequence, the typical event generator output is a set of discrete
phase-space points $(\check{\bbox{r}}_i,\check{t}_i,\check{\bbox{p}}_i)$
which one associates with the freeze-out positions of the final state
particles. This simulated event information  
$(\check{\bbox{r}}_i,\check{t}_i,\check{\bbox{p}}_i)$ lacks correlations
due to Bose-Einstein symmetrization and other types of final state
effects. 

We first discuss different schemes used to calculate {\it a
  posteriori} two-particle correlation functions for inputs of
discrete sets of phase-space points  
$(\check{\bbox{r}}_i, \check{t}_i,\check{\bbox{p}}_i)$.
We then turn to so-called shifting prescriptions which aim at
producing modified final state momenta with correct particle 
correlations. 

\subsection{Calculating $C({\protect\bbox{q}},{\protect\bbox{K}})$ 
            from event generator output}
\label{sec2f1}

The conceptual problem of determining particle correlations
from event generators is well-known~\cite{LS95,A96,LS97}: 
Bose-Einstein correlations
arise from squaring production amplitudes. They hence require
a description of production processes in terms of amplitudes.
Numerical event simulations, however, are formulated via
probabilities. This implies that various quantum
effects are treated only heuristically, if at all. Especially,
event generators do not take into account the quantum
mechanical symmetrization effects. In this sense, the event 
generator output is the result of an incomplete quantum 
dynamical evolution of the collision. The aim of Bose-Einstein 
weights is to remedy this artefact {\it a posteriori} by 
translating the phase-space information of 
$(\check{\bbox{r}}_i,\check{t}_i,\check{\bbox{p}}_i)$ 
into realistic momentum correlations. For a set of $N_{\rm ev}$
events of multiplicities $N_m$, this implies formally
  \begin{equation}
    \Bigg\lbrace
    \lbrace (\check{\bbox{r}}_i,\check{t}_i,\check{\bbox{p}}_i)
    \rbrace_{i\in [1,N_m]}
    \Bigg\rbrace_{m\in [1,N_{\rm ev}]}
    \Longrightarrow C(\bbox{q},\bbox{K})\, .
    \label{2.123}
  \end{equation}
The set $\lbrace{(\check{\bbox{r}}_i,\check{t}_i,\check{\bbox{p}}_i)}
\rbrace_{i\in [1,N_m]}$ denotes the phase-space emission points of the
$N_m$ like-sign pions generated in the $m$-th simulated event. 
The event generator simulates thus a classical phase-space 
distribution
 \begin{equation}
 \label{eq9}
   \rho_{\rm class}(\bbox{p},\bbox{r},t) =
   {1\over N_{\rm ev}} \sum_{m=1}^{N_{\rm ev}} \sum_{i=1}^{N_m}
   \delta^{(3)}(\bbox{r}-\check{\bbox{r}}_i)\,
   \delta^{(3)}(\bbox{p}-\check{\bbox{p}}_i)\, 
   \delta(t-\check{t}_i)\, .
 \end{equation}
Prescriptions of the type (\ref{2.123}) are not unique: a choice of
interpretation is involved in calculating two-particle correlations
from the event generator output. Here, we mention two different
interpretations of $(\check{\bbox{r}}_i,\check{t}_i,\check{\bbox{p}}_i)$, 
sometimes referred to as ``classical'' and ``quantum'' \cite{GEHW98}.

\subsubsection{``Classical'' interpretation of the event generator output}

In the ``classical'' interpretation~\cite{ZWSH97,WEHG98,GEHW98}
the distribution of phase-space points 
$\lbrace\, \lbrace (\check{\bbox{r}}_i,\check{t}_i,
\check{\bbox{p}}_i)\rbrace_{i\in [1,N_m]}\rbrace_{m\in [1,N_{\rm ev}]}$
is interpreted as a discrete approximation of the on-shell Wigner 
phase-space density $S(x,\bbox{p})$,
  \begin{equation}
    S(x,\bbox{p}) = \rho_{\rm class}(\bbox{p},\bbox{x},t)\, .
    \label{eq10}
  \end{equation}
The emission function is thus a sum over delta functions.
For practical applications, it is convenient to replace
the delta functions in momentum space by rectangular `bin 
functions'~\cite{ZWSH97} or by properly normalized 
Gaussians~\cite{WEHG98,GEHW98} of width $\epsilon \to 0$
(we denote both choices by the same symbol 
$\delta^{(\epsilon)}_{\check{\bbox{p}}_i,\bbox{p}}$)
 \begin{eqnarray}
   \delta^{(\epsilon)}_{\check{\bbox{p}}_i,\bbox{p}} &=& \left\{
   \begin{array}{r@{\quad:\quad}l}
     1/\epsilon^3 & p_j - {\epsilon\over 2} 
     \leq p_{i,j} \leq p_j + {\epsilon\over 2}
                \qquad (j=x,y,z) \\
     0 & {\rm else}\, ,
   \end{array}
   \right.
   \label{2.124}\\   
    \delta^{(\epsilon)}_{\check{\bbox{p}}_i,\bbox{p}}
    &=& {1\over (\pi\epsilon^2)^{3/2}}
      \exp \left( - (\check{\bbox{p}}_i - \bbox{p})^2/\epsilon^2\right)\, .
    \label{eq11}
  \end{eqnarray}
The one-particle spectrum and two-particle correlator then read 
\cite{ZWSH97,WEHG98,GEHW98}
  \begin{eqnarray}
    && E_p {dN\over d^3p} = \int d^4x\, S(x,\bbox{p}) =
       {1 \over N_{\rm ev}} \sum_{m=1}^{N_{\rm ev}}
       \sum_{i=1}^{N_m} 
       \delta^{(\epsilon)}_{\check{\bbox{p}}_i,\bbox{p}}\, ,
  \label{eq12} \\
    && C(\bbox{q},\bbox{K}) = 1 + { \sum_{m=1}^{N_{\rm ev}} \left[
       \left\vert \sum_{i=1}^{N_m} 
       \delta^{(\epsilon)}_{\check{\bbox{p}}_i,\bbox{K}}\, 
       e^{i(q^0\check{t}_i - \bbox{q}\cdot\check{\bbox{r}}_i)} 
       \right\vert^2 
       - \sum_{i=1}^{N_m} 
       \left(\delta^{(\epsilon)}_{\check{\bbox{p}}_i,\bbox{K}}
       \right)^2 \right]
   \over
       \sum_{m=1}^{N_{\rm ev}} \left[ \left(\sum_{i=1}^{N_m} 
       \delta^{(\epsilon)}_{\check{\bbox{p}}_i,\bbox{p}_1}\right)
       \left(\sum_{j=1}^{N_m} 
       \delta^{(\epsilon)}_{\check{\bbox{p}}_j,\bbox{p}_2}\right)
     - \sum_{i=1}^{N_m} 
       \delta^{(\epsilon)}_{\check{\bbox{p}}_i,\bbox{p}_1}
       \delta^{(\epsilon)}_{\check{\bbox{p}}_i,\bbox{p}_2}
   \right]}\, .
 \nonumber\\
 \label{eq13}
 \end{eqnarray}
The correlator (\ref{eq13}) is the discretized version of the
Fourier integrals in (\ref{2.24}). It does not invoke the smoothness 
approximation, in contrast to the popular earlier algorithm developed
by Pratt~\cite{P94,P97} (which includes final state interactions). The 
subtracted terms in the numerator and denominator remove the spurious
contributions of pairs constructed from the same particles~\cite{Weal97}. 

In general the result for the correlator at a fixed point 
$(\bbox{q},\bbox{K})$ will depend on the bin width $\epsilon$. Finite
event statistics puts a lower practical limit on $\epsilon$. Tests
have shown that accurate results for the correlator require smaller
values for $\epsilon$ (and thus larger event statistics) for more
inhomogeneous sources. In practice the convergence of the results must
be tested numerically~\cite{GEHW98}. 

\subsubsection{``Quantum'' interpretation of the event generator output}
\label{sec2.3b}

In the ``quantum'' interpretation \cite{Weal97,ZWSH97,W98,WEHG98,GEHW98} 
the event generator output 
$\rho_{\rm class}(\check{\bbox{r}},\check{t},\check{\bbox{p}})$ 
is associated with the centers of Gaussian wave packets (\ref{2.31}).
Neglecting multiparticle symmetrization effects, the corresponding
Wigner function according to (\ref{2.41}) reads
  \begin{equation}
    S(x,\bbox{K}) = 
    \int d^3\check{p}_i\, d^3\check{r}_i\, d\check{t_i}\,
    \rho_{\rm class}(\check{\bbox{r}}_i,\check{t_i},\check{\bbox{p}}_i)\, 
    s_0(\bbox{x}{-}\check{\bbox{r}}_i,t{-}\check{t}_i,
        \bbox{K}{-}\check{\bbox{p}}_i)\, .
    \label{eq7} 
  \end{equation}
The one- and two-particle spectra are~\cite{Weal97,GEHW98}
 \begin{eqnarray}
  && E_p {dN\over d^3p} = {1 \over N_{\rm ev}} \sum_{m=1}^{N_{\rm ev}}
     \nu_m(\bbox{p}) = {1 \over N_{\rm ev}} 
     \sum_{m=1}^{N_{\rm ev}} \sum_{i=1}^{N_m} s_i(\bbox{p})\, ,
 \label{2.128} \\
  && C(\bbox{q},\bbox{K}) =  1 + e^{-\sigma^2 \bbox{q}^2/2} 
     { \sum_{m=1}^{N_{\rm ev}} \left[ 
       \left\vert \sum_{i=1}^{N_m} s_i(\bbox{K})\,
        e^{i(q^0\check{t}_i - \bbox{q}\cdot\check{\bbox{r}}_i)} 
       \right\vert^2
      -\sum_{i=1}^{N_m} s_i^2(\bbox{K}) \right]
      \over 
        \sum_{m=1}^{N_{\rm ev}} \left[ \nu_m(\bbox{p}_a)\, \nu_m(\bbox{p}_b) 
      - \sum_{i=1}^{N_m} s_i(\bbox{p}_a)\,s_i(\bbox{p}_b) \right]}\, .
 \nonumber\\
 \label{2.129} 
 \end{eqnarray}
Again, the terms subtracted in the numerator and denominator
are finite multiplicity corrections which
become negligible for large particle multiplicities~\cite{Weal97}.

\subsubsection{Discussion}
\label{sec2.3c}

In both algorithms, the particle spectra are discrete functions of the
input $(\check{\bbox{r}}_i,\check{t}_i,\check{\bbox{p}}_i)$ but they
are continuous in the observable momenta $\bbox{p}_1$, $\bbox{p}_2$
and hence, no binning is necessary. Each of the sums in (\ref{eq13})
and (\ref{2.128}) requires only $O(N_m)$ manipulations. However, once
final state interactions are included, the number of numerical
operations increases quadratically with $N_m$ since the corresponding
generalized weights~\cite{AHR97} do no longer factorize. When also
accounting for multiparticle symmetrization effects, more than
$O(N_m^2)$ numerical manipulations are typically required~\cite{W98}.

The ``classical'' and ``quantum'' algorithms then differ in two points:
  \begin{enumerate}
    \item
      There is no analogue for the Gaussian prefactor
      $\exp\left(-\sigma^2\, \bbox{q}^2/2\right)$ of (\ref{2.129}) 
      in the ``classical'' algorithm. This is a genuine quantum 
      effect stemming from the quantum mechanical localization 
      properties of the wave packets.
    \item
      For the choice $\sigma = 1/\epsilon$, the bin functions
      $\delta^{(\epsilon)}_{\check{\bbox{p}}_i,\bbox{p}}$ are the
      classical counterpart of the Gaussian single-particle
      distributions $s_i(\bbox{p})$. Finite event statistics puts a
      lower practical limit on $\epsilon$, but in the limit $\epsilon
      \to 0$ the physical momentum spectra are recovered. In contrast,
      in the ``quantum'' algorithm $\sigma$ denotes the finite
      physical particle localization. In this case, the limit $\sigma
      \to \infty$ (corresponding to $\epsilon \to 0$) is not
      physically relevant: it amounts according to an emission
      function with infinite spatial extension, yielding
      $\lim_{\sigma\to\infty} C(\bbox{q},\bbox{K}) 
      = 1 + \delta_{\bbox{q},0}$~\cite{Weal97}. 
  \end{enumerate}

These algorithms have been shown to avoid certain inconsistencies
arising from the use of the smoothness approximation for sources 
with strong position-momentum correlations~\cite{MKFW96,ZWSH97}. 
Systematic studies indicate that violations of the smoothness
approximation occur only for emission functions $S(x,K)$ which
are inconsistent with the uncertainty relation, i.e., which cannot
be interpreted as Wigner densities. Pratt has shown that typical
source sizes in heavy-ion collisions are sufficiently large that
this problem can be neglected~\cite{P97}.

Extensions of these algorithms to include final state
interactions~\cite{AHR97} and multiparticle correlation 
effects~\cite{W98} were proposed but have not yet been implemented
numerically. 

\subsection{Shifting prescriptions}
\label{sec2f2}

In the previous subsection we reviewed algorithms which calculate
two-parti\-cle correlation functions from a discrete set of phase-space
points. The output of the algorithm is a correlator
$C(\bbox{q},\bbox{K})$ which denotes the probability of finding
particle pairs with the corresponding momenta; it is not a set of new
discrete momenta $\bbox{p}_j$ with the correct Bose-Einstein
correlations included. The latter is of interest e.g. for detector 
simulations which require on an event-by-event basis a 
simulated set of particle tracks to anticipate detector performance. 
Also, it could be used to investigate eventwise fluctuations
which is not possible with an ensemble averaged correlator.

The most direct way to achieve this goal would seem to use 
symmetrized amplitudes for the particle creation process. Such a scheme
has been developed in the context of the Lund string 
model~\cite{AH86,AR98} for the hadronization of a single string.
For more complicated situations there exist so far only algorithms
which shift after particle creation the generated momenta 
$\check{\bbox{p}}_j$ to their physically observed values $\bbox{p}_j$ 
  \begin{equation}
    \check{\bbox{p}}_j \stackrel{\footnotesize{
        \lbrace(\check{\bbox{r}}_i,\check{t}_i,\check{\bbox{p}}_i)
      \rbrace_{i\in [1,N]}}}{\Longrightarrow} 
    \bbox{p}_j\, .
    \label{2.130}
  \end{equation}
While the function $C(\bbox{q},\bbox{K})$ describes the two-particle
correlations only for the ensemble average, the 
set $\lbrace \bbox{p}_j\rbrace$ represents
all measurable momenta of a simulated {\it single}
event with realistic Bose-Einstein correlations.

Such a shifting prescription which employs the full phase-space
information of the simulated event was developed
by Zajc~\cite{Z84,Z87}. In Ref.~\cite{Z87} a self-consistent
Monte-Carlo algorithm is used for a simple Gaussian source to determine  
the shifts (\ref{2.130}) by sampling the momentum-dependent 
$N$-particle probability. Although technically feasible, this 
calculation of $N$-particle symmetrized weights involves
an enormous numerical effort.  

Another class of algorithms is used in event generators for high energy 
particle physics. Unlike (\ref{2.130}), they explicitly exploit 
only the momentum-space information of the simulated
events. Additionally, an ad hoc weight function is employed
which one may relate to the ensemble-averaged space-time 
structure of the source~\cite{LS95,LS97}:
  \begin{equation}
    Q \stackrel{\footnotesize{\lbrace(\check{\bbox{p}}_i)
      \rbrace_{i\in [1,N]}}}{\Longrightarrow} 
    Q + \delta Q\, .
    \label{2.131}
  \end{equation}
This shifting procedure involves only particle pairs and
is significantly simpler to implement numerically. By
decoupling the position and momentum information
one looses, however, possible correlations between the particle momenta
and their production points. Also, in an individual event this
shifting prescription is insensitive to the actual separation
of the particles in space-time. A further problem is that
the translation of $\delta Q$ from (\ref{2.131}) into a change 
of particle momenta is not unique. It changes the invariant mass of the 
particle pair and does not conserve simultaneously both energy 
and momentum. These deficiencies are repaired by a subsequent
rescaling of momenta; according to Ref.~\cite{LS97} 
the results show in practice little sensitivity to details of 
the implementation. 

A more sophisticated method~\cite{FW97,FW98,FWW98} attempts
to implement the full $N$-body symmetrization by a cluster
algorithm. Again the weights used in this algorithm, at least
in the present version, encode the space-time structure of the
source only via a single ensemble-averaged radius parameter.

\chapter{Gaussian parametrizations of the correlator}
\label{sec3}

In practice, the two-particle correlation function is usually
parametrized by a Gaussian in the relative momentum components, see
e.g. (\ref{1.5}). In this chapter we discuss different Gaussian
parametrizations and establish the relationship of the corresponding
width parameters (HBT radii) with the space-time structure of the
source.

This relation is based on a Gaussian approximation to the true
space-time dependence of the emission function
\cite{CSH95b,CSH95a,CNH95,WSH96,HTWW96} 
 \begin{eqnarray}
   S(x,K) &=& N(K)\,S(\bar x(K),K)
          \exp\left[ - {1\over 2} \tilde x^\mu(K)\,
            B_{\mu\nu}(K)\,\tilde x^\nu(K)\right]
        \nonumber \\
          &&
   + \delta S(x,K) \, .
 \label{3.1}
 \end{eqnarray}
For the present discussion, we neglect the correction term $\delta
S(x,K)$. We discuss in chapter~\ref{sec4} how it can be systematically
included. The space-time coordinates $\tilde{x}_{\mu}$ in (\ref{3.1})
are defined relative to the ``effective source centre'' $\bar x(K)$
for bosons emitted with momentum
$\bbox{K}$~\cite{CSH95b,HB95,WSH96,H96}  
 \begin{equation}
  \tilde x^\mu (K) = x^\mu - \bar x^\mu(K)\, , \qquad
  \bar x^\mu(K) = \langle x^\mu \rangle(K) \, ,
  \label{3.2} 
 \end{equation}
where $\langle \dots \rangle$ denotes an average with the 
emission function $S(x,K)$:
  \begin{equation}
  \langle f\rangle(K) = 
  { {\int d^4x\, f(x)\, S(x,K)}\over {\int d^4x\, S(x,K)} }\, .
  \label{3.3}
  \end{equation}
The choice
 \begin{equation}
  (B^{-1})_{\mu\nu}(K)
  = \langle \tilde x_\mu \tilde x_\nu \rangle(K)
  \label{3.5}
 \end{equation}
ensures that the Gaussian ansatz (\ref{3.1}) has the same
rms widths in space-time as the full emission function.
Inserting (\ref{3.1}) into the basic relation (\ref{1.4}) 
one obtains the simple Gaussian form for the correlator 
 \begin{equation}
   C(\bbox{q},\bbox{K}) = 1 + \exp\left[ - q_\mu q_\nu 
   \langle \tilde x^\mu \tilde x^\nu \rangle (\bbox{K}) \right]\, .
  \label{3.4} 
 \end{equation}
This involves the smoothness and on-shell approximations 
discussed in chapter~\ref{sec2} which permit to write the
space-time variances $\langle \tilde{x}_{\mu} \tilde{x}_{\nu}\rangle$
as functions of $\bbox{K}$ only. Note that the correlator depends only
on the relative distances $\tilde{x}^{\mu}$ with respect to the
source center. No information can be obtained about the absolute
position $\bar{x}(\bbox{K})$ of the source center in space-time. 

According to (\ref{3.4}) the two-particle 
correlator provides access to the rms widths of the effective source 
of particles with momentum $\bbox{K}$. In general, these width parameters 
do not characterize the total extension of the collision
region. They rather measure the size of the
system through a filter of wavelength $\bbox{K}$. In the
language introduced by Sinyukov~\cite{S95}, this size is the
``region of homogeneity'', the region from which particle
pairs with momentum $\bbox{K}$ are most likely emitted.
Space-time variances coincide with total source extensions
only in the special case that the emission function shows
no position-momentum correlation and factorizes,
$S(x,K) = f(x)\, g(K)$.

Relating (\ref{3.4}) to experimental data requires first the
elimination of one of the four relative momentum components
via the mass-shell constraint (\ref{2.27}).
Depending on the choice of the three independent  
components, different Gaussian parametrizations exist. 
In what follows, we focus on their interpretation in terms
of the space-time characteristics of the source.

\section{The Cartesian parametrization}
\label{sec3a}

\begin{figure}[ht]\epsfxsize=12cm 
\centerline{\epsfbox{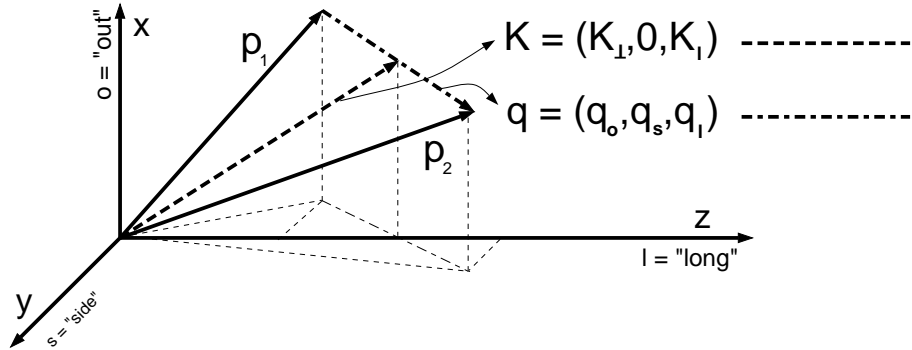}}
\caption{The {\it osl} coordinate system takes the longitudinal 
({\it long}) direction along the beam axis. In the transverse plane, the
{\it out} direction is chosen parallel to the transverse component
of the pair momentum $\bbox{K}_{\perp}$, the remaining Cartesian
component denotes the {\it side} direction.
}\label{fig1}
\end{figure}

The Cartesian parametrization~\cite{P83,P84,BGT88,CSH95b} is expressed
in the {\it out-side-longitudinal} ({\it osl}) coordinate system,
defined in Fig.~\ref{fig1}. It is based on the three 
Cartesian spatial components $q_o$ ({\it out}), $q_s$ ({\it side}),
$q_l$ ({\it long}) of the relative momentum $q$. The temporal 
component is eliminated via the mass-shell constraint (\ref{2.27})
  \begin{eqnarray}
    q^0 &=& \bbox{\beta}\cdot\bbox{q}\, ,\qquad
    \bbox{\beta} = \bbox{K}/K^0\, ,
    \label{3.6} \\
    \bbox{\beta} &=& \left(\beta_\perp, 0,\beta_l\right)
    \qquad \hbox{in the {\it osl}-system.}
    \label{3.7}
  \end{eqnarray}
This leads to a correlator of the form (\ref{1.5}) with $\sum_{ij}$
running over $i,j = o,s,l$.
In general this correlator $C(\bbox{q},\bbox{K})$ depends not only
on $K_\perp$ and $K_l$, but also on the azimuthal orientation $\Phi$ 
of the transverse pair momentum $\vert \bbox{K}_\perp\vert$. This angle, 
however, does not appear explicitly in the {\it osl}-system which 
is oriented for each particle pair differently by the angle $\Phi$ 
in the transverse plane.
$\Phi$ has to be defined with respect to some pair-independent
direction  in the laboratory system, e.g. relative
to the impact parameter $\bbox{b}$:
  \begin{equation}
        \Phi = \angle(\bbox{K}_\perp,\bbox{b})\, .
    \label{3.8}
  \end{equation}
In the following we discuss both the azimuthally symmetric
situation with impact parameter $\bbox{b} = 0$, when all
physical observables are $\Phi$-independent, and
the parametrization for finite impact parameter collisions. 
%

\subsection{Azimuthally symmetric collisions}
\label{sec3a1}

For central collisions, $\bbox{b} = 0$, the collision region is
azimuthally symmetric, and the emission function and correlator 
are $\Phi$-independent. In the {\it osl}-system, this
$\Phi$-invariance results in a reflection symmetry with respect
to the {\it side}-direction~\cite{CH94}:
  \begin{eqnarray}
    && S_{\rm lab}(x;K_{\perp},\Phi,K_l) =
             S_{\rm lab}(x;K_{\perp},\Phi+\delta\Phi,K_l)
             \nonumber \\
    && \Longleftrightarrow
    S_{\rm osl}(x,y,z,t;K_{\perp},K_l) 
    = S_{\rm osl}(x,-y,z,t;K_{\perp},K_l)\, .
    \label{3.9}
  \end{eqnarray}
We follow common practice in dropping now the subscript 
{\it osl}. Whenever azimuthal symmetry is assumed, the emission function
$S(x,K)$ is specified in the {\it osl}-system.

Due to the $y \to -y$ reflection symmetry of 
the emission function, $\overline{y} = 
\langle y\rangle = 0$ and the three space-time variances
$\langle \tilde{x}_\mu\tilde{x}_\nu \rangle(\bbox{K})$ 
linear in $\tilde{y}$ vanish. The symmetric tensor 
$B_{\mu\nu}(\bbox{K})$ then has only seven non-vanishing 
independent components. These combine to four non-vanishing 
HBT-radius parameters $R_{ij}^2(\bbox{K})$ which 
characterize the Gaussian ansatz (\ref{1.5}) for the 
correlator~\cite{CSH95b,HB95}:
  \begin{eqnarray}   
          R_s^2(\bbox{K}) &=& \langle \tilde{y}^2 \rangle(\bbox{K}) \, ,
        \label{3.10}\\
          R_o^2(\bbox{K}) &=& 
        \langle (\tilde{x} - \beta_\perp \tilde t)^2 \rangle(\bbox{K}) \, ,
        \label{3.11}\\
          R_l^2(\bbox{K}) &=& 
        \langle (\tilde{z} - \beta_l \tilde t)^2 \rangle(\bbox{K}) \, ,
        \label{3.12}\\
          R_{ol}^2(\bbox{K}) &=& 
        \langle (\tilde{x} - \beta_\perp \tilde t)
           (\tilde{z} - \beta_l \tilde t) \rangle(\bbox{K}) \, ,
        \label{3.13} \\
          R_{os}^2(\bbox{K}) &=& 0 \, ,
        \label{3.14} \\
          R_{sl}^2(\bbox{K}) &=& 0 \, .
        \label{3.15}
  \end{eqnarray}
Obviously, the more symmetries are satisfied by the emission
function, the simpler are the expressions obtained for the
HBT radius parameters. In this context we mention the case of
a longitudinally boost-invariant source showing Bjorken scaling.
Though not strictly satisfied by the finite sources created
in heavy-ion collisions, this can provide a simple intuitive 
picture of the collision dynamics near mid-rapidity. Longitudinal 
boost-invariance implies a $\tilde{z} \to -\tilde{z}$ reflection 
symmetry of the emission function. Thus, in addition to the 
space-time variances linear in $\tilde{y}$, now also those linear in 
$\tilde{z}$ vanish, and one is left with only 5 non-vanishing 
independent components of $B_{\mu\nu}(\bbox{K})$. In the 
longitudinally comoving system (LCMS), where $\beta_l = 0$, 
this leads to the further simplications
  \begin{eqnarray}
     && \qquad \qquad \qquad R_l^2(\bbox{K}) = 
        \langle \tilde{z}^2 \rangle(\bbox{K}) \, ,
  \label{3.16}\\
     && \qquad \qquad \qquad R_{ol}^2(\bbox{K}) = 0\, ,
  \label{3.17} \\
     && \hbox{for longitudinally boost-invariant sources in the LCMS.} 
        \nonumber
  \end{eqnarray}
The general relation between the symmetries of the system, the
number of its independent non-vanishing space-time variances, and the 
number of non-vanishing observable HBT parameters is summarized 
in the following table:
\vspace{0.4cm}\\
    \begin{tabular}{c|c|c}
      symmetry & $B_{\mu\nu}(\bbox{K})$ & $R_{ij}(\bbox{K})$ \\
      \hline
      none & 10 indep. fcts. & 6 indep. fcts.  \\
      & of $K_\perp$, $\phi$, $Y$ & of $K_\perp$, $\phi$, $Y$ \\
      \hline
      azimuthal & 7 indep. fcts. & 4 indep. fcts. \\
      & of $K_\perp$, $Y$ & of $K_\perp$, $Y$ \\
      \hline
      azimuthal & 5 indep. fcts. & 3 indep. fcts. \\
      + long. boostinv. & of $K_\perp$ & of $K_\perp$ \\
      in the LCMS & & \\
    \end{tabular}
\vspace{0.4cm}\\
In all cases, there are more independent space-time
variances $\langle x_\mu x_\nu\rangle(\bbox{K})$ than experimental
observables. This arises from the mass-shell constraint (\ref{2.27})
which leads to a mixing of spatial and temporal variances in the
observable HBT parameters.
One of the most important questions is therefore which other
properties of the expanding system can be exploited to further
disentangle spatial and temporal information about the
emission function.

For this, we note that independent of the particular
emission function, no direction is distinguished in the transverse
plane for $K_\perp = 0$. The {\it out}- and {\it side}-components
of all observables coincide in this kinematical limit. For the 
space-time variances, this implies
  \begin{eqnarray}
    \langle \tilde{x}^2 \rangle \Big\vert_{K_\perp = 0}
    &=& \langle \tilde{y}^2 \rangle \Big\vert_{K_\perp = 0}\, ,
        \label{3.18}\\
    \langle \tilde{z}\tilde{x} \rangle \Big\vert_{K_\perp = 0}
    &=& \langle \tilde{t}\tilde{x} \rangle \Big\vert_{K_\perp = 0} = 0\, .
        \label{3.19}
  \end{eqnarray}
As long as the limit $K_\perp \to 0$ of the emission function
$S(x,K)$ results in an azimuthally symmetric expression (an
exception is the class of opaque source models discussed in
section~\ref{sec5a2}), the above relations between the space-time
variances at $K_\perp = 0$ imply that the HBT radius parameters
satisfy
  \begin{eqnarray}
    \lim_{K_\perp \to 0} R_o^2(\bbox{K}) &=& 
    \lim_{K_\perp \to 0} R_s^2(\bbox{K})\, ,
    \label{3.20} \\
    \lim_{K_\perp \to 0} R_{ol}^2(\bbox{K}) &=& 0\, .
    \label{3.21}
  \end{eqnarray}
In phenomenological HBT analyses one very often exploits these
relations by distinguishing between an {\it implicit}
$\bbox{K}$-dependence (due to the $\bbox{K}$-dependence of the
space-time variances) and an {\it explicit} one (resulting from the
mass-shell constraint $q^0 = \bbox{q}\cdot\bbox{K}/K^0$). If the
emission function features no position-momentum correlation, then all
space-time variances are $\bbox{K}$-independent and
Eqs. (\ref{3.18}/\ref{3.19}) hold independently of $\bbox{K}$. The
difference between $R_o^2(\bbox{K})$ and $R_s^2(\bbox{K})$ at non-zero
$\bbox{K}$ is then only due to the explicit $\bbox{K}$-dependence in
(\ref{3.10}/\ref{3.11}), i.e. the term $\beta_\perp^2 \langle
\tilde{t}^2 \rangle$. This implies that the explicit
$\bbox{K}$-dependence dominates if the emission duration is
sufficiently large~\cite{RG96} or if the position-momentum
correlations in the source are sufficiently weak~\cite{B89,PCZ90},
  \begin{equation}
    R_o^2(\bbox{K}) - R_s^2(\bbox{K}) \approx \beta_\perp^2
     \langle \tilde{t}^2 \rangle\, .
     \label{3.22}
  \end{equation}
In this case, the difference between these two HBT radius parameters gives
direct access to the average emission duration $\langle \tilde{t}^2 \rangle$
of the source and allows to partially disentangle the spatial and temporal
information contained in (\ref{3.10})-(\ref{3.15}).

The gives rise to the following simple interpretation of the 
HBT radius parameters from the Cartesian parametrization: $R_s$ measures
the width of the emission region in the side direction, and $R_o$ 
measures the corresponding width in the out direction plus a 
contribution from the emission duration which can be extracted
according to (\ref{3.22}) under the assumption of a weak 
$K_\perp$-dependence of the emission function. The longitudinal 
radius $R_l$ finally describes the longitudinal extension of the 
region of homogeneity in the LCMS where $\beta_l = 0$. No easy 
intuitive interpretation exists for the out-longitudinal radius 
parameter $R_{ol}^2$. It is perhaps best understood in terms of the 
linear correlation coefficient~\cite{SSX98}
  \begin{equation}
    \rho_{ol}(\bbox{K}) = - {R_{ol}^2(\bbox{K})\over
                             R_o(\bbox{K})\, R_l(\bbox{K})}\, ,
    \label{3.23}
  \end{equation}
which can be positive or negative but is bounded by
$\vert  \rho_{ol}(\bbox{K})\vert \leq 1$ due to the Cauchy-Schwarz
inequality. This coefficient was shown~\cite{SSX98} 
to be of kinematical origin and useful for the interpretation of the
longitudinal momentum distributions. We shall see in section~\ref{sec3b}
that $R_{ol}$ plays a crucial role in the determination of the 
longitudinal velocity of the emitting source volume element.

\subsection{Collisions with finite impact parameter}
\label{sec3a2}

If the azimuthal symmetry of the particle emitting source is broken,
then the transverse one-particle spectrum depends on the azimuthal
direction of the emitted particles. This can be quantified in terms
of the harmonic coefficients $v_n(p_t,y)$,~\cite{O92,O93,VZ96}
  \begin{eqnarray}
    E{dN\over d^3p} &=&
    E {dN\over p_t\, dp_t\, dy\, d\phi} = \int d^4x\, S(x,p)
    \label{3.24}\\
    &=& {E\over 2\pi} {d^2N\over p_t\, dp_t\, dy}
        \left[ 1 + 2\sum_{n=1}^\infty v_n(p_t,y)\, \cos n(\phi -\psi_R) 
                \right]\, .
  \nonumber 
  \end{eqnarray}      
The size and momentum dependence of the lowest of these harmonic 
coefficients has been analyzed experimentally at both AGS and
CERN SPS energies~\cite{E87794,E87797,O97,PV98}. 
This allows to determine the orientation of
the reaction plane for semiperipheral collisions event by event
with an uncertainty of less than 30$^\circ$~\cite{VZ96,PV98}. 
Several attempts to extend this azimuthally sensitive 
analysis to two-particle correlation functions exist~\cite{VC96,F96,W97}.
The correlation measurements depend 
on the azimuthal direction $\Phi$ of the pair momentum, see (\ref{3.8}),
and hence
allow to provide additional azimuthally sensitive information.  
The corresponding Gaussian radius parameters can be written formally
in terms of space-time variances which are rotated via ${\cal D}_\Phi$
from the impact parameter fixed to the {\it osl} coordinate 
system~\cite{W97}
 \begin{eqnarray}
   && R_{ij}^2(\bbox{K}) 
    = \left\langle\left[\left({\cal D}_\Phi \tilde{x}\right)_i 
           - \left({\cal D}_\Phi \bbox{\beta}\right)_i\tilde{t}\right] 
           \left[\left({\cal D}_\Phi \tilde{x}\right)_j 
           - \left({\cal D}_\Phi \bbox{\beta}\right)_j\tilde{t}\right]
      \right\rangle\, ,
   \nonumber \\
   &&\left({\cal D}_\Phi \bbox{\beta}\right) = 
   \left(\beta_\perp,\, 0,\, \beta_l\right)\, .
   \label{3.25}
 \end{eqnarray}
We here differ from the notation adopted in the rest of this
review: the coordinates $x$, $y$ and $z$ are here given in the 
impact-parameter fixed system, not the {\it osl} one. 
As for the azimuthally symmetric case, the
HBT radius parameters show implict and explicit $\bbox{K}$-dependences.
Their $\Phi$-dependence thus has two different 
origins~\cite{VC96,W97}:
  \begin{eqnarray}
    && R_s^2(K_\perp,\Phi,Y) = \langle \tilde{x}^2\rangle \sin^2\Phi
                  + \langle \tilde{y}^2\rangle \cos^2\Phi
                  - \langle \tilde{x}\tilde{y}\rangle 
                       \sin 2\Phi \, ,
                      \nonumber \\
    && R_o^2(K_\perp,\Phi,Y) = \langle \tilde{x}^2\rangle \cos^2\Phi
                  + \langle \tilde{y}^2\rangle \sin^2\Phi
                  + \beta_\perp^2 \langle \tilde{t}^2\rangle
                      \nonumber \\
    && \qquad \qquad - 2\beta_\perp 
                       \langle \tilde{t}\tilde{x} \rangle \cos\Phi 
                     - 2\beta_\perp 
                       \langle \tilde{t} \tilde{y} \rangle \sin\Phi 
                     + \langle \tilde{x}\tilde{y}\rangle 
                       \sin 2\Phi \, ,
                      \nonumber \\
    && R_{os}^2(K_\perp,\Phi,Y) = 
                  \langle \tilde{x}\tilde{y}\rangle \cos 2\Phi 
                  + \textstyle{1\over 2} \sin 2\Phi 
                  (\langle \tilde{y}^2\rangle - \langle \tilde{x}^2\rangle)
                  \nonumber \\
    && \qquad \qquad + \beta_\perp \langle \tilde{t}
                       \tilde{x}\rangle \sin\Phi
                     - \beta_\perp \langle \tilde{t}
                       \tilde{y}\rangle \cos\Phi \, ,
                      \nonumber \\
    && R_{l}^2(K_\perp,\Phi,Y) = 
                  \langle (\tilde{z} -\beta_l\tilde{t})^2 \rangle \, ,
                      \nonumber \\
    && R_{ol}^2(K_\perp,\Phi,Y) = 
                  \langle (\tilde{z} -\beta_l\tilde{t})
                     (\tilde{x}\cos\Phi + \tilde{y}\sin\Phi 
                     - \beta_\perp\tilde{t}) \rangle\, ,
                      \nonumber \\
    && R_{sl}^2(K_\perp,\Phi,Y) = 
                  \langle (\tilde{z} -\beta_l\tilde{t})
                     (\tilde{y}\cos\Phi - \tilde{x}\sin\Phi) 
                     \rangle\, .
    \label{3.26}
  \end{eqnarray}
The {\it explicit} $\Phi$-dependence denoted here is a purely geometrical
consequence of rotating the $x$-axis from the direction of $\bbox{b}$ to
the direction of $\bbox{K}_\perp$. In addition, there is an 
{\it implicit} $\Phi$-dependence of the space-time variances,
$\langle \tilde{x}_{\mu}\tilde{x}_{\nu}\rangle$ 
$= \langle \tilde{x}_{\mu}\tilde{x}_{\nu}\rangle(K_\perp,\Phi,Y)$. 
This $\Phi$-dependence characterizes
the dynamical correlations between the size of the effective emission 
region (``region of homogeneity'')
and the azimuthal direction in which particles are emitted. Both
implicit and explicit $\Phi$-dependences are mixed in the harmonic 
coefficients
  \begin{eqnarray}
    {R_{ij,m}^c}^2 &=& {1\over 2\pi} 
           \int_{-\pi}^\pi R_{ij}^2\, \cos(m\Phi)\, d\Phi\, ,
    \label{3.27} \\
    {R_{ij,m}^s}^2 &=& {1\over 2\pi} 
           \int_{-\pi}^\pi R_{ij}^2\, \sin(m\Phi)\, d\Phi\, .
    \label{3.28}
  \end{eqnarray}
In models in which elliptic deformations dominate and higher than
second order harmonic coefficients can be neglected, these
coefficients satisfy the relations~\cite{W97}
  \begin{eqnarray}
    \alpha_1(K_\perp,Y) \approx {R_{s,1}^c}^2 
                        \approx \textstyle{1\over 3} {R_{o,1}^c}^2 
                        \approx -{R_{os,1}^s}^2 \, ,
    \label{3.29}\\
    \alpha_2(K_\perp,Y) \approx {R_{o,2}^c}^2 
                        \approx - {R_{s,2}^c}^2 
                        \approx - {R_{os,2}^s}^2\, . 
    \label{3.30}
  \end{eqnarray}
The anisotropy parameter $\alpha_1$ vanishes in the absence of
position-momentum correlations in the source and thus characterizes
dynamical anisotropies. On the other hand, $\alpha_2$ characterizes the
elliptical shape of the emission region. A violation of the 
relations (\ref{3.29})-(\ref{3.30}) would rule out a large 
class of model scenarios considered to be consistent with the 
present knowledge about the space-time evolution of the collision 
process.

Constraints of the type (\ref{3.29}/\ref{3.30}) on the harmonic 
coefficients lead to a minimal azimuthally sensitive parametrization 
of the two-particle correlator~\cite{W97}:
  \begin{eqnarray}
    C_{\psi_R}(\bbox{q},\bbox{K}) &\approx& 1 + 
        \lambda(\bbox{K})\, \,  C_{sym}(\bbox{q},\bbox{K})\,
                      C_1(\bbox{q},\bbox{K},\psi_R)
        \nonumber \\
        && \qquad \qquad \times
                C_2(\bbox{q},\bbox{K},\psi_R)\, ,
        \label{3.31} \\
    C_{sym}(\bbox{q},\bbox{K}) &=& \exp\lbrack
        -{R_{o,0}}^2\, q_o^2 - {R_{s,0}}^2\, q_s^2
        \nonumber \\
             && \qquad
             - {R_{l,0}}^2\, q_l^2 - 2\, {R_{ol,0}}^2\, q_o q_l \rbrack\, ,
        \label{3.32} \\
    C_1(\bbox{q},\bbox{K},\psi_R)    
        &=& \exp\big\lbrack -\alpha_1\,
                (3\, q_o^2 + q_s^2)\, \cos(\Phi-\psi_R)\,
        \nonumber \\
        && \qquad
               + 2\alpha_1\, q_o q_s \sin(\Phi -\psi_R) \big\rbrack \, ,
        \label{3.33} \\
    C_2(\bbox{q},\bbox{K},\psi_R)
        &=& \exp\big\lbrack -\alpha_2
                (q_o^2-q_s^2)\cos 2(\Phi-\psi_R)
        \nonumber \\
        && \qquad
                +2\,\alpha_2 q_oq_s \sin 2(\Phi-\psi_R) \big\rbrack\, .
    \label{3.34}
  \end{eqnarray}
In addition to the azimuthally symmetric part (\ref{3.32}) which
coincides with the Cartesian parametrization, this Gaussian
ansatz involves only two additional, azimuthally sensitive fit 
parameters. 
%
\begin{figure}[h]\epsfxsize=7cm 
\centerline{\epsfbox{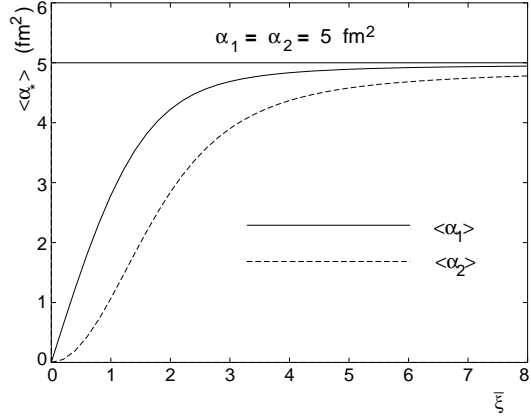}}
\caption{The HBT anisotropy parameters $\langle\alpha_i\rangle$ 
as a function
of the parameter $\bar{\xi}$ which characterizes the event-by-event
reconstruction uncertainty in the orientation of the reaction plane.
The parameters $\langle\alpha_1\rangle$, $\langle\alpha_2\rangle$ 
are determined by fitting (\protect{\ref{3.31}}) to an event sample 
(\protect{\ref{6.5}}) of correlators whose reconstructed reaction planes
fluctuate around the true impact parameter with the probability
distribution (\protect{\ref{6.3}}). The value $\bar{\xi} = 2$
corresponds to a reconstruction uncertainty of approximately $30^\circ$.
}\label{aniso}
\end{figure}

The major difficulty in determining the harmonic coefficients 
(\ref{3.27}/\ref{3.28}) from experiment is that they are expected to 
be statistically meaningful only for relatively large event samples
while their extraction is conditional upon the eventwise reconstruction 
of the reaction plane. For sufficiently high event multiplicities, the
probability distributions $W(v_1,\psi_R)$ of the experimentally
determined first harmonic coefficients and reaction plane around the
most likely, ``true'' values $(\bar{v}_1,\bar{\psi}_R)$ is given by a
Gaussian of variance $\eta^2$~\cite{O92,O93,VZ96}:
  \begin{equation}
    W(v_1,\psi_R) = {1\over 2\pi\eta^2}
    \exp\left(-{ {\bar{v}_1^2 + v^2_1 - 2\bar{v}_1v_1\, 
          \cos \psi_R} \over 2\eta^2} \right)\, .
    \label{6.3}
  \end{equation}
An event sample with oriented reaction plane should thus be compared
to a weighted average of the parametrization (\ref{3.31}),  
  \begin{equation}
    C_{\bar{\psi}_R}^{\rm eff}(\bbox{q},\bbox{K})
    = \int v_1\, dv_1\, d\psi_R\, W(v_1,\psi_R)\,
    C_{{\psi}_R}(\bbox{q},\bbox{K})\, .
  \label{6.5}
  \end{equation}
This effective correlator depends on the event averaged harmonic
coefficient $\bar{v}_1$ and the variance $\eta$ only via the ratio
$\bar{\xi} = \bar{v}_1/\eta$ which is a direct measure of the accuracy
for the reaction plane orientation~\cite{O92,O93,VZ96}. As can be seen
from Figure~\ref{aniso}, for an oriented event sample with the typical
30$^\circ$ uncertainty in the eventwise determination of the reaction
plane angle $\psi_r$ (which translates into $\bar{\xi} \approx
2$~\cite{VZ96}), more than 80 \% (50 \%) of the anisotropy signals
$\alpha_1$ ($\alpha_2$) survive in the actual experimental measurement.
Since the width $\eta$ is given by event statistics, it is the
possible to reconstruct the true values $\alpha_{1,2}$ from the
measure values $\langle \alpha_{1,2}\rangle$.

Furthermore calculating $C_{\bar{\psi}_R}^{\rm eff}(\bbox{q},\bbox{K})$
for $\bar{\xi} = 0$, i.e., for an azimuthally symmetric event
sample of finite impact parameter collisions, one can test to what
extent the azimuthally symmetric HBT radius parameters extracted
from a fit to such event samples will pick up contributions from 
non-zero $\alpha_1$ and $\alpha_2$. This effect is, however, expected to be 
small~\cite{VZ96,W97}.

\section{The Yano-Koonin-Podgoretski\u\i\ pa\-ra\-me\-tri\-za\-tion}
\label{sec3b}

The mass-shell constraint $K\cdot q = 0$, explicitly given in (\ref{3.6}), 
allows for different choices of three independent relative momenta. The 
Yano-Koonin-Podgoretski\u\i\ (YKP) parametrization, which assumes an
azimuthally symmetric collision region, uses the components $q_\perp =
\sqrt{q_o^2 + q_s^2}$, $q^0$ and $q_l$ and starts from the Gaussian 
ansatz~\cite{YK78,P83,CNH95,HTWW96}
 \begin{eqnarray}
   C(\bbox{q},\bbox{K}) &=&
   1 +  \lambda\,\exp\Bigl[ - R_\perp^2(\bbox{K})\, q_{\perp}^2 
                            - R_\parallel^2(\bbox{K}) (q_l^2 - {(q^0)}^2)
 \nonumber \\
    &&\qquad\qquad - \left(R_0^2(\bbox{K}) + R_\parallel^2(\bbox{K})\right)
                         \left(q{\cdot}U(\bbox{K})\right)^2
      \Bigr]  \, .
   \label{3.35}
 \end{eqnarray}
Here, $U(\bbox{K})$ is a ($\bbox{K}$-dependent) 4-velocity with only a 
longitudinal spatial component,
 \begin{eqnarray}
   U(\bbox{K}) &=& \gamma(\bbox{K}) \left(1, 0, 0, v(\bbox{K}) \right)\, ,
   \label{3.36} \\
   \gamma(\bbox{K}) &=& {1\over \sqrt{1 - v^2(\bbox{K})}}\, .
   \label{3.37}
 \end{eqnarray}
The combinations of relative momenta $(q_l^2 - {(q^0)}^2)$, 
$(q{\cdot}U(\bbox{K}))^2$ and $q_\perp^2$ appearing in (\ref{3.35})
are scalars under longitudinal boosts, and the three
YKP fit parameters $R_\perp^2(\bbox{K})$, $R_0^2(\bbox{K})$, and 
$R_\parallel^2(\bbox{K})$ are therefore longitudinally boost-invariant. 
In contrast to the Cartesian radius parameters, the values extracted 
for these YKP radius parameters do not depend
on the longitudinal velocity of the measurement frame. 
This is advantageous in fitting experimental data. 
The fourth YKP parameter is the Yano-Koonin (YK)
velocity $v(\bbox{K})$ which, as we will see, is closely
related to the velocity of the effective particle emitter. 
The corresponding rapidity
  \begin{equation}
    Y_{_{\rm YK}}(\bbox{K}) = {1\over 2} 
    \ln \left({ {1+v(\bbox{K})}\over {1-v(\bbox{K})} }\right)
    \label{3.38}
  \end{equation}
transforms additively under longitudinal boosts.

Since the ansatz (\ref{3.35}) uses four Gaussian parameters, it
is a complete parametrization for azimuthally symmetric collisions. 
These parameters can again be expressed in terms of the space-time
variances $\langle \tilde{x}_\mu \tilde{x}_\nu\rangle$~\cite{HTWW96,WHTW96}:
 \begin{eqnarray}
 \label{3.39}
    R_\perp^2(\bbox{K}) &=& R_s^2(\bbox{K}) = 
    \langle\tilde{y}^2\rangle(\bbox{K})\, ,
    \\
 \label{3.40}
   R_0^2(\bbox{K}) &=& A{-}v C , 
   \\
 \label{3.41}
   R_\parallel^2(\bbox{K}) &=& B{-}v C,
   \\
 \label{3.42}
   v(\bbox{K}) &=& {A+B\over 2C} \left( 1 - 
     \sqrt{1 - \left({2C\over A+B}\right)^2}
                       \right) \, ,
 \end{eqnarray}
where, with the notational shorthand 
$\tilde \xi \equiv \tilde x + i \tilde y$,  
 \begin{eqnarray}
 \label{3.43}
   A &=& \left\langle \left( \tilde t  
         - {\tilde \xi\over \beta_\perp} \right)^2 \right\rangle(\bbox{K}) \, ,
 \\
 \label{3.44}
   B &=&  \left\langle \left( \tilde z
         - {\beta_l\over \beta_\perp} \tilde \xi \right)^2 
     \right\rangle(\bbox{K}) 
   \, ,
 \\
 \label{3.45}
   C &=& \left\langle \left( \tilde t - {\tilde \xi\over \beta_\perp} \right)
                      \left( \tilde z - {\beta_l\over \beta_\perp} 
                             \tilde \xi \right) \right\rangle(\bbox{K}) \, .
 \end{eqnarray}
In these expressions, $\langle \tilde 
y\rangle = \langle \tilde x \tilde y \rangle = 0$ since we are dealing
with azimuthally symmetric sources. The kinematical limit $K_\perp \to 0$
is not free of subtleties, as one may guess by finding $\beta_\perp$
in the denominator of the above expressions. Indeed, for $K_\perp = 0$ the
mass-shell constraint (\ref{3.6}) reads $q^0 = \beta_l q_l$, and
the relative momenta $q^0$, $q_l$ and $q_\perp$ on which the
YKP ansatz (\ref{3.35}) is based are no longer independent. 
Hence, strictly speaking, the YKP parametrization exists only for 
$K_\perp \not= 0$. In practice this does not limit 
the applicability since the $K_\perp \to 0$-limit is well-defined
for all YKP parameters. 

Mathematically, the Cartesian and YKP parametrizations are equivalent
and differ only in the choice of the independent relative momentum
components. The Cartesian radius parameters can therefore be
expressed in terms of the YKP ones~\cite{HTWW96,WHTW96} via
 \begin{eqnarray}
 \label{3.46}
    R_s^2 &=& R_\perp^2 \, , \\
 \label{3.47}
   R_{\rm diff}^2 &=& R_o^2 - R_s^2 = \beta_\perp^2 \gamma^2 
             \left( R_0^2 + v^2 R_\parallel^2 \right) \, ,
 \\
 \label{3.48}
   R_l^2 &=& \left( 1 - \beta_l^2 \right) R_\parallel^2 
             + \gamma^2 \left( \beta_l-v \right)^2
             \left( R_0^2 + R_\parallel^2 \right)\, ,
 \\
 \label{3.49}
   R_{ol}^2 &=& \beta_\perp \left( -\beta_l R_\parallel^2 
             + \gamma^2 \left( \beta_l-v \right)
             \left( R_0^2 + R_\parallel^2 \right) \right)\, .
 \end{eqnarray}
This set of equations provides a useful consistency check for 
correlation data analyzed independently with both the
Cartesian and the YKP parametrizations. 
To invert them, one has to calculate 
 \begin{eqnarray}
   A &=& {1 \over \beta_\perp^2} R_{\rm diff}^2 , 
 \label{3.50} \\
   B &=& R_l^2 - {2\beta_l \over \beta_\perp} R_{ol}^2
         + {\beta_l^2 \over \beta_\perp^2} R_{\rm diff}^2  ,
 \label{3.51} \\
   C &=& - {1\over \beta_\perp} R_{ol}^2 
         + {\beta_l\over \beta_\perp^2} R_{\rm diff}^2 .
 \label{3.52}
 \end{eqnarray}
and insert them into (\ref{3.39})-(\ref{3.42}). These relations
imply in particular that the YK velocity $v(\bbox{K})$ can be
calculated from measured Cartesian HBT radii. In model studies
\cite{WHTW96}, it was demonstrated that this velocity follows closely
the velocity of the Longitudinal Saddle Point System (LSPS) which 
is the longitudinally comoving Lorentz frame at the point of 
highest particle emissivity for a given pair momentum $\bbox{K}$.
In this sense the YK velocity can be interpreted as the effective
source velocity. Note that in the Cartesian parametrization the
kinematical information associated with the YK velocity is contained
in the cross-term $R_{ol}^2$~\cite{P83,SSX98}.

While the values extracted for $R_0^2(\bbox{K})$ and
$R_\parallel^2(\bbox{K})$ are independent of the longitudinal velocity
of the observer system, their space-time interpretation is not. For
their analysis the so-called Yano-Koonin frame, which is pair momentum
dependent and defined by $v(\bbox{K}) = 0$, offers itself: in this
frame, the terms $\sim vC$ in (\ref{3.40})/(\ref{3.41}) vanish. 
For a class of Gaussian model emission functions including 
longitudinal and transverse flow it was shown that the spatio-temporal
interpretation of the fit parameters is then particularly
simple~\cite{CNH95,HTWW96}: 
 \begin{eqnarray}   
   R_\perp^2(\bbox{K}) &=& \langle \tilde{y}^2 \rangle(\bbox{K}) \, ,
 \label{3.53} \\
   R_\parallel^2(\bbox{K})  
     &\approx& \langle \tilde z^2 \rangle(\bbox{K}) \, ,
 \label{3.54} \\
   R_0^2(\bbox{K}) 
    &\approx& \langle \tilde t^2 \rangle(\bbox{K}) \, .
 \label{3.55}
 \end{eqnarray}
In other words, the three YKP radius parameters give directly
the transverse, longitudinal and temporal size of the effective
source in the rest frame of the emitter. Especially  the last
equation (\ref{3.55}) seems to imply that in the YKP parametrization
the emission duration $\langle \tilde t^2 \rangle$ can be accessed
directly. This, however, is model-dependent: in (\ref{3.54})/(\ref{3.55})
certain terms were omitted on the right hand side which can become
large in certain model scenarios. For example, opaque source models with 
strongly surface-dominated emission give a leading geometric contribution
to $R_0^2$~\cite{HV97,TH97}
  \begin{equation}
    R_0^2 \approx -{1\over \beta_\perp^2}
                  \left( \langle \tilde{x}^2 \rangle
                         - \langle \tilde{y}^2 \rangle \right)
          \qquad \hbox{for opaque sources.}
          \label{3.56}
  \end{equation}
Large geometric corrections were also observed for transversely
expanding sources with a box-shaped transverse density profile
\cite{T99}. As will be discussed in section~\ref{sec5d1}, first checks
indicate that opaque source models cannot reproduce the experimental
data consistently~\cite{TH97,WTH97}, but they play an important role
in understanding the range of validity of the approximations 
(\ref{3.54})/(\ref{3.55}).

It can happen \cite{TH98,TH98a}, especially for sources with $\langle
\tilde x^2 -\tilde y^2 \rangle < 0$, that the argument of the square
root in (\ref{3.42}) becomes negative. In this case the YKP parameters 
are not defined. For such situations a {\em modified YKP
  parametrization} was suggested in \cite{TH98,TH98a} which does not
have this potential problem. The corresponding modified YK velocity
still follows closely the fluid velocity at the point of highest
emissivity, i.e. also the modified YKP parametrization allows to
determine the effective source velocity. However, the interpretation
of the modified parameters ${R'_0}^2$, ${R'_\parallel}^2$ \cite{TH98a} 
is less straightforward than (\ref{3.54}/\ref{3.55}); in particular
the parameter ${R'_0}^2$ is in general {\em not} dominated by the
emission duration $\langle \tilde t^2 \rangle$.

So far, the YKP-parametrization has not been extended to collisions at
finite impact parameter.

\section{Other Gaussian parametrizations}
\label{sec3c}

A plethora of different Gaussian parametrizations can be found in 
the literature. They belong to either of two different
classes. The first class contains parametrizations which are 
equivalent to the ones discussed above. A typical example for
azimuthally symmetric sources is~\cite{M96}
 \begin{eqnarray}
   C(\bbox{q},\bbox{K}) &=&
       1 +  \lambda(\bbox{K})\,\exp\left[ - R_x^2(\bbox{K})\, q_x^2
                                - R_y^2(\bbox{K})\, q_y^2 \right.
                     \nonumber \\
       && \qquad \left.  - R_z^2(\bbox{K})\, q_z^2
                         - T^2(\bbox{K})\, {(q^0)}^2 \right]  \, .
   \label{3.57}
 \end{eqnarray}
It provides a perfectly valid azimuthally symmetric ansatz whose four
fit parameters are, after insertion of the mass-shell constraint
(\ref{3.6}), seen to be in one-to-one correspondence with the
Cartesian or YKP ones. One should keep in mind, however, that the
suggestive notation $T^2(\bbox{K})$ does not warrant a physical
interpretation in terms of a temporal extension; also the
$R_i^2(\bbox{K})$ do not only contain spatial information. The
interpretation of these parameters has to be established again on the
basis of space-time variances. Other equivalent parametrizations can
be found in the literature; the relations between the various radius
parameters are discussed in Refs.~\cite{TH98a,T99}.

The second class contains incomplete parametrizations: either certain
terms (e.g. the out-longitudinal cross-term in the Cartesian
parametrization) are neglected, or the ansatz is dimensionally
reduced. The prime example is the $q_{\rm inv}$-parametrization
($q_{\rm inv}^2 = \bbox{q}^2 - (q^0)^2$)
 \begin{equation}
   C(\bbox{q},\bbox{K}) =
       1 +  \lambda(\bbox{K})\,
        \exp\left[ - R_{\rm inv}^2(\bbox{K})\, q_{\rm inv}^2
             \right] \, .
   \label{3.58}
 \end{equation}
Here all the different spatial and temporal informations contained in
the space-time variances $\langle \tilde{x}_\mu \tilde{x}_\nu\rangle$
are mixed into one fit parameter $R_{\rm inv}^2(\bbox{K})$, and there is
no possibility to unfold them again. Furthermore, low-dimensional 
projections of a correlator which is 
well-described by a complete three-dimen\-sio\-nal Gaussian parametrization
in general deviate from a Gaussian shape. This is true in particular for 
projections on $q_{\rm inv}$; in fact, it was repeatedly observed that
$C(q_{\rm inv})$ is better described by an exponential or an inverse
power of $q_{\rm inv}$. For a space-time interpretation
of correlation data such incomplete parametrizations are not suitable. 
It is often argued that limited statistics forces one in practice to 
adopt dimensional reductions in the fit parameter space. But even
then it is preferable to bin the data in three independent $q$-components
first and to project this three-dimensional histogram onto different
one-dimensional directions for fitting purposes. The parameters
extracted this way can be compared to suitably averaged versions
of the HBT radius parameters (\ref{3.10})-(\ref{3.15}) or 
(\ref{3.39})-(\ref{3.42}).
%
\section{Estimating the phase-space density}
\label{sec3d}

As shown by Bertsch~\cite{B94} the correlation function can be
used to extract the average phase-space density at freeze-out.
In the present section we describe how this works.

The phase-space density $f(\bbox{x},\bbox{p},t)$ of free-streaming 
particles at time $t$ is obtained by summing up the particles
emitted by the source function up to this time along the 
corresponding trajectory:
  \begin{equation}
    f(\bbox{x},\bbox{p},t) = {(2\pi)^3\over E_p}\,
    \int_{-\infty}^t dt'\, S(\bbox{x}- \bbox{\beta}(t -t'),t';\bbox{p})\, .
    \label{3.59}
  \end{equation}
Here $\bbox{\beta}$ is the velocity of particles with momentum $\bbox{p}$.
For large times $t$, $f$ is normalized to the total 
event multiplicity, $\int d^3x\, d^3p\, f(\bbox{x},\bbox{p},t)/(2\pi)^3
= N$. According to Liouville's theorem, the spatial average of any
power of $f$ is time-independent after particle production has
ceased ($t>t_f$). This is in particular true for the average
phase-space density  
  \begin{equation}
    \langle f\rangle(\bbox{p}) = 
    {{\int d^3x\, f^2(\bbox{x},\bbox{p},t>t_f)}\over
    {\int d^3x\, f(\bbox{x},\bbox{p},t>t_f)}}\, .
     \label{3.59b}
  \end{equation}
This quantity can be obtained from the measured one- and two-particle
spectra. To this end one calculates, see (\ref{2.24}), 
  \begin{eqnarray}
        {\cal P}_1(\bbox{p}_1)\, {\cal P}_1(\bbox{p}_2)
        \bigl( C(\bbox{p}_1,\bbox{p}_2) - 1\bigr)
        &=& 
        {\cal P}_2(\bbox{p}_1,\bbox{p}_2) -       
        {\cal P}_1(\bbox{p}_1)\, {\cal P}_1(\bbox{p}_2)
        \nonumber \\
        &=& \left\vert  \int d^4x\, S(x,K)\, e^{iq\cdot x} 
        \right\vert^2
        \label{3.59c}
  \end{eqnarray}
and integrates over $q$ with the mass shell constraint
$q\cdot K = 0$. After substituting 
$\bbox{x}\to \bbox{x}+ \bbox{\beta}x^0$ (where $\bbox{\beta}=\bbox{K}/K^0$)
one obtains
  \begin{eqnarray}
        &&\int d^4q\, \delta(q\cdot K)\, 
        \left[ {\cal P}_1(\bbox{p}_1)\, {\cal P}_1(\bbox{p}_2)
        \bigl( C(\bbox{p}_1,\bbox{p}_2) - 1\bigr)\right] 
        \nonumber \\
        && = \int {d^3q\over K^0}
        \int d^3x\, d^3y\, e^{-i\bbox{q}\cdot (\bbox{x}-\bbox{y})}
        \nonumber \\
        && \qquad  \times
        \int dx^0\, S(\bbox{x}+ \bbox{\beta}x^0,x^0,K)
        \int dy^0\, S(\bbox{y} + \bbox{\beta}y^0,y^0,K)
        \label{3.59d}\\
        &&\approx { (2\pi)^3\over E_K}
        \int d^3x\, \Sigma^2(\bbox{x},\bbox{K})\, .
        \label{3.59e}
  \end{eqnarray}
In the last step we used the on-shell approximation 
$K^0 \approx E_K$ and introduced the time-integrated 
emission function
  \begin{equation}
        \Sigma(\bbox{x},\bbox{K}) = \int_{-\infty}^{\infty}
        dx^0\, S(\bbox{x}+\bbox{\beta}x^0,x^0,K)\, .
        \label{3.59f}
  \end{equation}
It is easy to show that
  \begin{equation}
        {(2\pi)^{3n}\over E_K^n} \
        \int d^3x\, \Sigma^n(\bbox{x},\bbox{K})
        = \int d^3x\, f^n(\bbox{x},\bbox{K},t>t_f)\, .
        \label{3.59g}
  \end{equation}
Combining this with (\ref{3.59b}) and (\ref{3.59e}) and using the
smoothness approximation ${\cal P}_1(\bbox{p}_1)\, 
{\cal P}_1(\bbox{p}_2) \approx \left({\cal P}_1(\bbox{K})\right)^2$
on the l.h.s. of (\ref{3.59d}), the phase-space density $f$
can be expressed in terms of observable quantities:
  \begin{equation}
    \langle f\rangle(\bbox{K}) \approx 
    {\cal P}_1(\bbox{K}) \int d^4q\, \delta(q\cdot K)\,
        \bigl( C(\bbox{q},\bbox{K}) - 1 \bigr)\, .
    \label{3.60}
  \end{equation}
Using the Cartesian parametrization of the correlator for 
zero impact parameter collisions as given in section~\ref{sec3a},
the r.h.s. takes the explicit form~\cite{B94,E87797b}
  \begin{eqnarray}
    \langle f\rangle(K_\perp,Y) &=& 
       {dN\over dY\, M_\perp dM_\perp\, d\Phi}\, 
        {1\over V_{\rm eff}(K_\perp,Y)}\, ,
    \label{3.61} \\
    V_{\rm eff}(K_\perp,Y) &=& {M_\perp\cosh{Y}\over \pi^{3/2}}
    R_s(\bbox{K})\sqrt{R_o^2(\bbox{K})\, R_l^2(\bbox{K}) - 
        (R_{ol}^2(\bbox{K}))^2}\, .
    \label{3.62}
  \end{eqnarray}
This expression assumes an intercept $\lambda = 1$ for the
correlator. In reality a considerable fraction of the observed
pions stems from resonance decays after freeze-out. The longlived
resonances affect the intercept parameter $\lambda$ 
(see section~\ref{sec5c5}),
and the corresponding decay pions should not be counted in the
average pion phase-space density near freeze-out. This can be
taken into account by substituting in (\ref{3.61})
  \begin{eqnarray}
    \langle f\rangle(\bbox{K}) &\longrightarrow& 
    \sqrt{\lambda(\bbox{K})}\,\,\, \langle f\rangle(\bbox{K})\, ,
    \label{3.63}\\
    \lambda(\bbox{K}) &=&   \left( 
     1 - \sum_{r} f_r(\bbox{K}) \right)^2\, ,
    \label{3.64}
  \end{eqnarray}
where the sum in (\ref{3.64}) runs over the resonance fractions
$f_r(\bbox{K})$ of longlived resonances contributing to the
one-particle spectrum at $\bbox{K}$.
%
\begin{figure}[ht]\epsfxsize=10.0cm 
\centerline{\epsfbox{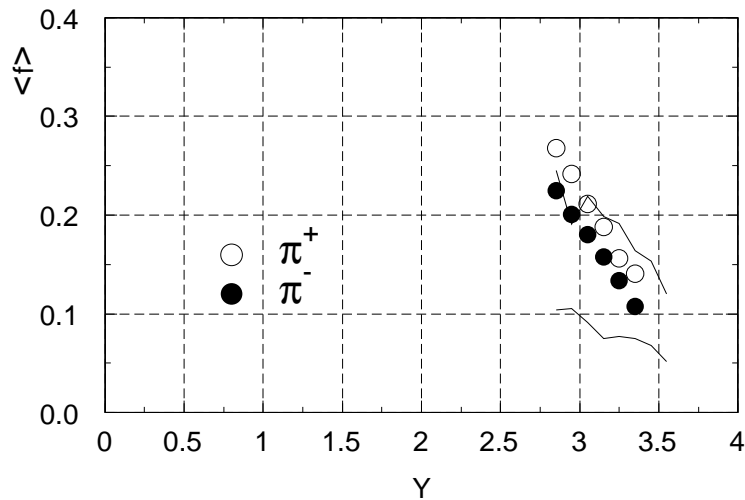}}
\caption{
Rapidity dependence of the average pion phase-space density 
(\ref{3.63}). The data points are for Au+Au collisions at 10.8 GeV/A
measured by E877 at the AGS. The statistical errors are smaller than
the symbols, the systematic normalization uncertainty is of the order
of 30 \%. The upper (lower) lines were obtained assuming a thermal 
distribution (\ref{3.67}) with a temperature extracted from the 
high (low) momentum part of the $\pi^-$-spectrum. (Figure taken
from~\protect\cite{M96}.) 
}\label{figdar}
\end{figure}

A first application of this approach was performed by the 
E877 experiment for Au+Au collisions at the AGS~\cite{M96,E87797b}.
The extracted average pion phase-space density in the forward 
pair rapidity region $2.7<Y<3.3$ is shown in
Figure~\ref{figdar} as a function of the pair rapidity,
averaged over $K_\perp$. As very forward rapidities are
approached, it decreases from $\approx 0.2$ to 
$\approx 0.1$. This suggests that 
the phase-space density is largest in the center of the
collision at mid rapidity (here: $Y = 1.57$) and decreases 
towards forward and backward rapidities. 

\begin{figure}[ht]\epsfxsize=10.0cm 
\centerline{\epsfbox{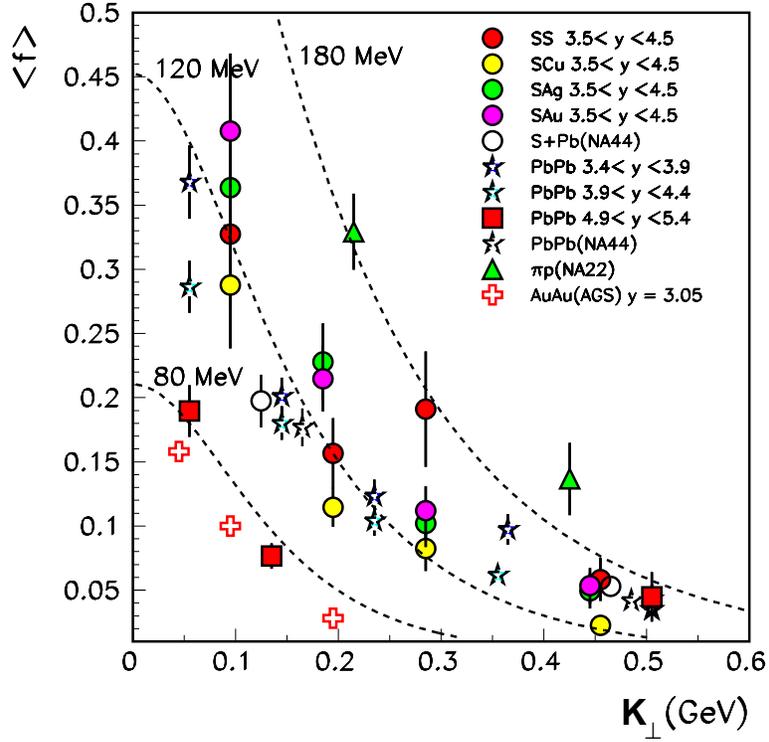}}
\caption{
Spatially averaged phase-space density in narrow rapidity windows
as a function of the transverse momentum $K_\perp$. The 
heavy-ion data span more than an order of magnitude in rapidity 
density $dN/d{\rm y}$, but the resulting freeze-out densities show much
less variation. The $K_\perp$-dependence of $\langle f\rangle$ can be 
well parametrized by an exponential function. The dashed lines show
Eq.~(\protect\ref{3.67}) for $T=80$, 120, and 180 MeV, respectively. 
(Figure taken from \protect\cite{FTH98}.)
}\label{figfth}
\end{figure}

More recently a large set of data, including $\pi$-$p$ collisions
at 250 GeV, S-nucleus collisions at 200 $A$ GeV, Pb+Pb collisions
at 158 $A$ GeV and Au+Au collisions at 10.8 $A$ GeV, was compiled
in Ref.~\cite{FTH98}. The resulting average phase-space densities
are shown in Figure~\ref{figfth}. They indicate a strong dependence
of the spatially averaged phase-space density on the transverse
momentum $K_\perp$ but very weak dependence on the size of the
collision system and on the pion rapidity density $dN/d{\rm y}$. 
This latter aspect can be interpreted as evidence for a universal
pion freeze-out phase-space density in heavy-ion collisions~\cite{FTH98}.
%

As a simple test of the thermalization assumption, one has 
compared~\cite{M96,FTH98} the measured phase-space densities in 
Figs.~\ref{figdar} and \ref{figfth} to that of a thermal Bose-Einstein 
equilibrium distribution 
  \begin{equation}
    f^{\rm BE}(K_\perp,Y) ={ 1\over {\exp\left[ M_\perp
          \cosh(Y-{\rm y}_s)/T\right] -1}}
    \label{3.67}
  \end{equation}
at pair rapidity $Y$ for a source rapidity ${\rm y}_s$, extracting 
the temperature from the measured single-particle spectra and 
two-particle correlations. Dynamical flow effects alter quantitative 
aspects of $f^{\rm BE}(\bbox{K})$, but qualitatively the phase-space 
density thus calculated compares surprisingly well with the 
experimentally determined one, see Figs.~\ref{figdar} and \ref{figfth}.
%
\chapter{Beyond the Gaussian parametrization}
\label{sec4}

The space-time variances characterizing the Cartesian HBT radius 
parameters (\ref{3.10})-(\ref{3.15}) can be written as second derivatives of
the correlator (\ref{3.4}) at $\bbox{q} = 0$:
  \begin{equation}
    \langle (\tilde{x}_i-{\beta}_i\tilde{t})
            (\tilde{x}_j-{\beta}_j\tilde{t})\rangle\,
            = - \left. { \partial^2 C(\bbox{q},\bbox{K})\over 
   \partial q_i\, \partial q_j}
        \right\vert_{ \bbox{q} = 0}\, .
 \label{4.1}
 \end{equation}
These curvature terms coincide with the experimentally determined
half widths of $C(\bbox{q},\bbox{K})$ only if the correlators is a
Gaussian in $\bbox{q}$.
Realistic two-particle correlation functions show, however, more or less 
significant deviations from a Gaussian shape. The consequences are
two-fold: the corresponding space-time variances do not agree
exactly with the fitted radius parameters, and the Gaussian 
radius parameters do not contain all the information contained
in $C(\bbox{q},\bbox{K})$. Nevertheless, qualtitatively all 
statements made above about mixing of spatial and temporal
information in the HBT radius parameters remain valid. The reason
is that the Fourier exponent $q\cdot x$ in (\ref{2.24}) can be
written as $\bbox{q}\cdot \left(\bbox{\beta}t\,- \bbox{x}\right)$.
Hence, the $\bbox{q}$ dependence of the correlator tests always
the same combinations $\left( \beta_i t\,- x_i\right)$
of spatial and temporal aspects of the emission function $S(x,K)$, 
irrespective of the shape of the two-particle correlator. 

If the correlator deviates from a Gaussian shape, one can either 
seek a more detailed characterization of $C(\bbox{q},\bbox{K})$ 
supplementing the Gaussian radius parameters by a larger set of
characteristic parameters, or one may proceed to reconstruct
from the measured correlations directly information about the
emission function without invoking a particular parametrization.
In the following we discuss both of these strategies.

%
\section{Imaging methods}\label{sec4a}
   
Instead of determining information about the emission function
$S(x,K)$, an alternative analysis strategy~\cite{BD97} aims at 
determining directly from the measured true correlator 
$C(\bbox{q},\bbox{K}) - 1$ 
the relative source function $S_{\bbox{K}}(\bbox{r})$ introduced in section 
(\ref{sec2b3a}):
  \begin{equation}
    C(\bbox{q},\bbox{K}) - 1 = \int d^3r\, K(\bbox{q},\bbox{r})\,
                             S_{\bbox{K}}(\bbox{r})\, .
    \label{4.2}
  \end{equation}
For the kernel, one usually chooses 
$K(\bbox{q},\bbox{r}) = \vert \phi_{\bbox{q}/2}(\bbox{r})\vert^2 - 1$
where $\phi_{\bbox{q}/2}$ describes the propagation of a pair, which
is created with a center of mass separation $\bbox{r}$ and detected
with relative momentum $\bbox{q}$. $\phi_{\bbox{q}/2}$ can include
two-particle final state interactions, cf. (\ref{2.98}) and (\ref{2.122});
for free particle propagation, 
$K(\bbox{q},\bbox{r}) = \cos(\bbox{q}\cdot \bbox{r})$.
 
Equation (\ref{4.2}) can be inverted uniquely, allowing in principle
for an unambiguous reconstruction of the relative source function 
$S_{\bbox{K}}(\bbox{r})$. In practice, finite measurement statistics
on the correlation function and the strongly oscillating nature of
the kernel $K(\bbox{q},\bbox{r})$ render the inversion problem
non-trivial. Brown and Danielewicz~\cite{BD97} suggest to parametrize
$S_{\bbox{K}}(\bbox{r})$ in terms of a finite number of basis functions
$g_j(\bbox{r})$,
  \begin{equation}
    S_{\bbox{K}}(\bbox{r}) = \sum_{j=1}^N S_j(\bbox{K})\, g_j(\bbox{r})\, ,
    \label{4.3}
  \end{equation}
and to determine the coefficients $S_j(\bbox{K})$ from the data.
In applications~\cite{BD97a,BD97} some insight has been gained
on how the reconstruction can be optimized by choosing 
suitable sets of functions $g_j$. 
%
\section{$q$-moments}\label{sec4b}

Rather than obtaining the HBT radius parameters from a Gaussian fit
to the measured correlation function,
a quantitative analysis of $C(\bbox{q},\bbox{K})$ can be based on
expectation values $\langle\!\langle g(\bbox{q})\rangle\!\rangle$ 
of the true correlator $C(\bbox{q},\bbox{K}) - 1$ in relative momentum 
space~\cite{WH96a,WH96b}:
   \begin{eqnarray}
     {\langle\!\langle q_i\, q_j \rangle\!\rangle} &=& 
     {\int d^3q\,\, q_i\, q_j\,\, 
       {\lbrack{ C(\bbox{q},\bbox{K}) - 1}\rbrack}
       \over \int d^3q\, {\lbrack{ C(\bbox{q},\bbox{K}) - 1}\rbrack}} 
     = {\textstyle{1\over 2}}\, {\left({ R^{-1}(\bbox{K})}\right)}_{ij},
     \label{4.7} \\
        R_{ij}(\bbox{K}) &=& \left(
        \begin{array}{ccc}
                R_o^2    & R_{os}^2 & R_{ol}^2 \\
                R_{os}^2 & R_s^2    & R_{sl}^2 \\
                R_{ol}^2 & R_{sl}^2 & R_l^2
        \end{array}\right)\,, \qquad i,j = o,s,l\, ,
     \label{4.8} \\
        \lambda(\bbox{K}) &=& \sqrt{\det R(\bbox{K})/\pi^3}
        \int d^3q\,
        {\lbrack{ C(\bbox{q},\bbox{K}) - 1}\rbrack}\, .
     \label{4.9} 
   \end{eqnarray}
Similar expressions exist for the YKP parameters~\cite{WH96a}. For a 
Gaussian correlator, the intercept parameter $\lambda$ and the
HBT radius parameters $R_{ij}^2$ obtained via these
$q$-moments coincide with the values extracted from a Gaussian
fit. For non-Gaussian correlators the HBT radius parameters and 
the intercept can be {\it defined} via (\ref{4.7})-(\ref{4.9}).

Deviations of the correlator from a Gaussian shape are then quantified 
by higher order $q$-moments, which in turn can be obtained as 
derivatives of the generating function $Z(\bbox{y},\bbox{K})$,
  \begin{eqnarray}
    \label{4.10} 
    Z(\bbox{y},\bbox{K}) &=& 
        \int d^3q\, e^{i\bbox{q}\cdot \bbox{y} }\,
                   {\lbrack{ C(\bbox{q},\bbox{K}) - 1 }\rbrack}\, ,\\
    \label{4.11}
     {\langle\!\langle{q_{i_1}\cdots q_{i_n}}\rangle\!\rangle} &=& 
     {(-i\partial)^n\over \partial y_{i_1}\cdots \partial y_{i_n}} 
     \ln Z(\bbox{y},\bbox{K})\bigg\vert_{\bbox{y}=0} \, .
  \end{eqnarray}
It is easy to see that $Z(\bbox{y},\bbox{K})$  
coincides with the relative source function (\ref{4.04})
  \begin{equation}
    Z(\bbox{y},\bbox{K}) = S_{\bbox{K}}(\bbox{y})\, .
    \label{4.12}
  \end{equation}
So far, only the uni-directional version of these expressions has been
used in theoretical and experimental investigations~\cite{lasiuk}: 
Restricting the correlator along one of the three Cartesian axes,
$\tilde C(\bbox{K},q_i) \equiv C(\bbox{K},q_i, q_{j\ne i}{=}0)$, the
corresponding HBT radius parameter and intercept are defined via 
the relations (we use the same notation as for the three-dimensional
$q$-moments) 
  \begin{eqnarray}
  \label{4.13}
     R_i^2(\bbox{K}) &=& {1\over 2\,\langle\!\langle q_i^2 
     \rangle\!\rangle}\, ,\qquad i=o,s,l,
  \\
  \label{4.14}
     \langle\!\langle q_i^2 \rangle\!\rangle(\bbox{K})
     &=& {\int dq_i\, q_i^2\, [\tilde C(\bbox{K},q_i)-1] \over
          \int dq_i\, [\tilde C(\bbox{K},q_i)-1] }\, ,
  \\
  \label{4.15}
     \lambda_i(\bbox{K}) &=& (R_i(\bbox{K})/\sqrt{\pi})
     \int dq_i\, [\tilde C(\bbox{K},q_i)-1] \, .
  \end{eqnarray}
In the case of deviations of $C(\bbox{q},\bbox{K})$ from a Gaussian 
form, the intercepts $\lambda_i$ from uni-directional fits
can differ in the different directions. A quantitative measure for the 
leading deviation of $C(\bbox{q},\bbox{K})$ from a Gaussian shape
is provided by the properly normalized fourth order moments, 
the kurtoses~\cite{WH96a,WH96b}
  \begin{equation}
  \label{4.16}
    \Delta_i(\bbox{K}) = { \langle\!\langle q_i^4 \rangle\!\rangle(\bbox{K})
                 \over
                3\, \langle\!\langle q_i^2 \rangle\!\rangle^2(\bbox{K})} - 1\, .
  \end{equation}
The uni-directional kurtosis $\Delta_i(\bbox{K})$ vanishes if the 
correlator at pair momentum $\bbox{K}$ is of Gaussian shape in 
the $q_i$-direction. The applications of $q$-moments to experimental 
data is limited by statistics: for higher order $q$-moments the
main contribution to the moment integral comes 
from larger relative momenta, while most
experimental information is concentrated at small $\bbox{q}$. 
%
\section{Three-particle correlations}\label{sec4c}
%
The complete set of $q$-moments (\ref{4.11}) provides a full 
characterization of the shape of the two-particle correlator. 
Alternatively, a complete characterization of this shape could be 
obtained in a Taylor expansion of $C(\bbox{q},\bbox{K})$ around 
$\bbox{q} = 0$. Extending equation (\ref{4.1}) to arbitrary orders, 
the corresponding Taylor coefficients read
  \begin{equation}
    \langle (\tilde{x}_{i_1}-{\beta}_{i_1}\tilde{t})\, 
            (\tilde{x}_{i_2}-{\beta}_{i_2}\tilde{t})\cdots
            (\tilde{x}_{i_n}-{\beta}_{i_n}\tilde{t})\rangle
            = \left. { i^n\, \partial^n C(\bbox{q},\bbox{K})\over 
   \partial q_{i_1}\, \partial q_{i_2}\cdots \partial q_{i_n}}
        \right\vert_{ \bbox{q} = 0}\, .
 \label{4.17}
 \end{equation}
However, due to finite momentum resolution, no detailed
information about the curvature of $C(\bbox{q},\bbox{K})$
at $\bbox{q} = 0$ is available from the data, and (\ref{4.17}) 
cannot be applied in practice. It illustrates, however, clearly
that the two-particle correlator $C(\bbox{q},\bbox{K})$ contains no
information about odd space-time variances: due to the
reflection symmetry $C(\bbox{q},\bbox{K}) \leftrightarrow 
C(\bbox{K},-\bbox{q})$, all odd space-time variances 
in (\ref{4.17}) vanish. 

The true two-particle correlator $R_2\equiv C - 1$ depends only on the 
modulus $\rho_{ij}$ of the Fourier transformed emission function but
not on its phase $\phi_{ij}$:
  \begin{eqnarray}
    R_2(i,j) &=& C(\bbox{p}_i,\bbox{p}_j) - 1
             = {\rho_{ij}^2\over \rho_{ii}\, \rho_{jj}}\, ,
             \label{4.17b} \\
    \rho_{ij}\,e^{i\phi_{ij}} &=& \int d^4x\, 
    S\left(x,\textstyle{1\over 2}(p_i + p_j)\right)\, 
    e^{i(p_i - p_j)\cdot x}\, .
    \label{4.18}
  \end{eqnarray}
Only the phase $\phi_{ij}$ contains information about 
odd space-time variances~\cite{HZ97}:
  \begin{equation}
    \phi_{ij} = q_{ij}\langle x\rangle_{ij}
                - \textstyle{1\over 6} 
                  \langle (q_{ij}\cdot\tilde{x}_{ij})^3\rangle_{ij}
                + O(q_{ij}^5)\, .
    \label{4.19}
  \end{equation}
Here $K_{ij}$, $q_{ij}$ denote the average and relative pair 
momentum for two particles with on-shell momenta $p_i$ and $p_j$,
respectively, and the space-time variances 
$\langle\, ..\,\rangle_{ij}$ are calculated with respect to the 
emission function $S(x,K_{ij})$. Three-particle correlations
give access to this phase as can be seen from~\cite{HZ97}
 \begin{eqnarray}
    r_3(\bbox{p}_1,\bbox{p}_2,\bbox{p}_3) &=&
    {R_3(\bbox{p}_1,\bbox{p}_2,\bbox{p}_3)\over
      \sqrt{ R_2(1,2)\, R_2(2,3)\, R_2(3,1)}}
    \nonumber \\
    &=& 2 \cos \left( \phi_{12} + \phi_{23} +\phi_{31} \right)\, ,
 \label{4.20}\\
    R_3(\bbox{p}_1,\bbox{p}_2,\bbox{p}_3) &=&
    C_3(\bbox{p}_1,\bbox{p}_2,\bbox{p}_3) - R_2(1,2)
    \nonumber \\
    && - R_2(2,3) - R_2(3,1) - 1\, .
 \label{4.21}
 \end{eqnarray}    
Here $R_3$  ($r_3$) is the true (normalized) three-particle correlation 
function. To investigate the space-time information contained in the 
phase combination on the r.h.s. of (\ref{4.20}) in more detail 
one can expand the emission function $S(x,K_{ij})$ in (\ref{4.18}) 
around  the average momentum of the particle triplet:
  \begin{eqnarray}
    \bar{K} &=& \textstyle{1\over 3}(p_1+p_2+p_3)
            = \textstyle{1\over 3}(K_{12}+K_{23}+K_{31})\, ,
    \label{4.22} \\
    K_{ij} &=& \bar{K} + \textstyle{1\over 6}(q_{ij} + q_{jk})\, ,
               \qquad i\not= j\not= k\, .
    \label{4.23}
  \end{eqnarray}
Using $q_{12}+q_{23}+q_{31} = 0$ one finds~\cite{HZ97,HV97}
  \begin{eqnarray}
    \Phi &=& \phi_{12} + \phi_{23} + \phi_{31} 
    \nonumber \\
    &=& {1\over 2}\, q_{12}^\mu\, q_{23}^\nu
                  \left[ {{\partial\langle x_\mu\rangle_3}
                          \over \partial \bar{K}^\nu}
                       - {{\partial\langle x_\nu\rangle_3}
                          \over \partial \bar{K}^\mu} \right]
    \nonumber \\
    && - {1\over 24}\, 
      \lbrack q_{12}^\mu\, q_{12}^\nu\, q_{23}^\lambda
              + q_{23}^\mu\, q_{23}^\nu\, q_{12}^\lambda\rbrack
                  \left[ {{\partial^2 \langle x_\mu\rangle_3}
                          \over \partial\bar{K}^\nu \partial\bar{K}^\lambda }
                       + {{\partial^2 \langle x_\nu\rangle_3}
                          \over \partial\bar{K}^\lambda \partial\bar{K}^\mu }
                       + {{\partial^2 \langle x_\lambda\rangle_3}
                          \over \partial\bar{K}^\mu \partial\bar{K}^\nu }
                        \right]
    \nonumber \\
    && - {1\over 2} q_{12}^\mu\, q_{23}^\nu\, 
       (q_{12} + q_{23})^\lambda\, 
       \langle \tilde{x}_\mu \tilde{x}_\nu \tilde{x}_\lambda\rangle_3 
       + O(q^4)\, .
       \label{4.24}
  \end{eqnarray}
Here the averages $\langle\, \dots\,\rangle_3$ have been calculated
with the emission function $S(x,\bar{K})$ taken at the momentum
$\bar{K}$ of the particle triplet.

The measurable phase $\Phi$ depends on the odd space-time variances
$\langle \tilde{x}^3\rangle$, etc., and on derivatives of the point of 
highest emissivity $\langle x\rangle_3$ with respect to $\bar{K}$. 
These reflect the asymmetries of the source around its center. In the 
Gaussian approximation (\ref{3.1}) they vanish. These considerations 
show that the true three-particle correlator contains additional 
information which is not accessible via two-particle correlations.

In practice, however, it is very difficult to extract this information. 
The leading contribution to $\Phi$ is of second order in the relative 
momenta $q_{ij}$, and in many reasonable models it even vanishes
\cite{HV97}. Therefore new information typically enters 
$r_3(\bbox{p}_1,\bbox{p}_2,\bbox{p}_3)$ at sixth order in $q$. The
measurement of the phase $\Phi$ is thus very sensitive to an accurate 
removal of all leading $q^2$-dependences by a proper determination and 
normalization of the two-particle correlator. These general arguments 
are supported by model studies which found that flow, resonance decay 
contributions or source asymmetries leave generically small effects on 
the phase~\cite{HV97}. 

On the other hand, it was pointed out that the intercept of the 
normalized true three-particle correlator $r_3$ in (\ref{4.20}) may 
provide a good test for the chaoticity of the source. Writing the 
emission function for a {\em partially coherent} source as 
$S = S_{\rm cha} + S_{\rm coh}$, the intercept $\lambda_3$ 
of $r_3$ is given in terms of the chaotic fraction $\epsilon(\bbox{p})$
of the single-particle spectrum \cite{BBMST90,HZ97}:
 \begin{eqnarray}
    \lambda_3(\bar{\bbox{K}}) &\equiv& 
    r_3(\bar{\bbox{K}},\bar{\bbox{K}},\bar{\bbox{K}})
    = 2\sqrt{\epsilon(\bar{\bbox{K}})}
      {{3-2\epsilon(\bar{\bbox{K}})}\over 
       \left(2-\epsilon(\bar{\bbox{K}})\right)^{3/2}}\, ,
 \label{4.25} \\
    && \qquad \epsilon(\bbox{p}) = 
    {\int d^4x\, S_{\rm cha}(x,p)\over \int d^4x\, S(x,p)}\, .
 \label{4.26}
 \end{eqnarray}
In contrast to the intercept 
$\lambda(\bbox{K})$ of the two-particle correlator, the intercept 
(\ref{4.25}) of the normalized three-particle correlator is not 
affected by decay contributions from long-lived resonances which cancel 
in the ratio (\ref{4.20}) \cite{HZ97}. 

Complete small-$q$ expansions of $R_2$ and $R_3$ which generalize the 
Gaussian parametrization  (\ref{3.4}) to the case of partially coherent 
sources and to three-particle correlations, improving on earlier
results \cite{BBMST90,PRW92}, can be found in \cite{HZ97}. 
In the framework of a
multidimensional simultaneous analysis of two- and three-pion correlations
they permit to separately determine the sizes of the homogeneity
regions of the chaotic and coherent source components as well as
the distance between their centers.
%
\chapter{Results of model studies}\label{sec5}

A completely model-independent reconstruction of the emission
function $S(x,K)$ from measured correlation data is not possible
since, due to the mass-shell constraint (\ref{2.27}), only certain 
combinations of spatial and temporal source characteristics are 
measurable. Only the time-integrated relative distance 
distribution in the pair rest frame $S_{\bbox{K}}(\bbox{r})$ 
can be determined uniquely from momentum correlations. This
corresponds to a whole class of emission functions $S(x,K)$, not a 
unique one.

In practice, the mass-shell constraint is not the only problem
for the reconstruction of the emission function. The
statistical uncertainties of experimental data turn 
even the reconstruction of the relative source function
$S_{\bbox{K}}(\bbox{r})$ into a complicated task~\cite{BD97}, where
additional model assumptions are employed to obtain a convergent 
numerical procedure.

Due to these fundamental and pragmatic problems, the
analysis of experimental correlation data starts
from a model of the emission function $S(x,K)$ from which
one- and two-particle spectra are calculated and compared to
the data. Here one can follow two alternative approaches:
either, one simulates directly the kinetic evolution of 
the reaction zone up to the freeze-out stage, calculates
the one- and two-particle momentum spectra and compares them
to the data. This is the approach followed in event generator
calculations. The space-time information is then extracted
from within the model simulation. Alternatively, one applies
the tools presented in the preceding chapters by using simple
parametrizations of the emission function and adjusting the
model parameters by a comparison to data. This approach makes
explicit use of the relation between the measured HBT parameters 
and the space-time variances of the emission function, trying 
to obtain a direct space-time interpretation of the measurements.
In this case the dynamical consistency of the extracted
space-time structure of the source with the preceding kinetic
evolution of the reaction zone must be established {\it a
posteriori}. We will here concentrate on the second approach;
reviewing a large body of model studies we illustrate the
extent to which momentum correlations have a generic 
space-time interpretation. The insights gained in these models
studies are summarized at the end of our review into an analysis 
strategy which allows for a simple determination of the relevant 
space-time aspects of the source from experimental data. 
This strategy is then applied to recent one- and two-particle spectra
for negatively charged particles measured by the NA49 collaboration
in Pb+Pb collisions at the CERN SPS.

\section{A class of model emission functions}
\label{sec5a}

We introduce a class of analytical models for the emission
function of a relativistic nuclear collision, starting from a
simple model and then discussing several dynamical and geometrical
refinements. Models of this class have been used extensively
in the literature~\cite{CL95,CL96,S95,AS96,CSH95a,CNH95,WSH96,WH96a,W97}. 

\subsection{The basic model}
\label{sec5a1}

Whatever the true particle phase-space distribution of the 
collision at freeze-out is, we expect that its main characteristics
can be quantified by its widths in the spatial and temporal
directions, a collective dynamical component (parametrized by
a collective flow field) which determines the strength
of the position-momentum correlations in the source, and a second, 
random dynamical component in momentum space
(parametrized by a temperature).

A parametrization which is sufficiently flexible to incorporate 
these features but still allows for an intuitive physical interpretation
of its model parameters assumes local thermalization 
prior to freeze-out at temperature $T$ and incorporates 
collective expansion in the longitudinal and transverse directions via
a hydrodynamic flow field $u_{\mu}(x)$. The source has a finite geometrical 
size in the spatial and temporal directions, encoded in transverse 
and longitudinal Gaussian widths $R$ and $\Delta \eta$ as well as 
in a finite particle emission duration $\Delta \tau$. Here 
$\tau = \sqrt{t^2 - z^2}$ denotes the longitudinal proper time and 
$\eta = \textstyle{1\over 2} \ln{\lbrack{ (t+z)/(t-z) }\rbrack}$
the space-time rapidity. The parametrization is optimized for
sources with strong, approximately boost-invariant longitudinal
expansion for which freeze-out occurs close to a hypersurface
of constant longitudinal proper time $\tau = \tau_0$. It
is thus more suitable for high than for low energy collisions.
The source is defined by an emission function 
for each particle species $r$~\cite{S95,AS96,CSH95a,CL96,WSH96,WH96a}: 
 \begin{eqnarray}
   S^{\rm dir}_r(x,p) &=& {2J_r + 1 \over (2\pi)^3 \pi\, \Delta\tau}\,
   m_\perp \cosh({\rm y}-\eta) 
   \exp\left[- {p \cdot u(x) - \mu_r \over T} \right]\,  
   \nonumber \\
   && \times \exp\left[ - {r^2\over 2 R^2} 
                     - {\eta^2\over 2 (\Delta\eta)^2}
                     - {(\tau-\tau_0)^2 \over 2 (\Delta\tau)^2}
                 \right] \, . 
 \label{5.1}
 \end{eqnarray}
For explicit calculations, it is helpful to express the particle 
four-momentum $p_\mu$ using the momentum rapidity ${\rm y}$ and the
transverse mass $m_\perp = \sqrt{p_\perp^2 + m^2}$.
This allows for a simple
expression of the Boltzmann factor in (\ref{5.1}),
  \begin{eqnarray}
    p_\mu &=& \left( m_\perp\cosh {\rm y},p_\perp,0,
     m_\perp\sinh {\rm y}\right)\, ,
    \label{5.2} \\
    {p\cdot u(x)\over T} &=& {m_\perp\over T}\cosh({\rm y}-\eta)\cosh\eta_t
                      - {p_\perp\over T}{x\over r}\sinh\eta_t\, .
    \label{5.3}
  \end{eqnarray}
For sharp freeze-out of the particles from the thermalized fluid
along a hypersurface $\Sigma(x) = (\tau_0\cosh\eta,x,y,\tau_0\sinh\eta)$,
we would have to choose the emission function proportional to
$p\cdot n(x)$, where 
  \begin{eqnarray}
    n_\mu(x) &=& \int_\Sigma d^3\sigma_\mu(x')\, 
                 \delta^{(4)}(x-x')\, ,
                 \label{5.4} \\
    p\cdot n(x) &=& m_\perp\cosh({\rm y}-\eta)\, \delta(\tau-\tau_0)\, .
                 \label{5.5}
  \end{eqnarray}
The four-vector $n_\mu(x)$ points normal to the freeze-out 
hypersurface. The term $m_\perp \cosh({\rm y}-\eta)$
in the emission function (\ref{5.1}) stems from this geometrical condition,
while we have replaced the $\delta$-function in (\ref{5.5}) 
by a properly normalized Gaussian to allow for a finite emission
duration $\Delta\tau$.
The factor $2J_r+1$ accounts for the spin degeneracy of the emitted
particle, and a chemical potential $\mu_r$ allows for separate
normalization of all particle yields. The ansatz implies
that all particles are assumed to freeze out with the same geometric 
characteristics and the same collective flow, superimposed by 
random thermal motion with the same temperature. 

For the flow profile we assume Bjorken scaling~\cite{BJ83} in the 
longitudinal direction, $v_l=z/t$; this identifies the flow
rapidity $\eta_{\rm flow} = \textstyle{1\over 2} \log 
\textstyle{ {1+v_l}\over {1-v_l}}$ with the space-time rapidity.
Assuming a linear transverse flow rapidity profile of strength 
$\eta_f$ in the transverse direction, the normalized flow field 
$u_\mu(x)$ reads
 \begin{eqnarray}
   u_\mu(x) &=& \left( \cosh\eta\, \cosh\eta_t, \textstyle{x\over r}
     \sinh\eta_t, \textstyle{y\over r} \sinh\eta_t, 
     \sinh\eta\, \cosh\eta_t\right)\, ,
     \nonumber \\
     && \eta_t(r) = \eta_f {r\over R}\, .
     \label{5.6}
 \end{eqnarray}
In spite of the longitudinal boost-invariance of the flow, 
the source as a whole is not boost-invariant unless the longitudinal
Gaussian width $\Delta\eta \to \infty$.

The model emission function (\ref{5.1}) is thus completely
specified by six common and one species-dependent model parameters:
  \begin{equation}
    T\, ,\eta_f\, ,R\, ,\Delta\eta\, 
    ,\Delta\tau\, ,\tau_0\, ,\mu_r\, .
    \label{5.7}
  \end{equation}
The number of model parameters can be reduced by assuming 
chemical equilibrium at freeze-out. This provides the following
constraint between the chemical potentials:
 \begin{equation}   
 \label{5.8}
   \mu_r = b_r \mu_B + s_r \mu_S\, .
 \end{equation}
Here $b_r$ and $s_r$ are the baryon number and strangeness of 
resonance $r$, and  $\mu_B$, $\mu_S$ are the two independent
chemical potentials required for baryon number and strangeness 
conservation in the reaction zone.

\subsection{Model extensions}
\label{sec5a2}

Comparative model studies investigate to what extent geometrical
or dynamical assumptions put into the emission function $S(x,K)$
leave traces in the observed one- and two-particle spectra,
and how this allows to distinguish between different
collision scenarios. To this end one compares, for example, the model 
(\ref{5.1}) with modified model assumptions about the particle 
production and emission processes in the collision region. 
Here we focus on a few possible extensions of the basic
model (\ref{5.1}) which all have a clear physical motivation
and which have been investigated in the literature. 
\subsubsection{Opaque sources}
    The particle production and freeze-out mechanisms in heavy
    ion collision are largely unknown. In particular,
    it is not settled whether hadronic freeze-out resembles more
    the surface evaporation of a hot water droplet or the
    simultaneous bulk freeze-out leading to the decoupling
    of photons and the transition from an opaque to a 
    transparent universe in the early stages of Big Bang
    cosmology. The parametrization (\ref{5.1}) reflects the
    second type of scenario, with a bulk transition from opaqueness
    to transparency at proper time $\tau_0 \pm \Delta\tau$.
    Surface emission can be included into (\ref{5.1}) by
    multiplying this emission function with an exponential absorption 
    factor~\cite{HV96a,TH98} 
    \begin{eqnarray}
      S_{\rm opaque}(x,p) &=& S(x,p)\, 
         \exp\left[-\sqrt{8/\pi} 
         \left( l_{\rm eff}/\lambda_{\rm mfp}\right) \right]\, ,
         \label{5.9} \\
      l_{\rm eff} &=& l_{\rm eff}(r,\phi)
      = e^{-{y^2\over 2R^2}} \int_x^\infty
        e^{-{x'^2\over 2R^2}}\, dx'\, ,
        \label{5.10}
    \end{eqnarray}
    where $y = r\, \sin\phi$, $x = r\cos\phi$. This extra factor
    suppresses exponentially emission from the interior of the
    emission region. The particle
    propagates in the $x$ (out) direction. The Gaussians
    in the expression (\ref{5.10}) parametrize the matter 
    density seen by the particle according to the geometrical
    source distribution in (\ref{5.1}). 
\subsubsection{Temperature gradients}
    The model emission function $S(x,p)$ presented in (\ref{5.1})
    assumes that all volume elements freeze out at the same
    temperature $T(x) = T$. Since freeze-out is controlled by
    a competition between the local expansion and scattering
    rates, and the latter have a very strong temperature
    dependence, this is not an unreasonable assumption~\cite{SH92,MH97}.
    With this assumption all position-momentum 
    correlations in the source in (\ref{5.1}) stem from the collective 
    dynamics characterized by the flow $u_\mu(x)$ in the Boltzmann
    factor. To contrast this scenario with models in which the
    $x$-$K$-correlations have a different origin, 
    model extensions with a particular
    temperature profile have been studied~\cite{CL96,TH97}:
    \begin{equation}
      {1\over T(x)} = {1\over T_0}
      \left( 1 + a^2{r^2\over 2R^2}\right)\, 
      \left( 1 + d^2 {(\tau - \tau_0)^2\over 2\tau_0^2}\right)\, .
      \label{5.11}
    \end{equation}
    Hereby transverse and temporal temperature gradients are introduced
    via two additional fit parameters
    $a$ and $d$. The model (\ref{5.11}) implies that the production
    of particles with larger $m_\perp$ is more strongly concentrated 
    near the symmetry axis and average freeze-out time. 
\subsubsection{Emission functions for non-central collisions}
    The emission function (\ref{5.1}) shows a 
    $y \leftrightarrow -y$ symmetry in the {\it osl}-coordinate
    system. It describes an azimuthally symmetric emission region
    which is adequate for zero impact parameter collisions.
    To investigate the new qualitative features introduced
    by collisions at finite impact parameter, one can 
    study azimuthally asymmetric extensions~\cite{W97,HH98,HL98} of the
    model emission function (\ref{5.1}). Here we take
    $x$ and $y$ to be defined in the laboratory system, with
    $x$ in the reaction plane. An elliptic geometric deformation
    of the source in the transverse plane, characterized 
    by an anisotropy parameter $\epsilon_s$,
    is obtained by replacing in (\ref{5.1})
    \begin{eqnarray}
      && \exp\left[ -{r^2\over 2R^2}\right] \longrightarrow
      \exp\left[ - {x^2 \over 2 \rho_x^2} - {y^2 \over 2 \rho_y^2}
           \right] \, ,
      \label{5.12} \\
      && \qquad \rho_x = R \sqrt{1-\epsilon_s}\, ,
      \qquad
      \rho_y = R \sqrt{1+\epsilon_s}\, .
      \label{5.13} 
    \end{eqnarray}
    An elliptic deformation of the transverse
    flow pattern can be introduced by
    \begin{eqnarray}
      u_\mu(x) &=& (\gamma_\perp\cosh\eta, u_x, u_y,
                         \gamma_\perp\sinh\eta )\, ,
                         \label{5.14} \\
      u_x &=& \eta_f \sqrt{1+\epsilon_f}\textstyle{x\over R}\, ,
      \qquad
      u_y = \eta_f \sqrt{1-\epsilon_f}\textstyle{y\over R}\, .
      \label{5.15}    \\               
      \gamma_\perp &=& \sqrt{1+u_x^2 + u_y^2}\, .
      \label{5.16}
    \end{eqnarray}
    These modifications implement in a simple way 
    some aspects of finite impact
    parameter collisions in the mid-rapidity
    region. They do not encode for the fact that in the
    fragmentation region the particle emission is peaked away from 
    the beam axis. Hence, the total angular momentum $\vec{L}$ 
    of the system,
    \begin{equation}
      L_i = \epsilon_{ijk} \langle\!\langle x_j\, p_k\, \rangle\!\rangle
          = \epsilon_{ijk} \int {d^3p\over E} \int d^4x\, x_j\, p_k\,
           S(x,p)\, ,
       \label{5.17}
    \end{equation}
    vanishes for the prescriptions (\ref{5.12})-(\ref{5.16})
    given above. A simple parametri\-za\-tion
    of the emission function in the fragmentation region which
    leads at least to a finite expression (\ref{5.17}), 
    reads~\cite{W97} 
    \begin{eqnarray}
          u_x^\chi &=& \eta_f \sqrt{1+\epsilon_f} 
        { {x+\chi {\rm y}}\over R}\, .
      \label{5.18} \\
      \rho_x &=& \left(R+\chi {\rm y}\cos\varphi\right)\, 
      \sqrt{1-\epsilon_s}\, ,
      \nonumber \\
      \rho_y &=& \left(R+\chi {\rm y}\cos\varphi\right)\, 
        \sqrt{1+\epsilon_s}\, .
      \label{5.19}
    \end{eqnarray}
    These dynamical (\ref{5.18}) and geometrical (\ref{5.19})
    assumptions effectively shift the center of the particle
    emission in the transverse plane as a function of the longitudinal
    particle rapidity ${\rm y}$ with some asymmetry strength $\chi$.
    They break explicitly the 180$^\circ$ rotation symmetry of the emission
    function $S_{\rm lab}$ in the transverse plane, which is left unbroken
    by (\ref{5.12})-(\ref{5.16}). 

\subsection{Resonance decay contributions}
\label{sec5a3}

A significant fraction of the most abundant candidates for
interferometric studies, charged pions, are produced by the
decay of unstable resonances after freeze-out. To a lesser
extent the same problem exists also for kaons. Longlived
resonances can escape to quite some distance from the original
freeze-out region before decaying. They then lead to 
HBT radius parameters which are larger than the width of the
particle production region. Furthermore, due to the resonance decay 
phase-space, secondary pions populate mainly the low momentum
region and can thus introduce an additional pair momentum 
dependence of the two-pion correlator. To obtain realistic
estimates for the geometry and dynamics of the particle emitting
source, resonance decay contributions therefore must be
analyzed quantitatively.

In the following discussion we focus on charged pions. We 
include all relevant resonance decay channels $r$ in the model 
emission function by writing~\cite{G77,Bolz93,S96,H96,WH96a}
  \begin{equation}
     S_{\pi}(x,p) = S_{\pi}^{\rm dir}(x,p) + 
     \sum_{r\ne\pi} S_{r\to \pi}(x,p)\, .
  \label{5.20}
  \end{equation}
The emission functions $S_{r\to \pi}(x,p)$ for the decay 
pions are calculated from the direct emission functions 
$S_r^{\rm dir}(X,P)$ for the resonances by taking into account 
the correct decay kinematics for two- and three-body decays.
Capital letters denote variables associated with the parent 
resonance, while lowercase letters denote pion variables.
In particular, $M_\perp$ and $\Phi$ here denote the transverse mass 
and azimuthal direction of
the parent resonance, in contrast to the rest of the review
where $M_\perp$ and $\Phi$ are associated with
the pair momentum $\bbox{K}_\perp$.

We follow the treatment in \cite{H63,SKH91,Bolz93,WH96a}.
The resonance $r$ is emitted with momentum $P$ at 
space-time point $X^\mu$ and decays after a proper time $\tau$ at 
$x^\mu = X^\mu + {P^\mu\over M} \tau$ into a pion of momentum $p$ and 
$(n-1)$ other decay products: 
  \begin{equation}
        r \longrightarrow \pi + c_2 + c_3 + ... + c_n \, .
  \label{5.21}
  \end{equation}
The decay rate at proper time $\tau$ is 
$\Gamma e^{-\Gamma\tau}$ where $\Gamma$ is the total decay width of 
$r$. Assuming unpolarized resonances with isotropic decay in their rest 
frame, $S_{r\to\pi}(x,p)$ is given in terms of the direct emission function 
$S_r^{\rm dir}(X,P)$ for the resonance $r$ by
 \begin{eqnarray}
   S_{r\to\pi}(x;p) &=& 
        M\, \int_{s_-}^{s_+} ds\, g(s)
        \int{d^3P \over E_{_P}}\, 
        {\delta}{\left({p\cdot P - E^* M}\right)}\,  
        \int d\tau \, \Gamma e^{-\Gamma\tau} 
   \nonumber \\
   &&  \times\, \int d^4X 
    \delta^{(4)}\left[ x - \left( X + {P\over M} \tau \right) \right]
        S_r^{\rm dir}(X,P)\, .
 \label{5.22}
 \end{eqnarray}
Variables with a star denote their values in the 
resonance rest frame, all other variables are given in the fixed 
measurement frame. 
$s = \left(\sum_{i=2}^n p_i \right)^2$ is the squared invariant 
mass of the $(n-1)$ unobserved decay products in (\ref{5.21});
it can vary between $s_- = \left( \sum_{i=2}^n m_i \right)^2$ and $s_+ 
= (M-m)^2$. $g(s)$ is the decay phase-space for the $(n-1)$ unobserved 
particles. For two-particle decays it reads
  \begin{equation}
    g(s) = {b\over 4\pi p^*} \delta \left( s - m_2^2 \right)\, ,
    \label{5.23}
  \end{equation} 
whereas the three-particle decay phase-space is given by
  \begin{eqnarray}
        g(s) &=& {M b\over 2\pi s} 
        {\sqrt{[s - (m_2 + m_3)^2][s - (m_2 - m_3)^2]}
        \over Q(M,m,m_2,m_3)}\, ,
        \label{5.24} \\
    Q(M,m,m_2,m_3) &=& \int_{s_-}^{s_+} {ds'\over s'}
        \sqrt{(M+m)^2 - s'}
        \nonumber \\
        && \times
        \sqrt{s_+ - s'}\sqrt{s_- - s'}
        \sqrt{(m_2-m_3)^2 - s'}\, .
        \label{5.25}
  \end{eqnarray}
Equation (\ref{5.22}) can be simplified considerably: for $p_\perp
\not= 0$, the energy-momentum conserving $\delta$-function constrains
the angle $\Phi$ of the resonance momentum $P_\mu = \left(M_\perp\cosh
  Y, P_\perp\cos\Phi, P_\perp\sin\Phi, M_\perp\sinh Y\right)$ to one
of two orientations if its decay product propagates in the
out-direction: 
  \begin{eqnarray}
    \delta(p \cdot P-E^*M) &=& \sum_{\pm} 
    {\delta(\Phi - \Phi_{\pm}) \over p_\perp P_\perp \sin\Phi_{\pm}}\, ,
    \label{5.26} \\
    \cos \tilde \Phi_\pm &=& { m_\perp M_\perp \cosh(Y-y) - E^* M
                       \over p_\perp P_\perp}\, .
  \label{5.27}
  \end{eqnarray}
This allows to do the $\Phi$-integration in (\ref{5.22}), leading to
  \begin{equation}
  \label{5.28}
        S_{r\to\pi}(x,p) = \sum_\pm \int_{\bf R} 
        \int_0^{\infty}{d\tau}\, \Gamma e^{-\Gamma\tau} 
        S_r^{\rm dir} \left( x -{P^\pm\over M} \tau,P^\pm \right) \, .
 \end{equation}
$\sum_\pm$ sums over the two azimuthal directions in (\ref{5.27}), 
and $\int_{\bf R}$ indicates the remaining integrations over the
resonance momenta; for details see Ref.~\cite{WH96a}.

\section{One-particle spectra}\label{sec5b}

The one-particle momentum spectrum, determined as the space-time
integral (\ref{1.3}) of the emission function $S(x,p)$, is sensitive
to the momentum distribution in $S(x,p)$ and thus allows to constrain
essential parts of the collision dynamics. It contains, however, no 
information about the space-time structure of the source. Statistical
errors on one-particle data are significantly smaller
than those on the two-particle spectra. In practice, exploiting the
temperature, flow and resonance mass dependence of the one-particle 
spectrum therefore allows to reduce the model parameter space 
significantly even before 
comparing model predictions to the measured two-particle correlations.

\subsection{Transverse one-particle spectrum}
\label{sec5b1}

We begin by discussing the rapidity-integrated transverse
momentum spectrum of the models discussed in the last subsection.
Both direct ``thermal'' pions produced in the collision region 
and those stemming from resonance decays contribute to the 
spectrum, see (\ref{5.28}):
  \begin{equation}
  \label{5.29}
    {dN_\pi\over dm_\perp^2} = {dN_\pi^{\rm dir}\over dm_\perp^2}
     + \sum_{r\ne\pi} 2 M_r \int_{\bf R} {dN_r^{\rm dir} \over dM_\perp^2}\, .
  \end{equation}
For the model (\ref{5.1}) this expression
takes a compact form~\cite{SKH91,SSH93,WH96a}: 
  \begin{eqnarray}
  \label{5.30}
    {dN_r^{\rm dir}\over dM_\perp^2} 
    &=& {2J_r+1 \over 4\pi^2} \, 
        (2 \pi R^2 \cdot 2 \tau_0 \Delta\eta)\,
        e^{\mu_r/T}\, M_\perp\,
        \int_0^\infty d\left({\xi^2\over 2}\right) e^{-\xi^2/2}
  \nonumber\\     
    &&\qquad \times \,  
        K_1\left( {\textstyle{M_\perp\over T}}\cosh\eta_t(\xi) \right)
    I_0\left( {\textstyle{P_\perp\over T}}\sinh\eta_t(\xi) \right) \, .
  \end{eqnarray}
$\xi = r/R$ is the rescaled transverse radius. Obviously the geometric
parameters $R$, $\Delta\eta$, $\tau_0$ of the source enter 
only in the normalization of the spectrum. The product of the
spatial and temporal extensions of the
thermal source determines the total particle yield, but not its
momentum dependence. To further constrain these parameters is not
possible without using two-particle correlations.
According to (\ref{5.29}/\ref{5.30}), the $m_\perp$-dependence 
of the pion spectrum is fully determined by the temperature $T$
(or $T(\xi)$ if $T$ is $r$-dependent), the rest masses $M_r$
and chemical potentials $\mu_r$ of the resonances, and 
the transverse flow profile $\eta_t(\xi)=\eta_f \xi^n$.

To study the dependence of one- and two-particle spectra on the
composition of the resonance gas, the resonance fractions $f_r$
can be computed~\cite{Bolz93,WH96a}:
  \begin{eqnarray}
    f_r(p) &=& {\int d^4x\, S_{r\to\pi}(x,p) \over
                     \sum_r \int d^4x\, S_{r\to\pi}(x,p)} 
                   = {dN_\pi^r/d^3p \over dN_\pi^{\rm tot}/d^3p}\, ;
    \nonumber \\
    && \sum_r f_r(p) = 1\, .
  \label{5.31}
  \end{eqnarray}
As we discuss in section~\ref{sec5c5},
these fractions play an important role in estimating the correlation
strength $\lambda(\bbox{K})$.
%
\begin{figure}[ht]\epsfxsize=13.5cm 
\centerline{\epsfbox{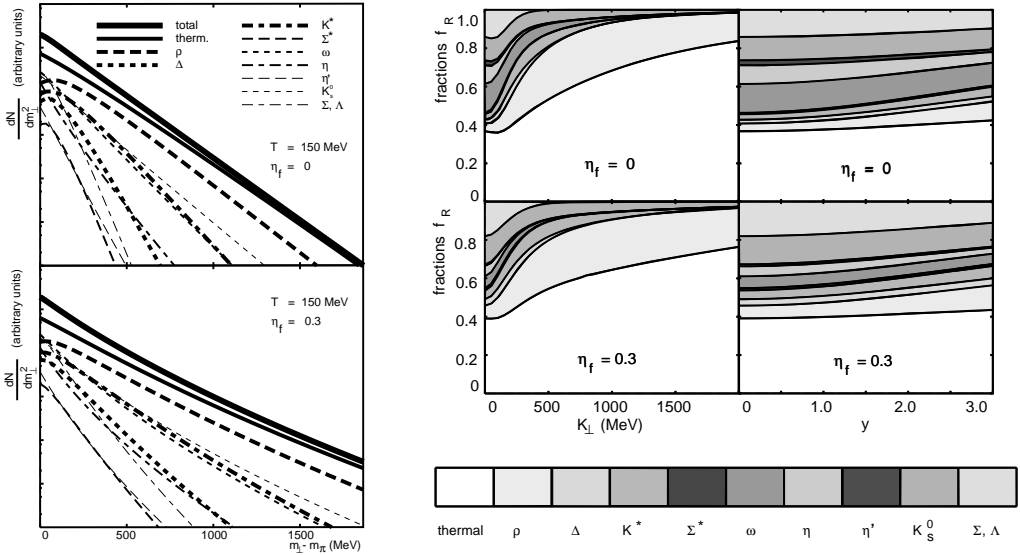}}
\vspace{0.5cm}
\caption{
Left: The single-pion transverse mass spectrum for the model 
(\protect\ref{5.1}) at $T=150$ MeV, transverse flow $\eta_f = 0$
(upper panel) or $\eta_f = 0.3$ (lower panel), and vanishing
chemical potentials $\mu_r=0$. The overall normalization is arbitrary, the
relative normalizations of the various resonance contributions are fixed
by the assumption of thermal and chemical equilibrium. 
Right: The resonance fractions $f_r({\rm y},p_\perp)$ according to 
Eq.~(\protect\ref{5.31}) for the same parameters. Left column: $f_r$ as
a function of transverse momentum at central rapidity. Right column: 
$f_r$ as function of rapidity at $p_\perp=0$.
}\label{fig2}
\end{figure}
%
In Fig.~\ref{fig2}, we plot the pion transverse mass spectrum 
$dN_\pi/ dm_\perp^2$ and the resonance fractions $f_r({\rm y},p_\perp)$
of the model (\ref{5.1}) for two sets of source parameters. 
All resonance decay contributions 
are shown separately. The resonances  $\omega$, $\eta$ and $\eta'$
contribute with 3-body decays whose decay pions are seen to be 
particularly concentrated at small $p_\perp$. Comparing the cases
of vanishing and non-vanishing transverse flow
one observes the well-known flattening of the transverse mass spectrum
by transverse radial flow \cite{SR79,N82,LHS90,SSH93,SH92}. The direct
pions reflect essentially an effective ``blueshifted'' 
temperature~\cite{LHS90}
  \begin{equation}
    T_{\rm eff} = T \sqrt{ {1 + \langle \beta_t \rangle \over 
        1-\langle\beta_t\rangle}}\, ,
    \label{5.32}
  \end{equation}
where the average transverse flow velocity $\langle \beta_t \rangle$
is directly related to $\eta_f$. This clearly does not allow to
separate thermal from collective motion. Deviations from (\ref{5.32})
are seen for the transverse mass spectra of heavier particles at
low $p_\perp$ and have been used to determine the temperature and
flow velocity separately~\cite{QM97,SQM98}.
One of the main goals of two-particle interferometry is to obtain
a more direct measure of the transverse expansion velocity at 
freeze-out.
%
\subsection{Rapidity distribution}
\label{sec5b2}

The rapidity distribution is given by 
  \begin{equation}
    {dN_\pi\over d{\rm y}} = \int m_\perp dm_\perp\, d\phi
                       \int d^4x\, S(x;p_\perp,\phi,{\rm y})\, .
    \label{5.33}
  \end{equation}
For the model (\ref{5.1}) its width is dominated by the
longitudinal width $\Delta\eta$ of the source. This is
a consequence of the assumed boost-invariant longitudinal
flow profile. Since the resonance decay fractions $f_r$ are 
essentially rapidity independent (see Fig.~\ref{fig2}), resonance 
decay contributions do not significantly affect the rapidity spectrum.

\subsection{Azimuthal dependence}
\label{sec5b3}

For collisions with non-zero impact parameter the triple-differential
one-particle spectrum (\ref{3.24}) contains information about
the orientation of the reaction plane.
The harmonic coefficients $v_n$ which characterize this
azimuthal dependence are given in terms of the 
Fourier transforms~\cite{VZ96,V97}
  \begin{eqnarray}
    \left( a_n,\, b_n  \right)
     &=& { {\int_0^{2\pi} E\, {dN\over d^3p} 
    \left(\cos(n\phi),\, \sin(n\phi)\right)\, d\phi} \over
    {\int_0^{2\pi} E\, {dN\over d^3p}\, d\phi} }\, ,
    \label{5.34} \\
    a_n &=& v_n\, \cos(n\psi_R)\, ,\qquad
    b_n = v_n\, \sin(n\psi_R)\, .
    \label{5.35}
  \end{eqnarray}
According to (\ref{5.34})
they are normalized to the azimuthally averaged double differential
particle distribution, and $v_0 = 1$. A symmetry argument similar
to that employed in (\ref{3.20}/\ref{3.21}) implies that
in the limit $p_\perp \to 0$ the $\phi$-dependent terms  
vanish, the emission probabilities in different azimuthal 
directions become equal, and 
  \begin{equation}
    \lim_{p_\perp\to 0} v_n(p_\perp) = 0\, \qquad
    \hbox{for all $n\geq 1$.}
    \label{5.36}
  \end{equation}
Furthermore, the odd harmonic coefficients $v_n$ vanish at midrapidity
for symmetric collision systems, due to the remaining
$\phi \to \phi + \pi$ symmetry in the transverse plane. 
A more explicit calculation depends on details of the model.
For the emission function (\ref{5.12})-(\ref{5.16}) one can show 
that $v_2 \propto p_\perp^2$ for small values of $p_\perp$.
For small transverse flow, the leading dependence on the
anisotropy parameters $\epsilon_s$ and $\epsilon_f$ is given by
  \begin{equation}
    v_2 \propto \eta_f^2\, { {2(\epsilon_f - \epsilon_s)}\over 
                   (1-\epsilon_f^2)\, (1-\epsilon_s^2) }\, .
    \label{5.37}
  \end{equation}
This describes correctly the main features of a numerical study 
of this model \cite{W97}. Both geometric and dynamical deformations
manifest themselves in the single-particle spectrum only for 
expanding sources with $\eta_f \ne 0$. 
The relative minus sign in the numerator
of (\ref{5.37}) reflects the different signs in the definitions
of $\epsilon_s$ and $\epsilon_f$ in (\ref{5.13}) and (\ref{5.15}).
Once this is taken into account, (\ref{5.37}) shows that an
increasing spatial deformation $\epsilon_s$ or an increasing 
flow anisotropy $\epsilon_f$ lead to similar effects on the 
azimuthal particle distribution. Therefore they cannot be separated 
without also using information from two-particle correlations.

\section{Two-particle correlator}\label{sec5c}

In this section we study the question how characteristics
of the emission function are reflected in the momentum-dependence
of the measured particle correlations. This was discussed already
in section ~\ref{sec3} in the context of the model-independent relations
between the space-time variances of $S(x,K)$ and the HBT radius
parameters of $C(\bbox{q},\bbox{K})$. According to this discussion
spatial and temporal geometric information about the source is contained
in the $\bbox{q}$-dependence of the correlator, while the pair
momentum dependence characterizes dynamical properties. These
statements can be made more explicit in the
context of specific model studies.

For quantitatively reliable studies of ``realistic'' (i.e. sufficiently
complex) emission functions the determination of the two-particle
correlator requires a numerical evaluation of the Fourier integral
in (\ref{1.4}). On the other hand, simple analytical approximations
for the HBT radius parameters allow to summarize the main physical
dependencies of the measurable quantities in an intuitive form
and are quite useful for a qualitative understanding. We give a
combined discussion of both the analytical approximations and
the exact numerical results for the model (\ref{5.1}).

\subsection{Saddle point approximation of HBT radius parameters}
\label{sec5c1}

Characterizing the emission function $S(x,K)$ by the Gaussian
widths (\ref{3.5}), $\left(B^{-1}\right)_{\mu\nu}(\bbox{K})$ 
$ = \langle \tilde{x}_{\mu}\tilde{x}_\nu\rangle(\bbox{K})$,
is more generally applicable than a saddle point approximation
around its center $\bar{x}(\bbox{K})$ which was earlier 
suggested~\cite{S95,CSH95a}. The tensor $B_{\mu\nu}(\bbox{K})$
characterizes essential features of the phase-space support 
of the emission function even if $S(x,K)$ is not differentiable 
or if its curvature at the saddle point does not represent
its average support sufficiently well. In all these cases the 
Gaussian widths $B_{\mu\nu}(\bbox{K})$ still translate directly 
into Gaussian radius parameters, as discussed in section~\ref{sec3}.
Nevertheless, a saddle point approximation of $S(x,K)$ can be 
technically useful for the evaluation of the integrals (\ref{3.3}) 
when approximating the HBT radius parameters as averages over the
emission function~\cite{S95,AS96,CSH95a,CL96,WSH96}. 

To illustrate the use and limitations of the analytical expressions 
thus obtained one can study a simplification of the model (\ref{5.1}) 
which is particularly amenable to analytical calculations: we neglect 
the resonance contributions and set $\Delta\eta = \infty$ and 
$\Delta\tau = 0$. This represents a source with exact longitudinal 
boost-invariance and a sharp freeze-out at time $\tau_0$
\cite{CSH95a,S95,WSH96}. For such a source the cross-term $R_{ol}$ 
vanishes in the LCMS (see (\ref{3.17})); it will be discussed in the 
following subsection. 

Expanding the exponent of the emission function $S(x,K)$ given 
in (\ref{5.1}) around $x_\mu = 0$ one obtains in this case
  \begin{equation}
    - {M_\perp\over T}
    \left( 1+ {\eta_f^2\over 2}{{x^2+y^2}\over R^2}\right)
    - {K_\perp\over T}\eta_f {x\over R} 
    - {{x^2+y^2}\over R^2}\, .
    \label{5.38}
  \end{equation}
In the saddle point approximation the terms 
bilinear in $y$ specify the (inverse of the) ``side'' radius parameter 
$R_s$, while the ``outward'' radius parameter receives an additional
contribution from the finite emission duration 
$\langle \tilde{t}^2\rangle = \tau_0^2 \langle \sinh^2\eta\rangle$
according to~\cite{CSH95a}
 \begin{eqnarray}  
    R_o^2(K_\perp) &=& R_s^2(K_\perp)
    + \textstyle{1\over 2} {\left({T\over M_{\perp}}\right)}^2 
    {\beta}_{\perp}^2 {\tau}_0^2 \, ,
    \label{5.39} \\
    R_s^2(K_\perp) &=& {R^2 \over 1 + {M_\perp \over T} \eta_f^2} \, ,
    \label{5.40} 
 \end{eqnarray}
These simple expressions illustrate several of the key concepts
employed in HBT interferometry: the overall size of 
the transverse radius parameters is determined by the transverse
Gaussian widths of the collision region, and the difference
$R_o^2 - R_s^2$ is proportional to the emission duration
$\beta_\perp^2 \langle \tilde{t}^2\rangle$. (Even a sharp
freeze-out at $\tau = \tau_0$ corresponds to a finite region
in $t = \tau_0 \cosh\eta$ since the source distribution is sampled
over a finite longitudinal range.) Most importantly, however, the radius
parameters $R_o^2$ and $R_s^2$ are sensitive to the transverse
flow strength $\eta_f$ of the source: the HBT radius shrinks for 
finite $\eta_f$ since a dynamically expanding
source viewed through a filter of wavelength $K$ is seen only
partially. This shrinking effect increases for larger values
of $M_\perp$ proportionally to the ratio $\eta_f^2/T$.
The $M_\perp$-dependence of $R_s$ is a consequence of transverse
position-momentum correlations in the source which here
originate from the transverse collective flow.

Saddle point integration also leads to simple expressions for
the longitudinal radius parameters. Due to boost-invariance,
$R_l$ is $\beta_l$-independent, and one has to 
calculate $R_l^2 = \langle \tilde{z}^2 \rangle$
$= \tau_0^2 \left( \langle\sinh\eta^2\rangle 
- \langle\sinh\eta\rangle^2\right)$. We summarize the
results for different approximation schemes:
  \begin{eqnarray}
    R_l^2 &\approx& {\tau}_0^2 {T\over M_{\perp}}\, , 
    \qquad \qquad\qquad \qquad \qquad \qquad 
    \left[\hbox{to}\, \, O\left(\textstyle{T\over M_\perp}\right)\, ,
      \cite{MS88}\right]
    \label{5.41} \\  
    R_l^2 &\approx& \tau_0^2 {T\over M_\perp} \,
    \left( 1 + 
    \left( {1 \over 2} + {1\over {1 + {M_\perp \over T} \eta_f^2}} \right)
          {T\over M_\perp} \right) \, ,
      \left[\hbox{to}\, \, O\left(\textstyle{T^2\over M^2_\perp}\right)\, ,
      \cite{CSH95a}\right] 
        \label{5.42} \\
   R_l^2 &=& \tau_0^2 {T\over M_\perp} \, 
   { K_2({M_\perp \over T}) \over K_1({M_\perp \over T}) }\, .
   \qquad\qquad \qquad
   \hbox{[exact for $\eta_f = 0$, \cite{HB95}]}
        \label{5.43} 
 \end{eqnarray}
In general the longitudinal radius is proportional to the 
average freeze-out time $\tau_0$ and falls off strongly with 
increasing transverse momentum. In this case the transverse 
momentum dependence signals {\it longitudinal} position-momentum
correlations in the source. Compared to the transverse directions,
it is much stronger since the source expands predominantly 
in the beam direction.

According to the Makhlin-Sinyukov formula (\ref{5.41}), fitting
an $1/ M_\perp$-hyper\-bola to $R_l^2$ (extracting the 
temperature e.g. from the one-particle spectrum), the source 
parameter $\tau_0$ can be determined. Early estimates of $\tau_0$
have been obtained by this argument. The improved calculations
(\ref{5.42}) and (\ref{5.43}) show that for realistic
temperatures of the order of the pion mass corrections cannot be
neglected. Similar remarks apply to the slopes of the
``side'' and ``out'' radius parameters (\ref{5.39}/\ref{5.40}). 
Especially, all the results presented here are based on an
expansion around $x_\mu = 0$, while the true saddle point 
$\bar{x}_{\mu}$ for finite $\eta_f$ is shifted in the out-direction;
this can be seen from (\ref{5.38}). In Ref.~\cite{WSH96} an 
approximation scheme was developed which takes these saddle point 
shifts into account and allows to systematically derive improved 
expressions for the HBT radii (\ref{5.39})-(\ref{5.42}).
The resulting expressions are involved and a numerical evaluation
of the radius parameters is often more convenient.
%
\begin{figure}[ht]\epsfxsize=11cm 
\centerline{\epsfbox{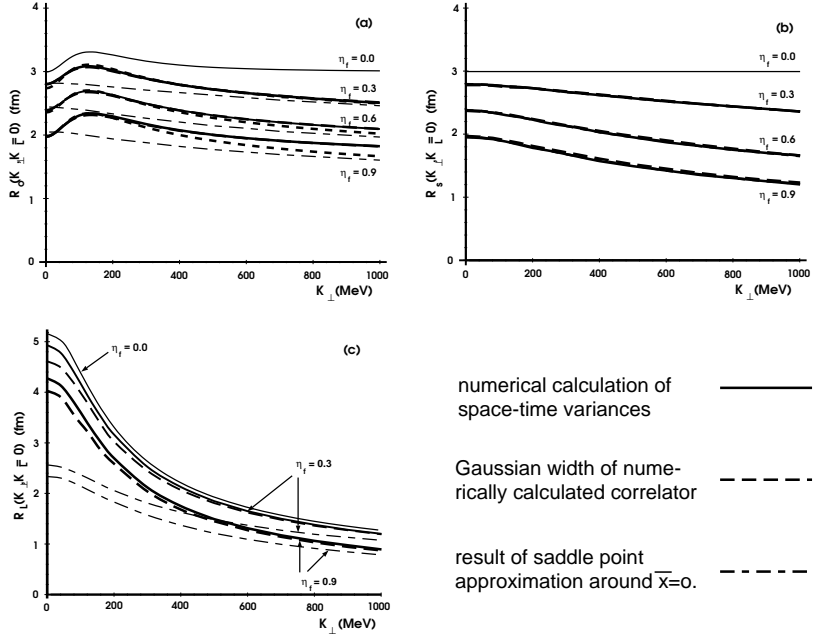}}
\caption{
$K_\perp$-dependence of the HBT radius parameters $R_o$ (a), 
$R_s$ (b) and $R_l$ (c) for the emission function 
(\protect\ref{5.1}) with $T=150$ MeV, $\tau_0 = 3$ fm/c, $R = 3$ fm,
$\Delta\eta = \infty$, $\Delta\tau = 0$. Curves for different values
of the transverse flow $\eta_f$ are shown. Differences between
numerically evaluated space-time variances and Gaussian widths
indicate deviations of $C(\bbox{q},\bbox{K})$ from a Gaussian shape.
Differences to the dash-dotted lines reflect the limited validity of
a naive saddle point approximation around $x_\mu = 0$.
}\label{fig3}
\end{figure}

The sources for quantitative uncertainties of the saddle point
approximation are two-fold: The determination of the saddle point 
$\bar{x}_\mu = \langle x_\mu \rangle$ is done only approximately in 
many model studies. The resulting inaccuracies turn out to be 
substantial and put severe limitations on the quantitative
applicability of the analytical expressions quoted above. This is
illustrated in Fig.~\ref{fig3}. Moreover, whenever 
the Gaussian widths of the correlator deviate significantly
from its curvature at $\bbox{q} = 0$, the 
relation between space-time variances and HBT-radii becomes
quantitatively unreliable, see chapter~\ref{sec4}. In the following
we therefore use the above simple analytical expressions only for
qualitative guidance, basing a quantitative discussion on
numerical results.

Equations (\ref{5.39})-(\ref{5.43}) have been used to extract
the transverse flow parameter $\eta_f$ and the ``freeze-out time''
$\tau_0$. This is not without danger. As pointed out in 
Ref.~\cite{CSH95a}, in this particular model the $M_\perp$-dependence 
of the HBT radius parameters reflects the
longitudinal and transverse flow velocity gradients. These are
responsible for the reduced homogeneity lengths compared to the
total source size. The parameter $\eta_f^2$ in (\ref{5.40}) 
really reflects the transverse flow velocity gradient~\cite{CSH95a},
and the parameter $\tau_0^2$ in (\ref{5.41})-(\ref{5.43}) 
similarly arises from the longitudinal velocity gradient at
freeze-out. From the general considerations of section~\ref{sec3a}
we know that the absolute position of the source in space and time
cannot be measured. Therefore, strictly speaking, $\tau_0$ cannot 
be directly associated with the absolute freeze-out time. Such
an interpretation of $\tau_0$ makes the additional dynamical
assumption that the longitudinally boost-invariant
velocity profile existed not only at the point of freeze-out
but throughout the dynamical evolution of the reaction zone.
If this were not the case and the collision region underwent,
for example, longitudinal acceleration before freeze-out, the
real time interval between impact and freeze-out would be
longer.

\subsection{The out-longitudinal cross-term}
\label{sec5c1b}
%
The cross-term $R_{ol}^2$ of the Cartesian parametrization vanishes
in the LCMS for longitudinally boost-invariant systems or in symmetric 
collisions at mid-rapidity. This follows from the corresponding 
space-time variance (\ref{3.13}) which vanishes in the LCMS 
if the source is symmetric under $\tilde{z} \to -\tilde{z}$. This
reflection symmetry is broken, however, in the forward and backward
rapidity regions for the systems with finite longitudinal extension.
%
\begin{figure}[ht]\epsfxsize=11cm 
\centerline{\epsfbox{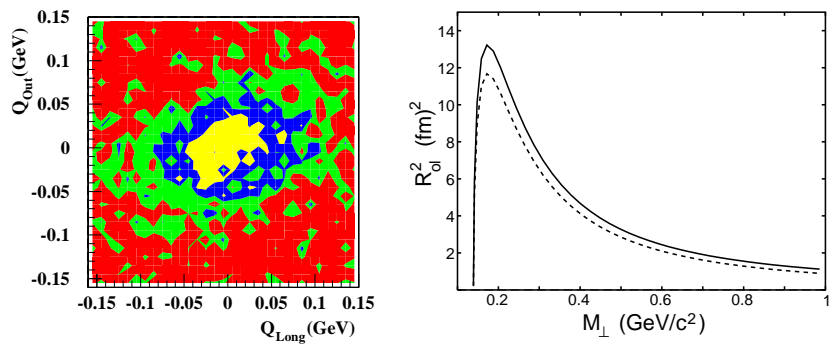}}
\caption{Left: Contour plot of the correlation function in the 
$q_o$-$q_l$-plane obtained by NA35 in S+Au collisions at 200 A GeV. 
The data are for forward rapidity pairs, $3.5 < Y < 4.5$, the momentum
differences are evaluated in a system at rapidity 3. (Figure taken
from~\protect\cite{alber}.) Right: $M_\perp$-dependence of the
out-longitudinal radius parameter for the model \protect(\ref{5.1}) at
$Y_{\rm cm} = 1.5$ in the LCMS. The curves are shown for
different values of the transverse flow, $\eta_f = 0.4$ (solid)
and $\eta_f = 1.0$ (dashed). (Figure taken from~\protect\cite{T99}.)
}\label{fig3b}
\end{figure}
%

As first observed by NA35~\cite{alber}, following a proposal of 
Chapman et al.~\cite{CSH95b,CSH95c}, the Gaussian correlator is then an 
ellipsoid in $\bbox{q}$-space whose main axes do not coincide with
the Cartesian ones: it is tilted in the out-longitudinal plane
(see the l.h.s. of Fig.~\ref{fig3b}). This tilt is parametrized
by the size and sign of $R_{ol}^2$. In the forward rapidity
region, $R_{ol}^2$ is positive in the LCMS and negative in the
CMS~\cite{CSH95c,appels,schoen}. The sign is reversed for the backward
rapidity region.

Also, $R_{ol}^2$ has a characteristic transverse momentum dependence.
At vanishing $K_\perp$, the $\tilde{x} \to -\tilde{x}$ symmetry of 
the source is restored and $R_{ol}^2$ vanishes, see (\ref{3.21}) and 
the l.h.s. of Fig.~\ref{fig3b}. At large $K_\perp$, on the other hand, 
the homogeneity region peaks sharply around the point of highest 
emissivity, and the $\tilde{z} \to -\tilde{z}$ symmetry is again 
approximately restored \cite{WHTW96}. The resulting generic $M_\perp$ 
dependence is seen in Fig.~\ref{fig3b}: $\vert R_{ol}^2\vert$ rises 
sharply at small $K_\perp$, reaches a maximum and then decreases again.

\subsection{The Yano-Koonin velocity}
\label{sec5c2}

The Yano-Koonin-Podgoretski\u\i\ parametrization
\cite{YK78,P83,HTWW96,WHTW96} describes the correlator of an
azimuthally symmetric source by a longitudinal velocity $v(\bbox{K})$
and three Gaussian radius parameters $R_\perp^2(\bbox{K})$,
$R_\parallel^2(\bbox{K})$, and $R_0^2(\bbox{K})$. The Yano-Koonin
velocity $v(\bbox{K})$ contains important information about the
longitudinal expansion of the source. 
To investigate this, a detailed study~\cite{WHTW96} was done within
the framework of the class of models (\ref{5.1}), of 
the relation among the following different longitudinal reference 
frames:
\begin{itemize}
\item
  {\bf CMS}: The centre of mass frame of the fireball, specified by
        $\eta_0 = 0$.
\item
  {\bf LCMS} (Longitudinally CoMoving System \cite{CP91}): a
        pair-dependent fra\-me, specified by $\beta_l = Y =0$.
        In this frame, only the transverse velocity component of the
        particle pair is non-vanishing.
\item
  {\bf LSPS} (Longitudinal Saddle-Point System \cite{CL96}):
        The longitudinally moving rest frame of the point of
        maximal emissivity for a given pair momentum. In general,
        the velocity of this frame depends on the momentum
        of the emitted particle pair. For symmetric sources
        the point of maximal emissivity (``saddle point'')
        coincides with the ``source centre'' $\bar x(\bbox{K})$
        defined in (\ref{3.2}). In this approximation, for a
        source like (\ref{5.1}), the LSPS velocity is given
        by the longitudinal component of $u^\mu(\bar x(\bbox{K}))$.
\item
  {\bf YK} (Yano-Koonin frame \cite{HTWW96}): The frame for which the
        YKP velocity parameter vanishes, $v(\bbox{K})=0$. Again, this
        frame is in general pair momentum dependent.
\end{itemize}
The velocities (or rapidities) of the CMS and LCMS frames can be determined
experimentally, the first from the peak in the single particle rapidity
distribution, the second from the longitudinal momentum of the measured
pion pair. The velocity of the LSPS is determined by the flow velocity
at $\bar x(\bbox{K})$ and hence, it is not directly measurable (neither
the one- nor the two-particle spectra depend on $\bar x(\bbox{K})$).
However, the YK- and LSPS-systems coincide as long as the particle
emission is symmetric around $\bar x(\bbox{K})$~\cite{P83,WHTW96}.
%
\begin{figure}[ht]\epsfxsize=9.5cm 
\centerline{\epsfbox{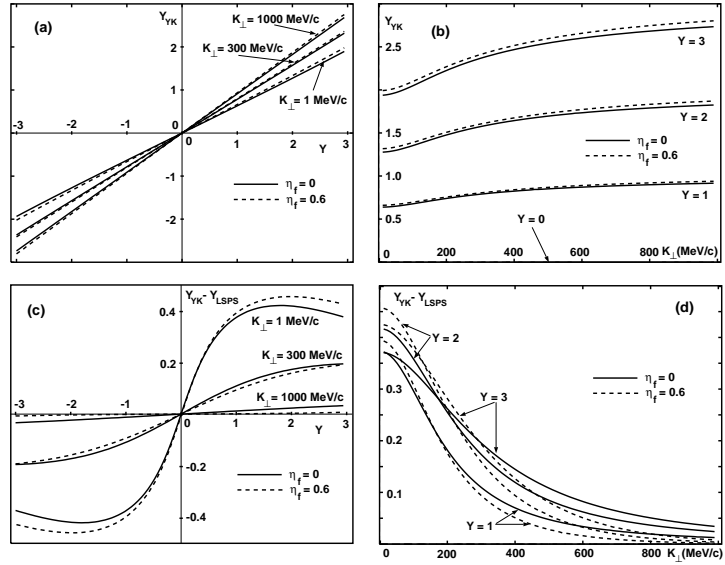}}
\caption{
 Calculation of the Yano-Koonin rapidity for the model 
 (\protect\ref{5.1}) without resonances. Source parameters:
 $R = 3$ fm, $\tau_0 = 3$ fm/c, $\Delta\tau = 1$ fm/c,
 $\Delta\eta = 1.2$ and $T=140$ MeV.
 (a) YK rapidity as a function of the pion 
     pair rapidity $Y$ (both measured in the CMS frame of the source), for 
     various values of the transverse pair momentum $K_\perp$ and 
     for two values of the transverse flow rapidity $\eta_f$.
 (b) Same as (a), but shown as a function of $K_\perp$ for different 
     values of $Y$. The curves for negative $Y$ are obtained by 
     reflection along the abscissa.  
 (c) The difference $Y_{_{\rm YK}} - Y_{_{\rm LSPS}}$ between the 
     rapidity of the YK frame and the longitudinal rest system of the 
     saddle point, plotted in the same way as (a). 
 (d) Same as (c), but shown as a function of $K_\perp$ for different 
     values of $Y$. 
}\label{fig4}
\end{figure}
%
In model studies based on the parametrization (\ref{5.1}) 
asymmetries are found to be small~\cite{WHTW96}.
This is seen in the difference between 
the corresponding rapidities, $Y_{_{\rm YK}} - Y_{_{\rm LSPS}}$, plotted
in Figure~\ref{fig4}c,d as a function of $K_\perp$ and $Y$ respectively.
The difference is generally small, especially for large transverse 
momenta~\cite{HTWW96,WHTW96} where thermal smearing can be neglected:
  \begin{equation}
    v(\bbox{K}) \approx v_{_{\rm LSPS}}(\bbox{K})\, .
    \label{5.44}
  \end{equation}
The observable YK velocity thus tracks the unobservable
LSPS velocity. This is important since longitudinal expansion 
of the source leads to a characteristic dependence of the LSPS 
velocity $v_{_{\rm LSPS}}$ on the pair rapidity which - using
(\ref{5.44}) -  one can confirm by measuring the YK-velocity. 
Two extreme examples illustrate this: for a static source 
without position-momentum correlations the rapidity of the 
LSPS is independent of the pair rapidity $Y$ and identical 
to the rapidity of the CMS:
  \begin{equation}
    Y_{_{\rm LSPS}} = \frac 12 \ln {1+v_{_{\rm LSPS}}\over
    1-v_{_{\rm LSPS}}} = {\rm const.}\
    \ \ \hbox{ for a static source.}
    \label{5.45}
  \end{equation}
In contrast, for a longitudinally boost-invariant model (\ref{5.1}) with 
$\Delta\eta = \infty$, the longitudinal saddle point lies at
$\eta = Y$. In this case, the LSPS and the LCMS coincide,
  \begin{equation}
    Y_{_{\rm LSPS}} = \eta = Y\,
    \qquad \hbox{ for a longitudinal boost-invariant source.}
    \label{5.46}
  \end{equation}
For the model (\ref{5.1}) the measurable YK rapidity $Y_{_{\rm YK}}$
is plotted in Figure~\ref{fig4}a,b as a function of the longitudinal
pair rapidity $Y$ and transverse momentum $K_\perp$. The plot
confirms a linear relation
  \begin{equation}
    Y_{_{\rm YK}} \approx {\rm const.}\, \times\, Y\, ,
    \qquad \hbox{ for the model~(\ref{5.1}).}
    \label{5.47}
  \end{equation}
with a proportionality constant which approaches unity for large
$K_\perp$. Since $Y_{_{\rm YK}} \approx Y_{\rm LSPS}$, this linear
relation between the rapidity $Y_{_{\rm YK}}$ of the Yano-Koonin
frame and the pion pair rapidity $Y$ is a direct reflection of 
the longitudinal expansion flow. The linear relation shown in 
Fig.~\ref{fig4}a has been confirmed 
subsequently by experiment~\cite{appels}. There are also first
experimental hints for the $K_\perp$-dependence shown 
in~\ref{fig4}a. 

We emphasize that the observation of a linear relation
$Y_{_{\rm YK}} = Y$ cannot be generally interpreted as
evidence for {\it boost-invariant} longitudinal expansion,
as was suggested in Refs.~\cite{HTWW96,WHTW96,He96}. In
Fig.~\ref{fig4}a one sees that for the model (\ref{5.1})
this linear relation is the better satisfied the larger
the transverse momentum of the particle pair. This can be
easily understood as a consequence of the reduced thermal
smearing at large $K_\perp$. Whenever the random component 
in the momentum distribution becomes small, the observed pair 
velocities track directly the flow velocity of the emitting volume 
element, irrespective of the actual flow velocity profile. The 
generic behaviour shown in Fig.~\ref{fig4}a therefore indicates 
longitudinal expansion which is sufficiently strong to overcome 
the thermal smearing, but not necessarily boost-invariant.

\subsection{Yano-Koonin-Podgoretski\u\i\ radius parameters}
\label{sec5c3}

We discussed in section~\ref{sec3b} that, as long as certain 
asymmetries of the source are negligible, the YKP-radius parameters
have a particularly simple space-time interpretation. 
The transverse radius parameter always coincides 
with the source width in the side-direction, 
$R_\perp^2(\bbox{K}) = \langle \tilde{y}^2\rangle$, 
and~\cite{HTWW96,WHTW96}
 \begin{eqnarray}   
   R_\parallel^2(\bbox{K}) &=& 
   \left\langle \left( \tilde z - {\beta_l\over\beta_\perp} \tilde x
                \right)^2 \right \rangle   
     - {\beta_l^2\over\beta_\perp^2} \langle \tilde y^2 \rangle 
     \approx \langle \tilde z^2 \rangle \, ,
     \label{5.48} \\
   R_0^2(\bbox{K}) &=& 
   \left\langle \left( \tilde t - {1\over\beta_\perp} \tilde x
                \right)^2 \right \rangle 
    - {1\over\beta_\perp^2} \langle \tilde y^2 \rangle 
    \approx \langle \tilde t^2 \rangle \, .
 \label{5.49}
 \end{eqnarray}
%
\begin{figure}[ht]\epsfxsize=9.5cm 
\centerline{\epsfbox{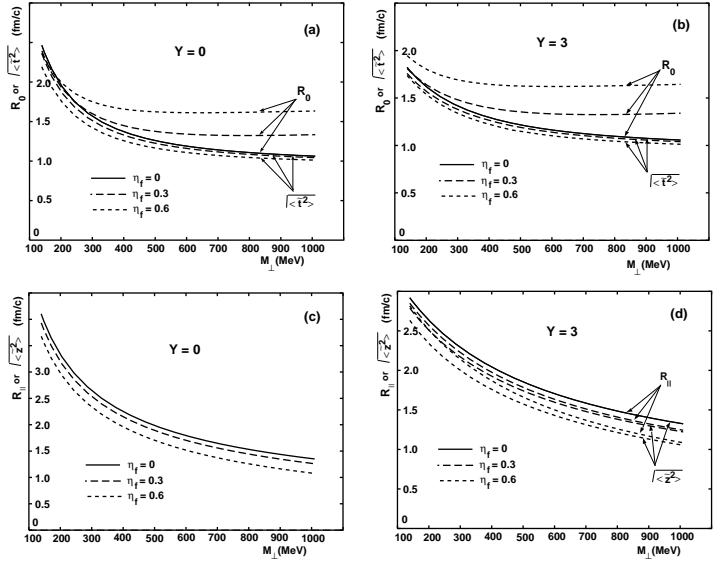}}
\caption{
 YKP radii for the same model as in Fig.~\ref{fig4}.
 (a) $R_0$ and $\protect \sqrt{\langle \tilde t^2 \rangle}$,
     evaluated in the YK frame, as a  
     function of $\protect M_\perp$ for three values of 
     the transverse flow rapidity $\eta_f$, for pion pairs with 
     CMS rapidity $Y=0$.  
 (b) Same as (a), but for pions with CMS rapidity $Y=3$.
 (c) and (d): Same as (a) and (b), but for $R_\parallel$ and the 
     longitudinal length of homogeneity $\protect \sqrt{\langle \tilde 
     z^2 \rangle}$, evaluated in the YK frame. For $Y=0$, $R_\parallel$ and 
     $\protect \sqrt{\langle \tilde z^2 \rangle}$ agree exactly 
     because $\beta_l=0$ in the YK frame.
}\label{fig5}
\end{figure}
%

The approximation in these equations results from dropping terms 
proportional to $\langle \tilde z \tilde x\rangle$, $\langle \tilde x
\tilde t \rangle$, and $\langle \tilde x^2 - \tilde y^2 \rangle$.
The first two of these terms vanish if the source is symmetric around
its point of highest emissivity $\bar{x}(\bbox{K})$. In Figure~\ref{fig5}
the effective emission duration $\sqrt{\langle\tilde{t}^2\rangle}$
and longitudinal size of homogeneity $\sqrt{\langle\tilde{z}^2\rangle}$
are compared to $R_0$ and $R_\parallel$ for the model (\ref{5.1}). 
The approximations (\ref{5.48}/\ref{5.49}) become exact for vanishing 
transverse flow while differences occur especially in $R_0$ for large 
$K_\perp$ or significant transverse flow. These can be traced to
the terms 
$-2\langle \tilde x \tilde t \rangle/\beta_\perp + \langle \tilde
x^2 - \tilde y^2 \rangle/\beta_\perp^2$ which are neglected in
(\ref{5.49}). For the model (\ref{5.1})
the space-time variances $\langle \tilde x \tilde t \rangle$
and $\langle \tilde{x}^2 - \tilde y^2 \rangle$ indeed vanish 
for $K_\perp \to 0$ where the azimuthal $x$-$y$-symmetry of
the source is restored. However, when divided by powers of $\beta_\perp$,
the corresponding terms in (\ref{5.49}) result in small
but finite contributions even for $K_\perp \to 0$. 
For opaque sources (\ref{5.9}) the term
$-2\langle \tilde x \tilde t \rangle/\beta_\perp + \langle \tilde
x^2 - \tilde y^2 \rangle/\beta_\perp^2$ 
can be the dominant contribution; this can
lead to large negative values of $R_0^2(K_\perp)$. 
The approximation (\ref{5.49}) thus breaks down for such opaque
sources~\cite{HV96a,TH97,TH98}, and the leading $K_\perp$-dependence
can be recast in the approximate expression (\ref{3.56}).
The experimental data exclude large negative values for 
$R_0^2(K_\perp)$ and thus rule out certain opaque emission
functions~\cite{TH97,WTH97,TH98}.

\subsection{Azimuthal dependence of HBT radius parameters}
\label{sec5c4}

Both the calculation of HBT-radius parameters in the saddle-point
approximation and the numerical calculation have been 
extended~\cite{W97} to the finite impact parameter model described 
in (\ref{5.12})-(\ref{5.16}). For this model the first
harmonics vanish, and the zeroth and second harmonics 
can be written in the saddle point approximation in terms of
an average size $\bar{R}$ and a dimensionless
anisotropy parameter $\tilde{\alpha}_2$:
%
\begin{figure}[ht]\epsfxsize=10cm 
\centerline{\epsfbox{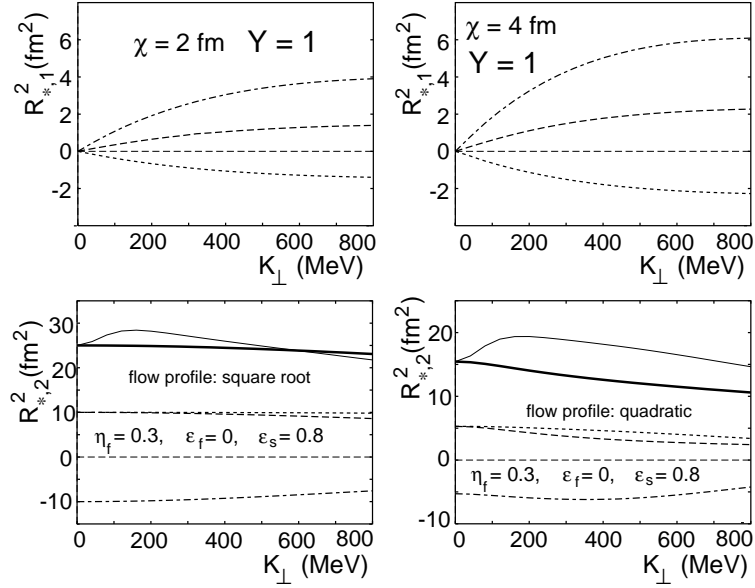}}
\caption{
First (upper panel, $R_{*,1}^2$) and second (lower panel, $R_{*,2}^2$) 
harmonic coefficients of the transverse  HBT radius parameters of 
the model (\protect\ref{5.18})/(\protect\ref{5.19}) 
and the model (\protect\ref{5.14})/(\protect\ref{5.15}),
respectively. The subscript ``$*$'' in $R_{*,1}^2$, $R_{*,2}^2$
stands for the out (dash-dotted), 
side (dashed) and out-side (dotted) components. Solid thin
and thick lines denote the zeroth harmonics of the out and side
radius parameters, respectively.
All calculations are for $T = 150$ MeV, $\tau_0 = 5$ fm/c, 
$\Delta\tau = 1$ fm/c,  $\Delta\eta = 1.2$ and $R = 5$ fm.
}\label{fig6}
\end{figure}
%
  \begin{eqnarray}
    {R_{o,0}}^2 &=& \bar{R}^2 
           + \beta_\perp^2 \langle \tilde{t}^2\rangle\, , \qquad
    {R_{s,0}}^2 = \bar{R}^2\, ,
    \label{5.50} \\
    {R_{o,2}^c}^2 &=& - {R_{s,2}^c}^2 = - {R_{os,2}^s}^2
                   = \tilde{\alpha}_2 \bar{R}^2\, . 
    \label{5.51}
  \end{eqnarray}
$\tilde{\alpha}_2$ is related to the parameter $\alpha_2$ in
(\ref{3.30}) by $\alpha_2 = \tilde{\alpha}_2 \bar{R}^2$. 
According to (\ref{5.51}) the relations (\ref{3.30}) are
exact in the saddle point approximation, with
  \begin{eqnarray}
  \bar{R}^2 &=& { R^2 \left(
       {1 + \textstyle{M_\perp\over T}\eta_f^2\, (1-\epsilon_s^2)}\right)
               \over
               {1 + 2\textstyle{M_\perp
                     (1-\epsilon_s\epsilon_f) \over T}\eta_f^2 
                  + \textstyle{M_\perp^2(1-\epsilon_s^2)\, (1-\epsilon_f^2)
                    \over T^2}\eta_f^4} }\, ,
    \label{5.52}\\
    \tilde{\alpha}_2 &=& -{\epsilon_s\over 2} { 
      { 1 + \textstyle{\epsilon_f\over\epsilon_s}\, 
            \textstyle{M_\perp\over T}\eta_f^2\, 
                   (1-\epsilon_s^2)}
               \over
               {1 + \textstyle{M_\perp \over T}\eta_f^2 
                  (1-\epsilon_s^2)}}\, .
    \label{5.53} 
  \end{eqnarray}
For vanishing anisotropy $\epsilon_s = \epsilon_f = 0$
the side radius parameter in (\ref{5.52}) reduces to the
corresponding expression (\ref{5.40}). For non-zero 
$\epsilon_s$ or $\epsilon_f$ the average size $\bar{R}$ depends
only weakly on these anisotropies.

In Figure~\ref{fig6} we show the results of a numerical study of 
the model (\ref{5.14}/\ref{5.15}) which qualitatively confirms 
the results of the saddle-point approximation. 
The deviations from the leading dependence 
${R_{o,2}^c}^2\, :\, {R_{s,2}^c}^2\, :\, - {R_{os,2}^s}^2$
$= 1\, :\, -1\, :\, -1$ of (\ref{5.51}) are seen to be small.
Similarly, the numerical study of the model
(\ref{5.18}/\ref{5.19}) which includes directed transverse flow
in the forward rapidity region allows to confirm the relation 
(\ref{3.29}) between the first harmonic coefficients,
${R_{o,1}^c}^2\, :\, {R_{s,1}^c}^2 \, :\,{R_{os,1}^s}^2$ 
$= 3\, :\,1\, :\,-1$.

\subsection{Resonance decay contributions}
\label{sec5c5}

\begin{figure}[ht]\epsfxsize=12cm 
\centerline{\epsfbox{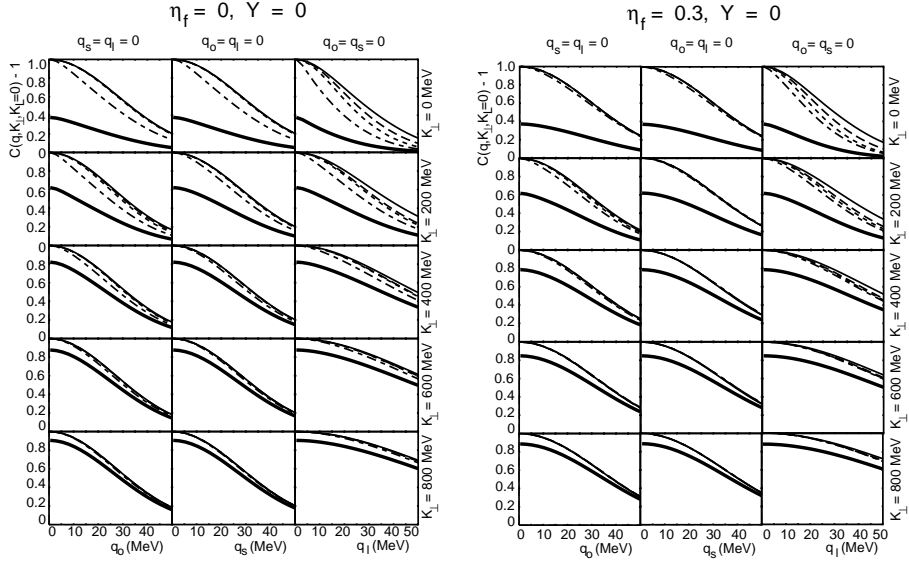}}
\caption{
Two-pion correlations for the model (\ref{5.1}) without transverse 
flow ($\eta_f = 0$), calculated according to (\ref{5.20}).
Curves show the correlator without resonance contributions (thin
solid lines), including pions from $\rho$ decays (long-dashed), 
other shortlived resonances $\Delta$, $K^*$, $\Sigma^*$ (short-dashed), 
from the $\omega$ (dash-dotted), and including pions from the longlived 
resonances $\eta$, $\eta'$, $K_S^0$, $\Sigma$, $\Lambda$ (thick solid 
lines). 
}\label{fig7}
\end{figure}

\begin{figure}[ht]\epsfxsize=8cm 
\centerline{\epsfbox{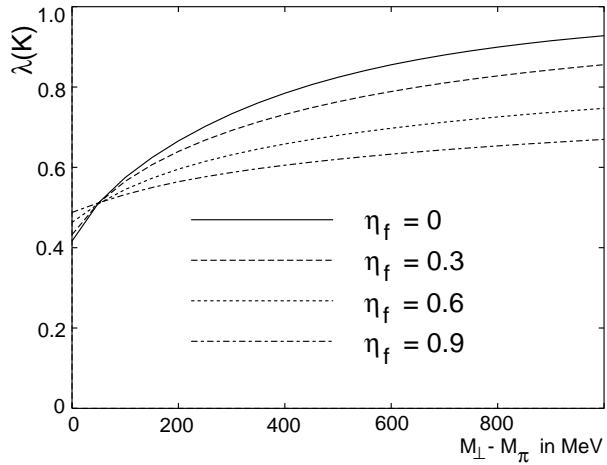}}
\caption{
The intercept parameter as a function of $M_\perp$ for 
different transverse flow strengths $\eta_f$. We use the 
model (\ref{5.1}) with temperature $T=150 MeV$.  
The calculation is not based on a proper fit of the two-particle
correlation function, but on (\ref{5.55}),
including in the sum over resonances the contributions from 
$\omega$, $\eta$, $\eta'$, and $K_S^0$. 
}\label{fig7b}
\end{figure}

Resonance decay pions affect the two-particle correlator 
by reducing its intercept $\lambda(\bbox{K})$ and by changing its 
$\bbox{q}$-dependence. Typical examples are
shown in Fig.~\ref{fig7}. The modifications of $C(\bbox{q},\bbox{K})$
due to resonance decays are a consequence of the exponential decay
law in (\ref{5.28}), which provides the emission function $S_{r\to\pi}$
with a non-Gaussian tail in coordinate space.
%
%
The latter is reflected in a non-Gaussian
shape of the correlator. The details depend on the lifetime of the
corresponding parent resonances~\cite{WH96a}: 
 \begin{itemize}
 \item
  {\it Short-lived resonances, $\Gamma > 30$} MeV: 
  In the rest frame of the particle emitting fluid element these 
  resonances decay very close to their production point, especially
  if they are heavy and have only small thermal velocities.
  This means that the emission function $S_{r\to\pi}$ of the daughter
  pions has a very similar spatial structure as that of the parent 
  resonance, $S_r^{\rm dir}$, although at a shifted momentum and shifted
  in time by the lifetime of the resonance. 
  Since the Fourier transform of the direct emission function is rather 
  Gaussian and the decay pions from short-lived resonances appear close  
  to the emission point of the parent, they maintain the Gaussian features of 
  the correlator.
 \item
  {\it Long-lived resonances, $\Gamma \ll 1$} MeV: 
  These are mainly the $\eta$ and $\eta'$, with lifetimes $c\tau\approx
  17.000$ and $1000$ fm, respectively, and the weak decays of $K_S^0$ and 
  the hyperons which on average propagate several cm. 
  Even with 
  thermal velocities these particles travel far outside the direct 
  emission region before decaying, generating a daughter pion emission 
  function $S_{r\to\pi}$ with a very large spatial support. The Fourier 
  transform $\tilde{S}_{r\to\pi}(q,K)$ thus decays very rapidly for
  $q\ne 0$, giving no contribution in the experimentally accessible 
  region $q >1$ MeV. (This lower limit in $q$ arises from the finite 
  two-track resolution in the experiments.) The decay pions do, however,
  contribute to the single particle spectrum $\tilde{S}_{r\to\pi}(q=0,K)$
  in the denominator and thus ``dilute'' the correlation. In this way
  long-lived resonances decrease the correlation strength $\lambda$
  without affecting the shape of the correlator where it can 
  be measured.
 \item
  {\it Moderately long-lived resonances,} $1$ MeV $< \Gamma < 30$ MeV. 
  There is only one such resonance, the $\omega$ meson. It is not 
  sufficiently long-lived to escape detection in the correlator, and 
  thus it does not affect the intercept parameter $\lambda$. Its 
  lifetime is, however, long enough to cause a long exponential tail in
  $S_{\omega\to\pi}(x,K)$. This seriously distorts the shape of the 
  correlator and destroys its Gaussian form.
 \end{itemize}
The main effects of resonance decay contributions on the 
$\lambda$ intercept parameter are mimicked by the 
``core-halo-model''~\cite{CLZ96} which assumes that the
emission function can be written as a sum of two contributions
  \begin{equation}
    S(x,K) = S_c(x,K) + S_h(x,K)\, .
    \label{5.54}
  \end{equation}
Here, the ``halo'' source function $S_h$ is regarded as the sum
over the longlived resonance contributions which are wide enough
to be unresolvable by HBT measurements. $S_h$ thus affects only the 
intercept parameter. The ``core'' emission function describes the 
contributions
from the direct pions and shortlived resonance decay pions which
are emitted from the same central region. The model (\ref{5.54})
is thus a simplified version of (\ref{5.20}) and provides a
simple qualitative picture for the intercept:
   \begin{equation}
     \label{5.55}
     \lambda(K) \approx \left( 
     1 - \sum_{r = {\rm longlived}} f_r(K) \right)^2\, .
   \end{equation}
The core-halo model neglects contributions from 
moderately longlived resonances, essentially the $\omega$,
for which the distinction into
core and halo does not apply. Due to their non-Gaussian
shape these affect the fit parameters for the intercept $\lambda$
considerably and can lead to quantitative corrections of (\ref{5.55})
of up to 10 \%~\cite{WH96a}. An important consequence following
already from (\ref{5.55}) is the transverse flow dependence of the
intercept parameter, depicted in Figure~\ref{fig7b}. With 
increasing transverse flow, the $K_\perp$-dependence of the
$\lambda$-parameter becomes flatter~\cite{VCK98}, though it
does not vanish (see Fig.~\ref{fig7b}). This is important since 
current measurements are consistent with a $K_\perp$-independent intercept 
parameter~\cite{Beker95,appels,schoen}. 
%

Aside from this (model-independent) lifetime effect, which generically
increases the effective pion emission region, various 
model-dependent features can affect the degree to which resonance
decay contributions change the shape of the correlator. In the
model (\ref{5.1}) for example, the size of the effective emission region 
in the transverse plane shrinks with increasing transverse flow $\eta_f$ 
approximately according to (\ref{5.40}). The simplified core-halo
model~\cite{CLZ96} neglects the proper resonance decay kinematics and thus 
does not describe this effect.
As a consequence, the direct resonance emission function $S_r^{\rm dir}$
in (\ref{5.1}) has a smaller effective emission region than that
of the thermal pions, due to the larger transverse mass $M_\perp$ of
the parent resonance (``transverse flow effect''). For finite transverse
flow this reduces the size of the resonance emission 
region and counteracts the lifetime effect. Other models of heavy 
ion collisions~\cite{SX96,S97} do not show this behaviour
and lead to significantly different $\bbox{q}$-dependences of the correlator. 
%
\subsection{Kurtosis of the correlator}
\label{sec5c6}

If, as in the presence of resonance decays, the correlator deviates
from a Gaussian shape, the
characterization of $C(\bbox{q},\bbox{K})$ via HBT radius parameters
is not unambiguous. Fit results then depend on the relative
momentum region covered by the data, on the statistical
weights of the different $\bbox{q}$-bins, and on details of the
fitting procedure. 

\begin{figure}[ht]\epsfxsize=12cm 
\centerline{\epsfbox{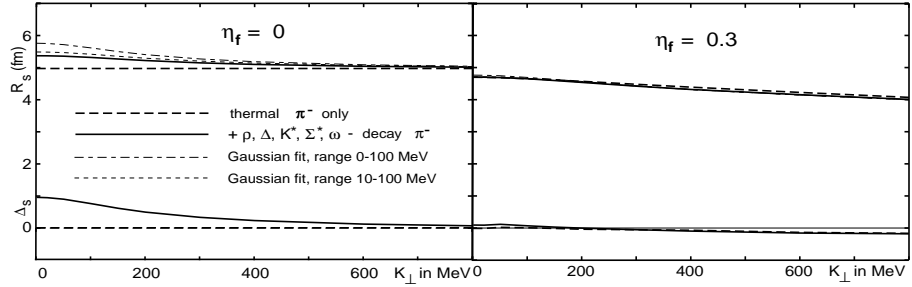}}
\caption{
The inverted uni-directional second $q$-variance $R_s$ of (\protect\ref{4.13})
and the kurtosis $\Delta_s$ of (\protect\ref{4.16}) as a function
of $K_\perp$ at mid rapidity for the model (\protect\ref{5.1}) 
with $T = 150$ MeV, $R = 5$ fm, $\Delta\eta = 1.2$, $\tau_0 = 5$ 
fm/c, $\Delta\tau = 1$ fm/c and vanishing chemical potentials. 
Left: $\eta_f = 0$ (no transverse flow). Right: $\eta_f = 0.3$.
The difference between the dashed and solid curves is entirely
dominated by $\omega$-decays.
}\label{fig8}
\end{figure}

The $q$-moments discussed in section~\ref{sec4b} allow to quantify
deviations of the correlator from a Gaussian shape~\cite{WH96b,lasiuk}. 
They are sensitive to differences 
between model scenarios which cannot be 
distinguished on the basis of Gaussian radius parameters.
To illustrate this point we show in Figure~\ref{fig8} both 
the side radius parameter $R_s(\bbox{K}_\perp)$ and the corresponding
one-dimensional kurtosis $\Delta_s(\bbox{K}_\perp)$, calculated
according to (\ref{4.13}) and (\ref{4.16}). In the present case,
the side radius parameter calculated from the inverted second
$q$-moment coincides rather accurately with the radius parameters 
extracted from a fit to
  \begin{equation}
    \label{5.56}
   \tilde C(q_i,\bbox{K}) = 1 \, + \, \lambda\, e^{- R_i^2 q_i^2}\, ,
   \qquad\, i = o,s,l
  \end{equation}
in the range $q_i \leq 100$ MeV. For a transverse flow between
$\eta_f = 0$ and $\eta_f = 0.3$, $R_s$ shows approximately the 
same $K_\perp$-slope. Hence, once resonance decays are taken into 
account, scenarios with and without transverse flow cannot be 
distinguished unambiguously on the basis 
of (\ref{5.40}). The physical origin of the $K_\perp$-slope of
$R_s$ is, however, different in the two situations, and this shows up 
in the kurtosis of the correlator. Without transverse flow,
resonance decay contributions increase $R_s$ due to the lifetime effect. 
For non-zero transverse flow, on the other hand, the $K_\perp$-slope 
arises from the $K_\perp$-dependent shrinking (\ref{5.40}) of the 
effective transverse emission region which is more prominent for
resonances than for thermal pions. $S_\pi^{\rm dir}$ is spatially more 
extended in the transverse plane than $S_R^{\rm dir}$, and thus ``covers'' 
a substantial part of the exponential tails of $S_{R\to\pi}$~\cite{WH96a}. 
As a consequence, the total emission function (\ref{5.20})  
shows much smaller deviations from a Gaussian shape for the scenario 
with transverse flow and results in a more Gaussian correlator.
This explains why the kurtosis $\Delta_s(K_\perp)$ plotted in 
Fig.~\ref{fig8} provides a clearcut distinction between the two 
scenarios.
%
%
\section{Analysis strategies for reconstructing the source in 
heavy-ion collisions}\label{sec5d}

A realistic emission function of heavy-ion collisions should 
simultaneously reproduce the spectra and particle yields of all observable 
particle species. Every model implies certain constraints between
all these observables.
For instance, in the model (\ref{5.1}) the assumption of 
resonance production with thermal abundances inside the {\it same} 
space-time geometry relates the production of different particle 
species. While this may be sufficient to model gross properties 
of particle production, extensions involving additional model 
parameters may be needed to account for finer details (for example,
partial strangeness saturation or rapidity dependent chemical 
potentials~\cite{SSH95}). In this review, we will stay with a
simple model and try to describe only the rough features of the
freeze-out process. We will restrict our discussion to the single-particle
spectra of negatively charged particles and to two-pion correlations,
setting all chemical potentials to zero. We aim to extract from the data the 
phase-space properties of the pion production region, characterized
by the model parameters in (\ref{5.1}).
%
\begin{figure}[ht]\epsfxsize=13.5cm 
\centerline{\epsfbox{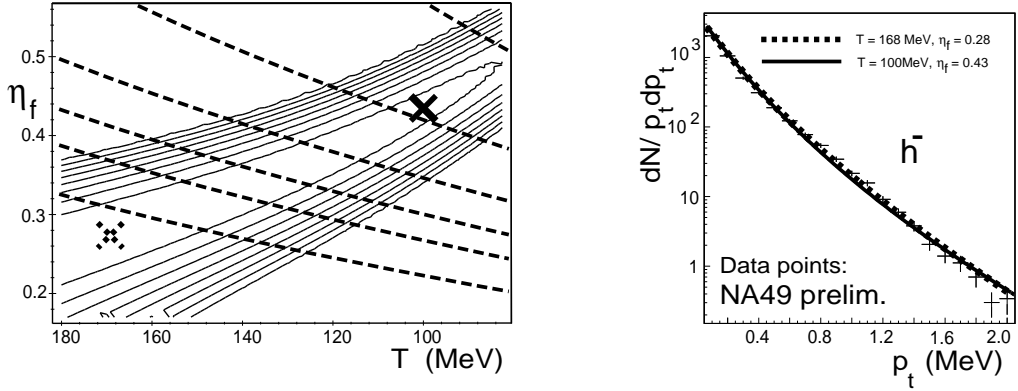}}
\caption{LHS: $\chi^2$ contour plot of a fit to the NA49 
$h^-$-spectrum~\protect\cite{J96}.
Dashed lines are for constant values of $\eta_f^2/T$. RHS: different
combinations of temperature $T$ and transverse flow $\eta_f$ can account
for the same one-particle slope.
}\label{fig9}
\end{figure}
%
\subsection{Determining the model parameters of analytical emission
functions}
\label{sec5d1}

The model parameters of an analytical emission function
should be determined by a multi-parameter fit of the corresponding 
one- and two-particle spectra (\ref{1.3})-(\ref{1.5}) to the data. 
For sufficiently complicated emission functions, where no accurate
analytical approximations of (\ref{1.3})-(\ref{1.5})
are available, this is a significant numerical task. 
In a model like (\ref{5.1}), however,
certain model parameters are almost exclusively determined by 
particular properties of the measured particle spectra. Based on
this observation we outline here a strategy for the comparison of
(\ref{5.1}) to data, which uses heavily the model studies presented
in sections~\ref{sec5b} and ~\ref{sec5c}. It allows to determine 
the model parameters $T$, $\eta_f$, $R$, $\Delta\eta$, $\Delta\tau$, 
$\tau_0$ by a significantly simpler method, according to the following
steps~\cite{WTH97}:
  \begin{enumerate}
    \item
      {\it The transverse single-pion spectrum $dN/dm_{\perp}^2$ determines
       the blue-shif\-ted effective temperature $T_{\rm eff}$.}\\
      For pions with $T \approx m_\pi$, the slope of $dN/dm_{\perp}^2$ 
      is essentially given by
      $T_{\rm eff} = T \sqrt{ (1 + \langle \beta_t\rangle)/ 
      (1 - \langle \beta_t\rangle)}$,
      as argued in section~\ref{sec5b1} and Fig.~\ref{fig2}.
      Resonance decay contributions 
      affect the local slope of $dN/dm_{\perp}^2$ and thereby the
      fit parameter~\cite{SSH93}. They have to be properly taken into account.
      Then a fit to the corresponding one-particle spectrum determines
      a ``valley'' of parameter pairs $T$ and $\eta_f$ 
      (related to $\langle\beta_t\rangle$) all of which can 
      account for the same data. This is clearly seen in the 
      $\chi^2$-plot of a fit to recent NA49 $h^-$-spectra~\cite{J96}, 
      presented in Fig.~\ref{fig9}. 
    \item
      {\it Combining the single-particle spectrum $dN/dm_{\perp}^2$ 
        \underline{and} the transverse HBT radius parameter 
        $R_s(M_{\perp}) = R_{\perp}(M_{\perp})$ 
        disentangles temperature $T$ and transverse flow $\eta_f$. 
        $R_{\perp}$ then fixes the transverse extension $R$.}\\
      In the saddle-point approximation (\ref{5.40}) the 
      $M_{\perp}$-slope of the transverse radius $R_{\perp}$ 
      is proportional to $\eta_f^2/T$. In Fig.~\ref{fig9}
      we have superimposed lines of constant $\eta_f^2/T$ onto
      the fit results for $dN/dm_{\perp}^2$. Due to the different
      correlation between $T$ and $\eta_f$, the additional 
      information provided by $R_{\perp}(M_{\perp})$ 
      allows to disentangle temperature $T$ and transverse
      flow effects~\cite{schoen}. Once the transverse flow is fixed, the
      overall size of $R_{\perp}(M_{\perp})$ determines the
      Gaussian width $R$ of the source according to (\ref{5.40}).
      Quantitative details change if the saddle-point approximation 
      is abandoned, but the qualitative argument survives. In a 
      numerical calculation including resonance decay 
      contributions~\cite{WTH97} 
      one extracts then from the combination 
      of $R_{\perp}(M_{\perp})$ and $dN/dm_{\perp}^2$ the
      following values for the model parameters, see Fig.~\ref{fig10}:
      $\eta_f \approx 0.35$, $T \approx 130$ MeV, $R \approx 7$ fm.
    \item
      {\it The single-particle rapidity distribution $dN/d{\rm y}$ fixes
      the longitudinal source extension $\Delta\eta$.} \\
      The single-particle rapidity distribution (\ref{5.33})
      is determined by $\Delta\eta$, the only parameter which
      breaks the longitudinal boost-invariance of the source
      (\ref{5.1}). Fig.~\ref{fig9b} shows
      that the $h^-$ rapidity spectrum is a Gaussian with a width
      of 1.4 rapidity units. In the parametrization of the present
      model, this translates into $\Delta\eta = 1.2$.
%
\begin{figure}[ht]\epsfxsize=7.0cm 
\centerline{\epsfbox{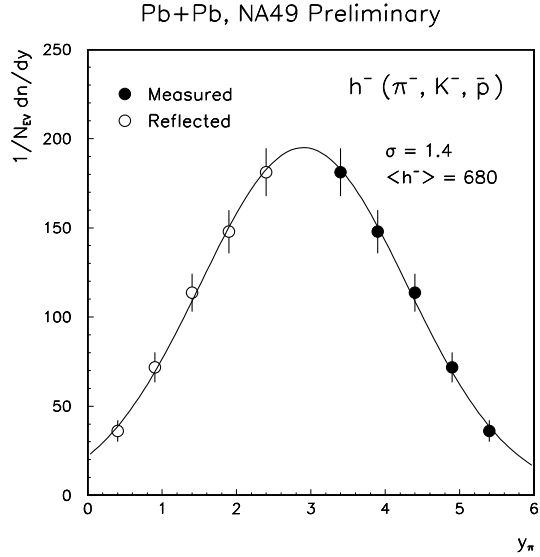}}
\caption{The rapidity distribution of primary negative hadrons. 
The preliminary NA49 Pb+Pb data are taken from ~\protect\cite{J96}.
}\label{fig9b}
\end{figure}
    \item
      {\it $R_{\parallel}$ determines $\tau_0$.} \\
      In principle, $R_{\parallel}$ and $R_l$
      depends on $\tau_0$, $\Delta\eta$ and $\Delta\tau$~\cite{WSH96}. 
      Model calculations~\cite{TH98} show, however,  
      that for (\ref{5.1}) the dependence on the emission 
      duration $\Delta\tau$ is weak. As argued above, $\Delta\eta$
      can be fixed from the single-particle rapidity distribution.
      The data presented in Fig.~\ref{fig10}b then clearly favour a
      value of $\tau_0 \approx 9$ fm/c. In this plot $\Delta\tau$
      was chosen to $\Delta\tau = 1.5$ fm/c. From the arguments given
      at the end of section~\ref{sec5c1}, this value of $\tau_0$ 
      is likely to provide
      a lower estimate for the total lifetime of the collision
      region.
    \item
      {\it $R_0$ discards opaque sources.}\\
      For the model (\ref{5.1}) the YKP-parameter $R_0$ is
      mainly sensitive to the mean emission duration of the source, 
      since for this model the approximation (\ref{3.55}) is
      satisfied. The large statistical
      uncertainties of the NA49 data for $R_0$ do not allow to 
      constrain the model parameter space further, see 
      Fig.~\ref{fig10}c. Certain models of opaque 
      sources, however, which include the opacity factor (\ref{5.10}),
      lead according to (\ref{3.56}) to a negative radius parameter 
      $R_0^2$, and can be  
      excluded~\cite{WTH97,TH98} already by the present data.
  \end{enumerate}    
%
\begin{figure}[ht]\epsfxsize=13.5cm 
\centerline{\epsfbox{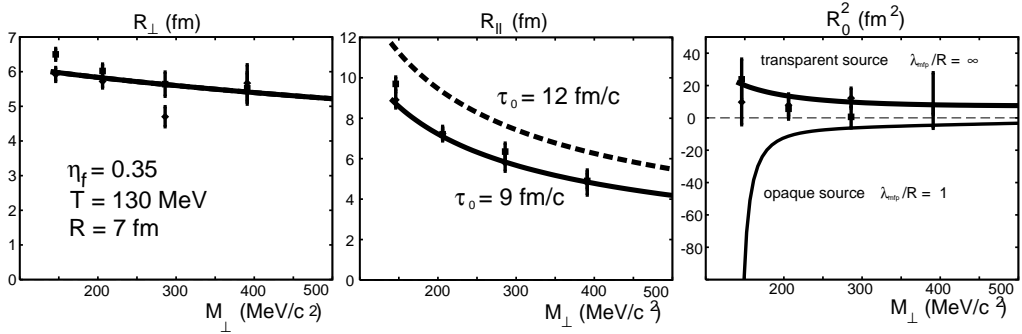}}
\vspace{0.5cm}
\caption{ Yano-Koonin-Podgoretski\u\i\ HBT-radius parameters.
The figure uses preliminary NA49 Pb+Pb data for $h^+h^+$ (squares) and
$h^-h^-$ (diamonds) correlations~\protect\cite{appels}. Final data have 
since been published in~\protect\cite{NA49corr}.
}\label{fig10}
\end{figure}
%
\subsection{Uncertainties in the reconstruction program}
\label{sec5d2}

We now list some sources of uncertainties in the 
reconstruction program described above:
  \begin{enumerate}
   \item
   {\it Particle Identification:}\\ The above analysis is based
   on $h^-$ (all negative hadrons) and $h^+$ (all positive 
   hadrons) spectra and correlations. These are
   dominated by the corresponding charged pion contributions.
   The effect of other particle species (e.g. kaons or (anti)protons)
   on the one-particle spectra can be included in (\ref{5.29})
   and has a negligible effect on the shape of the fit (which is
   what matters) shown in Fig.~\ref{fig2}.
   For the two-particle correlations misidentified particles lead
   to the counting of ``wrong pairs'' which do not show Bose-Einstein 
   correlations. This decreases the intercept parameter whose value
   was, however, not used in the reconstruction program. There are no 
   indications that the problem of particle misidentification
   affects the $K_\perp$-dependence of the HBT parameters significantly.
   Nevertheless, it would be preferable to use spectra of identified 
   particles. This would additionally allow to compare different 
   effective source sizes e.g. from pion-pion and kaon-kaon 
   correlations and thereby investigate the question whether all
   particle species are emitted from the same source volue. Such identified
   two-particle correlations were measured at the
   AGS~\cite{M96} and by the NA44 Collaboration at the CERN
   SPS~\cite{Fr96}. Both of these data sets are, however, restricted
   to narrow regions in $K_\perp$ and $Y$ which limits their usefulness
   for the above reconstruction program. If the different acceptance of 
   the two experiments is properly taken into account, the 
   $\pi^-\, \pi^-$-correlations from NA44 and the 
   $h^-\, h^-$-correlations from NA49 are compatible within
   statistical errors (P. Seyboth and J.P. Sullivan,
   private communication). 
   \item
   {\it Coulomb Corrections:}\\
   Coulomb corrections affect the size of the HBT radius parameters 
   as well as their 
   $K_\perp$-dependence~\cite{appels,schoen,lasiuk}. For example, the
   difference $R_o^2-R_s^2$ in NA35 correlation data was found 
   to be non-zero only after changing from a naive Gamow
   correction based on (\ref{2.99}) to an effective correction
   (\ref{2.100}) which takes the finite source size into 
   account~\cite{NA35newcoulomb}. Furthermore, in the NA49 experiment
   a proper treatment of the Coulomb correction proved essential for
   a successful check of the consistency relations (\ref{3.46})-(\ref{3.52})
   between the Cartesian and YKP HBT parameters~\cite{appels}.
   \item
   {\it Uncertainties affecting the $K_\perp$-dependence of
    HBT radius parameters:}\\
   Particle misidentification and Coulomb correction are not the
   only possible sources of uncertainty which may affect the momentum 
   slope of HBT radius parameters. Other sources on the
   theoretical side are for example
   multiparticle symmetrization effects~\cite{CZ97,W98}, 
   the dependence on a central Coulomb charge~\cite{B96}, the
   mixing of anisotropy effects into the HBT radius parameters
   resulting from collisions with non-zero impact parameter in
   the central collision sample~\cite{VZ96,W97}, or
   multiparticle final state interactions. On the experimental
   side, residual correlations in the mixed event sample used 
   to normalize the two-particle correlator or the shape of the
   experimental acceptance in the $K_\perp$-$Y$-plane can affect
   the presented $K_\perp$-dependence of HBT-radius parameters.
  \end{enumerate} 
One usually argues that these uncertainties are small  
for the size and $K_\perp$-dependence of the HBT-radius parameters. 
However, already small differences in the transverse slope 
of $R_\perp$, for example, affect significantly the optimal 
combination of fit parameters $(T,\eta_f)$ in Fig.~\ref{fig9}a. 
We therefore consider the systematical error in the slope of 
$R_\perp(K_\perp)$ to be
the most important uncertainty in the reconstruction program
of section~\ref{sec5d1}.
%
\subsection{Dynamical interpretation of model parameters}
\label{sec5d3}

Pions, as most other hadrons, rescatter during the expansion stage of a 
heavy-ion collision. Their phase-space distribution $S_{\pi}(x,p)$ 
characterizes the geometrical and dynamical properties of the final
freeze-out stage after their last strong interaction. The emission
function does {\it not} contain direct information about the hot 
and dense earlier stages of the collision. 
However, the emission function, as reconstructed from the spectra
and correlation data, provides an experimentally justified
starting point for a dynamical extrapolation back towards the 
earlier stages. To illustrate this point, we discuss here 
the values of the geometrical and dynamical parameters, 
extracted for the model (\ref{5.1}) from preliminary
NA49 data:
  \begin{eqnarray}
    R &\approx& 7\, \, \hbox{fm}\, ,
    \label{5.57}\\
    T &\approx& 130\, \, \hbox{MeV}\, ,
    \label{5.58}\\
    \eta_f &\approx& 0.35\, ,
    \label{5.59}\\
    \tau_0 &\approx& 9\, \, \hbox{fm/}c\, ,
    \label{5.60}\\
    \Delta\eta &\approx& 1.3\, ,
    \label{5.61}\\
    \Delta\tau &\approx& 1.5\, \, \hbox{fm/}c\, .
    \label{5.62}
  \end{eqnarray}
To obtain from these data a dynamical picture of the collision
process, we compare first the two-dimensional rms width obtained
from the transverse width $R \approx 7$ fm,
  \begin{equation}
    r_{\rm rms}^{\rm source} = 
    \sqrt{ \langle \tilde{x}^2 + \tilde{y}^2\rangle }
    = \sqrt{2}\, R \approx 10\, \, \hbox{fm}\, ,
    \label{5.63}
  \end{equation}
with the two-dimensional rms widths of a cold lead nucleus. The
nuclear hard sphere radius $R_{\rm hs} = r_0\, A^{1/3}$
with $r_0 = 1.2$ fm for lead is $R_{\rm hs}^{\rm Pb} = 7.1$ fm. 
The corresponding two-dimensional transverse rms width
  \begin{equation}
    r_{\rm rms}^{\rm cold\, Pb} = 
    \sqrt{ \langle \tilde{x}^2 + \tilde{y}^2\rangle_{\rm Pb} }
    =   \sqrt{3/5}\, \, R_{\rm hs} 
    \approx 4.5\, \, \hbox{fm}\, .
    \label{5.64}
  \end{equation}
From this we conclude that during the collision the system has
expanded by a factor $\approx 2$ from the transverse size of the 
overlapping cold lead
nuclei to the transverse extension at freeze-out. For not quite central
collisions the initial nuclear overlap region is in fact expected to 
be somewhat smaller than given by (\ref{5.64}). 
With an average transverse flow velocity of about $0.35\, c$ matter
can travel over $\approx 4$ fm in a time of $9$ fm/$c$. This is barely
enough to explain the observed expansion. This indicates that indeed,
as argued before, the parameter $\tau_0$ is a lower estimate of the
total duration of the collision. The rather low thermal freeze-out
temperature of 130 MeV (other analyses indicate even lower 
values~\cite{KPPHS97,T99}) differs significantly from the 
chemical freeze-out temperature ($\approx 170$ MeV) needed to describe
the observed particle ratios in these thermal models~\cite{BGS98},
consistent with a long expansion stage. A first attempt to extrapolate
the final state characterized by (\ref{5.57})-(\ref{5.62}) all the
way to the beginning of the transverse expansion~\cite{H98} has
led to an estimate for the average energy density at this point of 
$\epsilon \approx 2.5$ GeV/fm$^3$.

\chapter{Summary}\label{sec6}

In this work we reviewed the underlying concepts, calculational
techniques and phenomenological uses of Hanbury Brown/Twiss particle 
interferometry for relativistic heavy-ion collisions. Compared to the
astrophysical applications of HBT interferometry, its use for  
relativistic nuclear collisions is substantially complicated by (i) 
the time-dependence and short lifetime of the particle emitting source, 
(ii) the position-momentum gradients in the source resulting from the strong 
dynamical expansion of the collision region, and (iii) other 
dynamical origins of particle momentum correlations which have to be 
subtracted properly to make a space-time interpretation of the 
measured correlation data possible.

The Wigner phase-space density (``emission function'') $S(x,K)$, 
interpreted as the probability that a particle with momentum $K$ 
is emitted from a space-time point $x$ in the collision 
region, provides the appropriate starting point for the analysis 
of measured HBT correlations from relativistic heavy
ion collisions. It accounts for the time-dependence 
and the position-momentum gradients of the source. Moreover,
other contributions to the momentum correlations between pairs
of identical particles, for example from
final state Coulomb interactions, multiparticle 
symmetrization effects, or resonance decay contributions, can be 
calculated once the emission function is given. In chapter~\ref{sec2} 
we discussed this in detail, after deriving the basic relation 
(\ref{1.4}) between the phase-space emission function $S(x,K)$ and 
the measured two-particle momentum correlation $C(\bbox{q},\bbox{K})$.

The key to a geometric and dynamical understanding of the measured 
two-particle correlations are the model-independent relations 
between the space-time variances (Gaussian widths) of the emission 
function and the HBT radius parameters which are extracted from Gaussian 
fits (\ref{1.5}) to the two-particle correlator. These were
derived in chapter~\ref{sec3}. In general, the HBT radius 
parameters $R_{ij}^2(\bbox{K})$ do not measure the total geometric 
source size, but the size of regions of homogeneity 
in the source from which most particles with momentum $\bbox{K}$ are 
emitted. These homogeneity regions typically decrease in the presence
of temperature or flow velocity gradients which lead to characteristic
position-momentum correlations in the source. Generically, the
homogeneity regions are smaller for pair with larger transverse 
momentum. The $\bbox{K}$-dependence of the
two-particle correlator thus gives access to dynamical characteristics
of the collision region. 

Expressing the HBT radius parameters
in terms of space-time variances also shows explicitly how spatial and 
temporal information about the source is mixed in the measured 
momentum correlations. We explained this first for the 
Cartesian parametrization of $C(\bbox{q},\bbox{K})$ and its extension 
to collisions at non-zero impact parameter. We then introduced
the YKP parametrization, an alternative Gaussian parametrization
of $C(\bbox{q},\bbox{K})$ which is particularly well adapted for 
collision systems with strong longitudinal expansion:
(i) Three of the four YKP fit parameters are invariant under 
longitudinal Lorentz boosts, i.e. their value do not depend
on the observer frame. (ii) The fourth
fit parameter is the Yano-Koonin velocity which measures the 
longitudinal velocity of the particle emitting source element.
Its rapidity dependence allows to determine
the strength of the longitudinal expansion in the collision region.
(iii) In the particular observer frame in which the Yano-Koonin
velocity vanishes, the YKP radius parameters for a large class
of emission functions cleanly separate the longitudinal, transverse 
and temporal aspects of the source.
This considerably simplifies the space-time interpretation of the Gaussian
fit parameters considerably.

An important aspect of HBT interferometry is that from a combined 
analysis of the single-particle spectra and HBT correlation radii
an estimate of the average phase-space density of the source at
freeze-out can be obtained. Applied to heavy-ion data at the 
AGS and SPS, this method provided evidence for a universal 
freeze-out phase-space density for pions. Its transverse momentum
dependence is in rough agreement with expectations based on models
assuming thermalization prior to freeze-out.

The connection between HBT radii and space-time
variances of the emission function is based on Gaussian
parametrizations of the source and the measured particle correlations. 
Deviations of the two-particle correlator from a Gaussian shape
can contain additional space-time information which is not contained 
in the HBT radii. In chapter~\ref{sec4} we described
refined techniques which give access to such 
additional information. The imaging method and the method
of $q$-moments discussed in sections~\ref{sec4a} and~\ref{sec4b}
characterize in different ways the relative source function
$S_{\bbox{K}}(\bbox{r})$ which is defined as a time averaged distribution
of relative distances in the source. Identical three 
particle correlations, discussed in section~\ref{sec4c}, give 
access to odd orders of the space-time variances of $S(x,K)$ which 
drop out in identical two-particle correlations. 

The main goal of particle interferometry for relativistic
heavy-ion collisions is to extract from the measured momentum 
spectra as much information as possible 
about the emission function $S(x,K)$. 
We explained why a completely model-independent reconstruction
of $S(x,K)$ is not possible.
In practice, one therefore must take recourse to a model dependent
approach. This was illustrated in chapter~\ref{sec5} for a class of 
analytical emission functions. Comprehensive model studies allowed
to separate the generic from the more model-dependent features. 
Simple approximate expressions for the HBT radius parameters were
given which provide a qualitative understanding of the dominant
geometrical and dynamical effects. Their quantitative accuracy
was tested by numerical means.

The size and momentum dependence of the different HBT radius
parameters and the transverse momentum slopes of the one-particle
spectra were shown to depend in general only on one or two of the
model parameters of the emission function. This allows for a simple
analysis strategy for the reconstruction of the emission function.
It was illustrated in section~\ref{sec5d} in an application to
preliminary data from the 158 $A$ GeV lead beam experiment NA49 
at the CERN SPS. The extracted source parameters are consistent with
the creation of a highly dynamical system which after impact expanded
in the transverse direction over a time of at least $9$ fm/$c$ with
approximately one third of the velocity of light before emitting
particles at a temperature of around $130$ MeV. This information about
the hadronic emission region provides a starting point for a dynamical
back extrapolation into the hot and dense early stage of the
collision, and it can be directly compared with the output of
numerical event simulations of relativistic heavy-ion
collisions. Compared to an analysis based on momentum space
information only, the additional space-time information obtained from
particle interferometry provides severe constraints for our
understanding of heavy-ion collision dynamics. 

Remaining uncertainties in the approach were discussed. Better
experimental statistics and further progress in our quantitative 
understanding of effects which influence the pair momentum dependence
of the two-particle correlator are expected to lead in the near future
to considerable improvements in the analyzing power of this method.

\subsubsection*{Acknowledgements}

This report grew out of the habilitation thesis submitted by one of us
(UAW) to the University of Regensburg in April 1998. The present
review would not have been possible without the continuous influx of
ideas resulting from fruitful discussions with many friends and
colleagues. We would like to mention explicityly 
J. Aichelin,
D. Anchishkin,
H. Appelsh\"auser,
F. Becattini,
G. Bertsch,
A. Bialas,
H. B{\o}ggild,
P. Braun-Mun\-zin\-ger,
D.A. Brown,
S. Chapman, 
J. Cramer,
T. Cs\"org\H{o},
P. Danielewicz,
J. Ellis, 
H. Feldmeier,
D. Ferenc,
P. Filip,
P. Foka, 
M. Ga\'zdzicki,
K. Geiger, 
M. Gyulassy,
H. Heiselberg,
T. Humanic,
B. Jacak,
K. Kadija, 
B. K\"ampfer,
H. Kalechofsky,
B. Lasiuk,
R. Lednick\'y,
B. L\"orstad,
M. Martin,
U. Mayer,
D. Mi\'skowiec,
B. M\"uller,
R. Nix,
S. Padula,
Y. Pang, 
T. Peitzmann,
D. Pelte,
J. Pi\v s\'ut,
M. Pl\"umer,
S. Pratt, 
P. Renk,
D. R\"ohrich,
G. Roland,
R. Scheibl,
B.R. Schlei, 
S. Sch\"onfelder,
P. Scotto, 
C. Slotta, 
P. Seyboth, 
E. Shuryak,
Yu. Sinyukov,
T. Sj\"ostrand,
S. Soff,
J. Sollfrank,
J. Stachel,
R. Stock,
B. Tom\'a\v{s}ik, 
S. Vance,
A. Vischer,
S. Voloshin,
Y.-F. Wu,
N. Xu,
W.A. Zajc,
K. Zalewski,
Q.H. Zhang and
J. Zim\'anyi.
This work was supported by DFG, GSI, BMBF and DOE (contract no.
DE-FG02-93ER40764). 


\end{document}